\documentclass[fleqn,usenatbib]{mnras}

\usepackage{amsfonts}

\usepackage[T1]{fontenc}
\usepackage{ae,aecompl}
\usepackage[position=top]{subfig}
\usepackage{float}
\usepackage{comment}

\usepackage{grffile}
\usepackage{etoolbox}
\makeatletter 
  \patchcmd{\NAT@citex}
    {\@citea\NAT@hyper@{%
      \NAT@nmfmt{\NAT@nm}%
      \hyper@natlinkbreak{\NAT@aysep\NAT@spacechar}{\@citeb\@extra@b@citeb}%
      \NAT@date}}
    {\@citea\NAT@nmfmt{\NAT@nm}%
    \NAT@aysep\NAT@spacechar\NAT@hyper@{\NAT@date}}{}{}

  \patchcmd{\NAT@citex}
    {\@citea\NAT@hyper@{%
      \NAT@nmfmt{\NAT@nm}%
      \hyper@natlinkbreak{\NAT@spacechar\NAT@@open\if*#1*\else#1\NAT@spacechar\fi}%
        {\@citeb\@extra@b@citeb}%
      \NAT@date}}
    {\@citea\NAT@nmfmt{\NAT@nm}%
    \NAT@spacechar\NAT@@open\if*#1*\else#1\NAT@spacechar\fi\NAT@hyper@{\NAT@date}}
    {}{}
\makeatother

\usepackage{etoolbox}
 \makeatother
\usepackage{graphicx}	
\usepackage{amsmath}	
\usepackage{amssymb}	
\usepackage{pifont}
\newcommand{\cmark}{\ding{51}}%
\newcommand{\xmark}{\ding{55}}%
\newcommand\HII{{H\,\textsc{ii}}} 
\newcommand{\uLW}{\mathrm{LW}}%
\newcommand{\uH}{\mathrm{H}}

\title[GWs from Pop~III-seeded BBHs]{Gravitational waves from Population~III binary black holes formed by dynamical capture}

\author[B. Liu, V. Bromm]{Boyuan Liu\textsuperscript{\href{https://orcid.org/0000-0002-4966-7450}{\includegraphics[width=2.5mm]{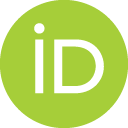}}\,}\thanks{E-mail: boyuan@utexas.edu}$^{1}$, 
and Volker Bromm$^{1}$
\\
$^{1}$Department of Astronomy, University of Texas, Austin, TX 78712, USA\\
}

\date{Accepted XXX. Received YYY; in original form ZZZ}

\pubyear{2020}

\begin{document}
\label{firstpage}
\pagerange{\pageref{firstpage}--\pageref{lastpage}}
\maketitle

\begin{abstract}
We use cosmological hydrodynamic simulations to study the gravitational wave (GW) signals from high-redshift binary black holes (BBHs) formed by dynamical capture (ex-situ formation channel). We in particular focus on BHs originating from the first generation of massive, metal-poor, so-called Population~III (Pop~III) stars. An alternative (in-situ) formation pathway arises in Pop~III binary stars, whose GW signature has been intensively studied. In our optimistic model, we predict a local GW event rate density for ex-situ BBHs (formed at $z> 4$) of $\sim 0.04\ \mathrm{yr^{-1}\ Gpc^{-3}}$. This is comparable to or even higher than the conservative predictions of the rate density for in-situ BBHs $\sim 0.01-0.1\ \mathrm{yr^{-1}\ Gpc^{-3}}$, indicating that the ex-situ formation channel may be as important as the in-situ one for producing GW events. We also evaluate the detectability of our simulated GW events for selected planned GW instruments, such as the Einstein Telescope (ET). For instance, we find the all-sky detection rate with signal-to-noise ratios above 10 to be $\lesssim 100\ \rm{yr^{-1}}$ for the xylophone configuration of ET. However, our results are highly sensitive to the sub-grid models for BBH identification and evolution, such that the GW event efficiency (rate) is reduced by a factor of 4 (20) in the pessimistic case. The ex-situ channel of Pop~III BBHs deserves further investigation with better modeling of the environments around Pop~III-seeded BHs. 
\end{abstract}



\section{Introduction}

The detection of gravitational waves (GWs) from merging compact objects, such as black holes (BHs) and neutron stars, has opened a new observational window in astrophysics, cosmology and fundamental physics (reviewed by, e.g. \citealt{barack2019black}). The statistics of GW events will place new constraints on a variety of astrophysical processes, such as cosmic star formation, the BH mass distribution, the formation and evolution of compact binaries, and cosmology (e.g. \citealt{fishbach2018does,vitale2019measuring,
perna2019constraining,safarzadeh2019measuring,farr2019future,
adhikari2020binary,tang2020dependence,safarzadeh2020branching}). 
The population of binary black holes (BBHs), as detected by the Laser Interferometer Gravitational-wave Observatory (LIGO) collaboration \citep{abbott2019gwtc}, is dominated by massive systems ($\gtrsim 10\ \mathrm{M}_{\odot}$), including binaries with inferred total masses above $\sim 40\ \mathrm{M}_{\odot}$ (e.g. GW170729 and GW170502), which indicates that they originate from massive progenitor stars over a range of redshifts. However, it remains a mystery how such massive BBHs are formed and evolved to merge across cosmic time. Numerous scenarios have been proposed (e.g. \citealt{belczynski2016first,rodriguez2016dynamical,dicarlo2019,conselice2019gravitational}), involving a variety of astrophysical processes, such as the evolution of binary stars and mergers of ultra-dwarf galaxies. 

Over the next decades, more advanced GW instruments will come into operation, including high-frequency ground-based detectors for stellar-mass BHs (SBHs), such as improved versions of LIGO 
and Virgo 
 \citep{abbott2018prospects,abbott2019search}, as well as the Kamioka Gravitational Wave Detector 
(KAGRA). Intermediate-mass BHs (IMBHs) will be targeted with the third-generation instruments, such as the Einstein Telescope 
(ET; \citealt{punturo2010einstein,gair2011imbhet}), the Cosmic Explorer 
(CE; \citealt{abbott2017exploring}), and the Decihertz-class Observatories 
(DOs; \citealt{dechihertz,tiango}). Finally, low-frequency arrays in space will probe the high-mass end of the BBH range, such as the Laser Interferometer Space Antenna 
(LISA; \citealt{robson2019construction}), and the TianQin observatory \citep{tianqin}. These facilities will cover a large range in frequency and mass ($\sim 1-10^{7}\ \mathrm{M}_{\odot}$), thus providing us a clear portrait of BH populations over cosmic time, in a wide spectrum of host systems. Combined with theoretical predictions, GW observations will be a powerful probe of early structure formation \citep{sesana2009observing,sesana2011reconstructing,
Fragione2018gw,jani2019detectability}. Furthermore, the GW window ideally complements the electromagnetic (EM) one. The latter is biased towards massive/luminous systems at high redshifts, as EM signals decay rapidly with distance ($\propto d_{L}^{-2}$). While the amplitudes of GWs decay slower with distance ($\propto d_{L}^{-1}$), and BBHs formed in the early Universe can merge at lower redshifts, reflecting their long delay times.

On the theory side, the goal is to self-consistently predict the GW signals of high-$z$ BBHs, originating from different channels of BH seeding and growth, as well as addressing the rich physics of BBH formation and evolution. A specific challenge is that physical processes on vastly different scales are involved, many of which are poorly understood. Currently, there are three main models for high-$z$ BH seeds: (i) remnants of the first generation of massive, metal-free Population~III (Pop~III) stars with seed masses $M_{\mathrm{BH}}\sim 40-140\ \mathrm{M}_{\odot}$ (e.g. \citealt{bond1984evolution,schneider2000gravitational,
madau2001massive,bromm2011first,hirano2014one}), (ii) runaway collisions in dense star clusters with $M_{\mathrm{BH}}\sim 10^{3}-10^{4}\ \mathrm{M}_{\odot}$ (e.g. \citealt{devecchi2009formation,katz2015seeding}); and (iii) rapid infall of primordial gas in peculiar environments leading to direct-collapse BHs (DCBHs) with $M_{\mathrm{BH}}\gtrsim 10^{4}\ \mathrm{M}_{\odot}$ (e.g. \citealt{BrommLoeb2003,volonteri2010formation,johnson2016early, maio2019early,smith2019supermassive,inayoshi2019}). The first two models produce relatively light seeds dominating in number, while seeds in the third class are believed to be rare, but important to explain the observed luminous quasars at high-$z$, powered by supermassive BHs (SMBHs) with masses up to $\sim 10^{9}\ \mathrm{M}_{\odot}$ (e.g. \citealt{schleicher2013massive,dunn2018dcbh,
becerra2018assembly,wise2019formation,regan2019emergence,SMBHmf}).

For high-$z$ BBHs in turn, there are two formation channels: (i) \textit{in-situ} formation from binary Pop~III stars\footnote{Binary supermassive stars can also be formed during direct collapse of primordial gas, which leads to in-situ formation of binary DCBHs (e.g. \citealt{latif2020}). However, since DCBH formation only happens in rare peculiar environments, we expect the Pop~III-seeded BBHs to dominate the in-situ channel.}, and (ii) \textit{ex-situ} formation by dynamical capture of two BHs, either born into one dense star cluster, or from two originally separate formation sites (e.g. during galaxy mergers). 
The first channel has been intensely studied before and after the first LIGO detection (e.g. 
\citealt{kinugawa2014possible,kinugawa2015detection,dvorkin2016metallicity,
hartwig2016,inayoshi2016gravitational,
belczynski2017likelihood,mapelli2019properties}). The local ($z\sim 0$) intrinsic merger rate density of in-situ BBHs is predicted to be $\sim 0.1-100\ \mathrm{yr^{-1}\ Gpc^{-3}}$, where the disagreement among different studies is dominated by uncertainties in the initial binary parameters and evolution models for Pop~III binary stars (e.g. \citealt{stacy2013constraining,kinugawa2014possible,belczynski2017likelihood}). It remains unclear whether such in-situ Pop~III-seeded BBHs contribute a significant fraction to the LIGO estimate of $9-240\ \mathrm{yr^{-1}\ Gpc^{-3}}$ \citep{abbott2019gwtc}. 

The second channel 
has been studied with semi-analytical models (e.g. \citealt{sesana2009observing,sesana2011reconstructing,dayal2019hierarchical}) in the context of cosmic structure formation. For instance, the detection rate for ET is predicted to be $\lesssim 2\ \mathrm{yr^{-1}}$ \citep{sesana2009observing} with signal-to-noise ratios (SNRs) above 6, while that for LISA (with $\mathrm{SNR}>7$ at $z>4$) is $\sim 3-5\ \mathrm{yr^{-1}}$ \citep{dayal2019hierarchical}. However, in principle the dynamical capture process can only be modelled properly with cosmological simulations, as complex dynamics of BHs embedded in gas and stars is involved, which has been shown non-trivial in previous studies (e.g. \citealt{tremmel2015off,rovskar2015orbital,
tamfal2018formation,pfister2019erratic,ogiya2019}). 

In light of this, we use high-resolution cosmological hydrodynamic simulations to study the GW signals from ex-situ BBHs formed in the early Universe. We only consider the ex-situ BBHs involving two BHs from separate formation sites (i.e. star-forming clouds for Pop~III seeded BHs), and defer the \textit{in-cluster} scenario to future studies\footnote{In the local Universe, the \textit{in-cluster} scenario is particularly relevant for globular clusters (e.g. \citealt{haster2016n,Fragione2018gw,Fragione2018tidal,rodriguez2018redshift,kremer2019post}) and nuclear star clusters (e.g. \citealt{oleary2009gwnsc,petrovich2017greatly,hoang2018black}), which lead to local merger rate densities of $n_{\mathrm{GW,GC}}\sim 1-20\ \mathrm{yr^{-1}\ Gpc^{-3}}$ and $n_{\mathrm{GW,NSC}}\sim 5-15\ \mathrm{yr^{-1}\ Gpc^{-3}}$, respectively.}. 
We particularly focus on Pop~III-seeded BHs in the mass range $M_{\mathrm{BH}}\sim 40-600\ \mathrm{M}_{\odot}$, as they dominate the number counts. Besides, previous studies have shown that such light seeds can hardly grow via accretion at high redshifts (e.g. \citealt{johnson2007aftermath,alvarez2009accretion,hirano2014one,smith2018growth}), so that it is challenging to observe them as quasars with EM signals, and GW detection may be the only available probe. A technical reason is that light seeds originate from small-scale structures (i.e. minihaloes), for which a small simulation volume ($V_{C}\sim 100\ \mathrm{Mpc}^{3}$) is sufficient to provide a valid cosmological representation of the high-$z$ Universe ($z\gtrsim 4$), so that achieving high resolution is not computationally prohibitive. This work nicely complements the existing studies on the in-situ BBH formation channel, which also predominantly involves BH seeds of similar masses.

Our simulations are equipped with customized sub-grid models for Pop~III and Population~II (Pop~II) star formation and feedback, as well as Pop~III BH seeding, accretion, dynamical friction, capture and feedback. For completeness, we also adopt a sub-grid model to identify direct-collapse black hole (DCBH) candidates, similar to that used in the \textsc{romulus} simulations \citep{tremmel2017romulus}. Any DCBH candidates in our simulations, however, may not be representative, considering our limits on volume, resolution and feedback modelling. Here, we carry out our simulations within the standard $\Lambda$CDM cosmology. It is also interesting to investigate the GW signals in alternative dark matter (DM) models, which we defer to future studies.

The paper is structured as follows. Section~\ref{s2} describes our simulation setup and sub-grid models for stars and BHs. In Section~\ref{s3}, we compare our simulation results with observational constraints in the EM window, specifically the star formation and BH accretion histories, as well as halo-stellar-BH mass scaling relations, thus justifying our sub-grid models and choice of simulation parameters. In Section~\ref{s4}, we describe our model for ex-situ binary evolution, and the resulting GW detection rates from such ex-situ BBH mergers for selected future instruments. In Section~\ref{s5}, we summarize our findings and discuss potential caveats, as well as promising directions for future work.



\begin{table*}
    \centering
    \caption{Simulation parameters. $V_{C}$ is the co-moving volume of the (target) simulation region in $\mathrm{Mpc}^{3}$. $m_{\mathrm{gas}}$, $m_{\mathrm{DM}}$ and $m_{\mathrm{\star}}$ are the masses of simulation particles for gas, dark matter (DM) and stars in $\mathrm{M}_{\odot}$. $\epsilon_{\mathrm{gas/DM}}$ and $\epsilon_{\mathrm{\star/BH}}$ are the (co-moving) gravitational softening length for gas/DM and stellar/BH particles in $h^{-1}\mathrm{kpc}$. The last column is the flag \texttt{FDBKPopII}, indicating whether Pop~II feedback is included.}
    \begin{tabular}{ccccccccc}
    \hline
        &  Run & $V_{C}$ $[\mathrm{Mpc}^{3}]$ & $m_{\mathrm{gas}}$ $[\mathrm{M}_{\odot}]$ & $m_{\mathrm{DM}}$ $[\mathrm{M}_{\odot}]$ & $m_{\mathrm{\star}}$ $[\mathrm{M}_{\odot}]$ & $\epsilon_{\mathrm{gas/DM}}$ $[h^{-1}\mathrm{kpc}]$ & $\epsilon_{\mathrm{\star/BH}}$ $[h^{-1}\mathrm{kpc}]$ & \texttt{FDBKPopII} \\
    \hline
        & \texttt{FDzoom} & 10.9 & $9.4\times 10^{3}$ & $5.2\times 10^{4}$ & 586 & 0.2 & 0.02 & \cmark \\
        & \texttt{NSFDBKzoom} & 10.9 & $9.4\times 10^{3}$ & $5.2\times 10^{4}$ & 586 & 0.2 & 0.02 & \xmark \\
        & \texttt{FDbox} & 205.9 & $9.4\times 10^{3}$ & $5.2\times 10^{4}$ & 586 & 0.2 & 0.02 & \cmark\\
   	\hline
        & \texttt{FDzoomHR} & 10.9 & $1.2\times 10^{3}$ & $6.5\times 10^{3}$ & 586 & 0.1 & 0.02 & \cmark \\
    \hline
    \end{tabular}
    \label{t1}
\end{table*}

\begin{figure*}
\hspace{-5pt}
\centering
\subfloat[Gas temperature]{\includegraphics[width= 1.065\columnwidth]{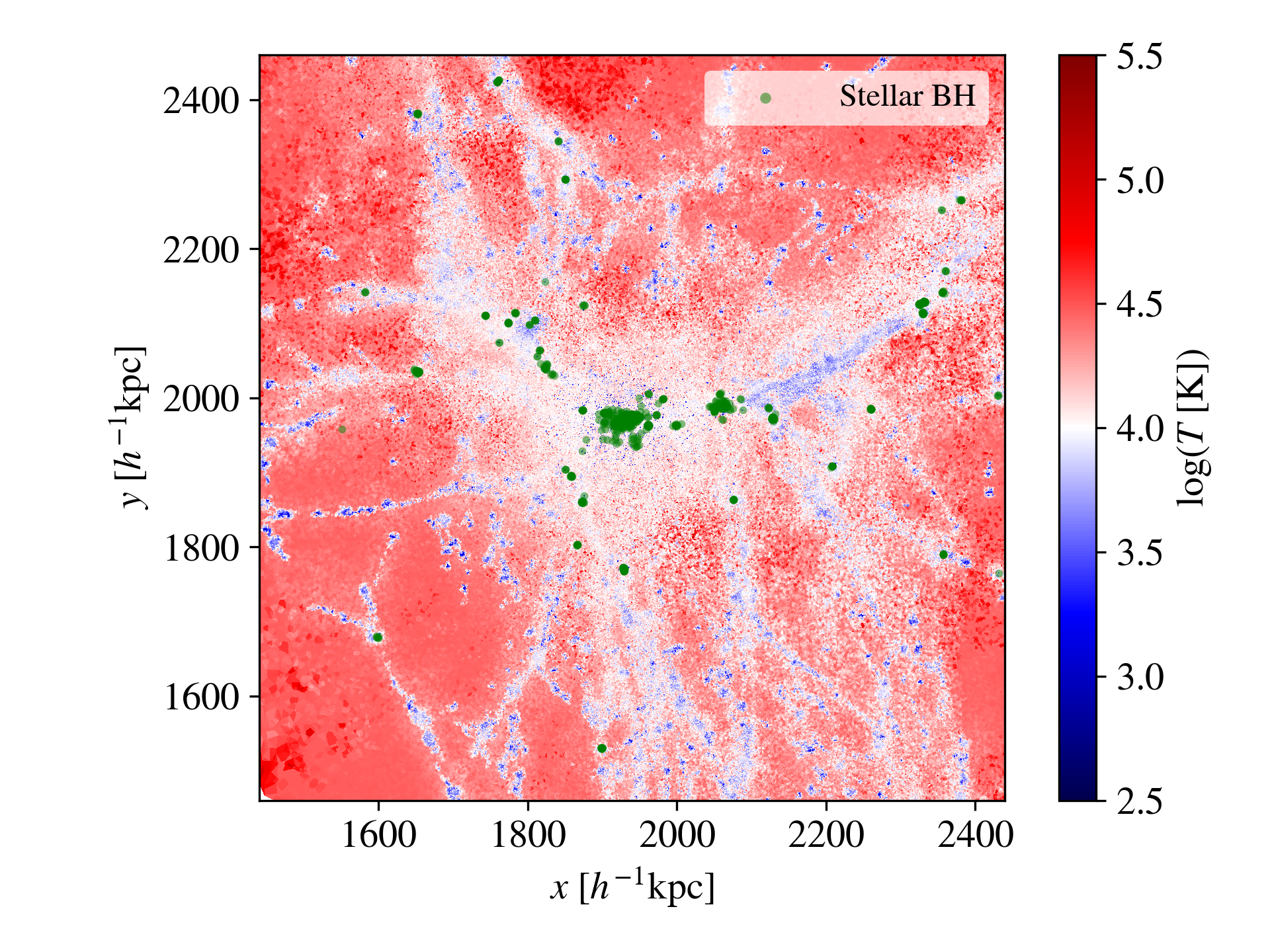}}
\subfloat[DM distribution]{\includegraphics[width= 1.065\columnwidth]{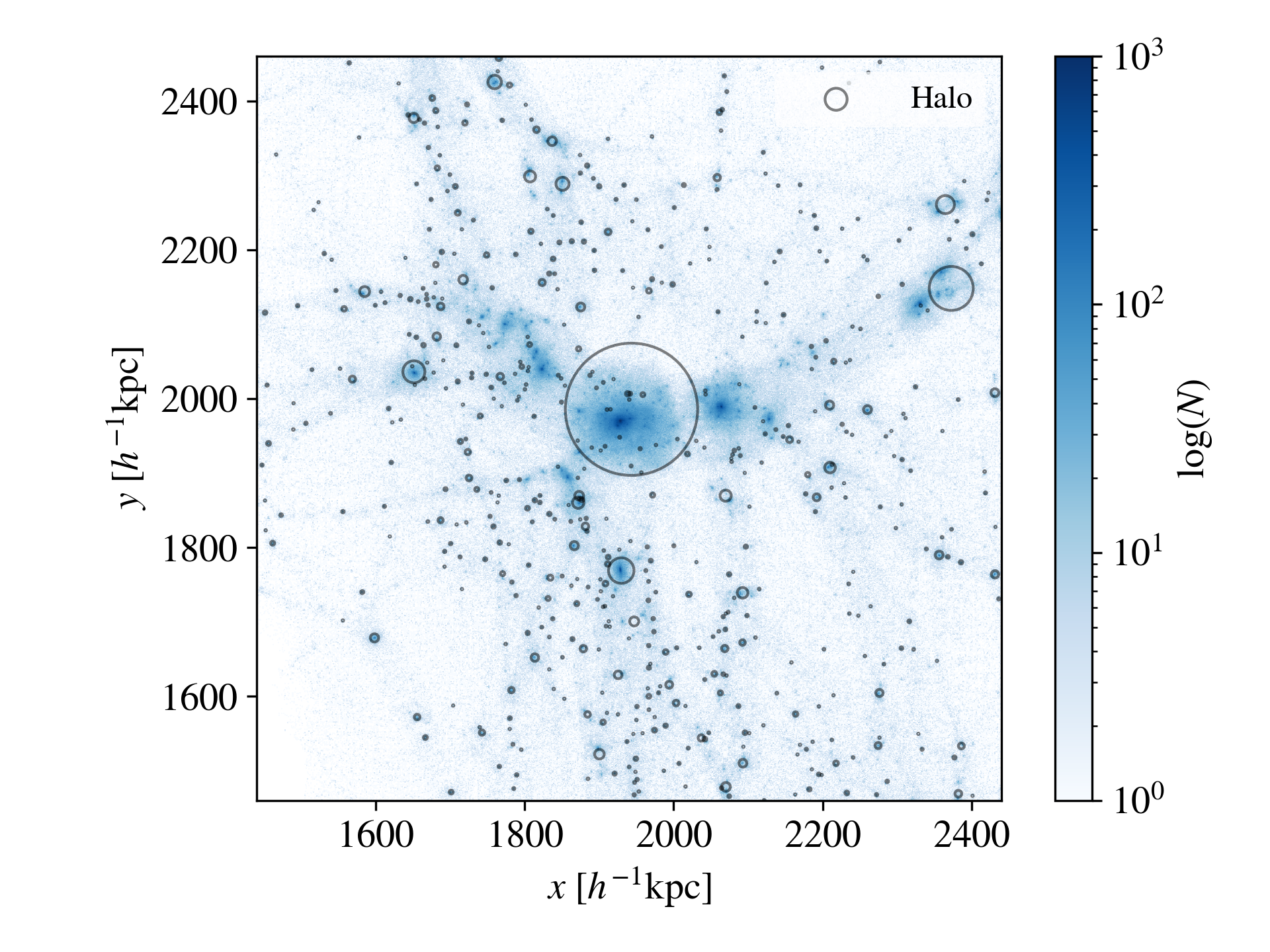}}
\vspace{-10pt}
\caption{Gas temperature distribution (left) and projected DM density field (right) for a slice of thickness $400\ h^{-1}\mathrm{kpc}$ from \texttt{FDzoom\_Lseed} (see Sec.~\ref{s2.3.1} for the meaning of \texttt{Lseed}) at $z=4$. In the left panel, stellar BHs from Pop~III progenitors are labelled with green dots. In the right panel, the DM haloes above the $\mathrm{H_{2}}$ cooling threshold (identified with the standard friends-of-friends method) are shown by circles whose size scales with halo mass. $N$ denotes the number of DM particles in a pixel of size $1\ h^{-2}\mathrm{kpc^{2}}$, such that column density is given in units of $\Sigma_{U}=2.4\times 10^{4}\ \mathrm{M_{\odot}\ kpc^{-2}}$.}
\label{cswb}
\end{figure*}

\section{Methodology}
\label{s2}
We use the \textsc{gizmo} code \citep{hopkins2015new}, which couples new hydrodynamic algorithms with the parallelization and gravity solver of \textsc{gadget-3} \citep{springel2005cosmological}. We here adopt the Lagrangian meshless finite-mass (MFM) version of \textsc{gizmo} (with a number of neighbours $N_{\mathrm{ngb}}=32$), which is a hybrid of smoothed particle hydrodynamics (SPH) and grid-based hydro solvers. For physics beyond gravity and hydrodynamics, our simulations include the primordial chemistry, cooling and metal enrichment model from \citet{jaacks2018baseline}, as well as a modified version of the star formation (SF) and stellar feedback model in \citet{jaacks2018baseline,jaacks2018legacy}, further discussed in Sec.~\ref{s2.2}. Besides, we have implemented customized sub-grid models for the seeding, dynamical capture, accretion and feedback of BHs formed from Pop~III stellar populations (Sec.~\ref{s2.3})\footnote{We did not model Pop~II-seeded BHs as they are typically less massive ($M_{\mathrm{BH}}\lesssim 10\ \mathrm{M_{\odot}}$) and suffer from strong SN natal kicks, such that their accretion, mergers and feedback are inefficient. We also did not consider X-ray binaries whose effect on Pop~III star formation has been found negligible, although their feedback may affect early BH accretion and reionization (e.g. \citealt{jeon2014radiative,ryu2015formation}).}, based on the BH model in \textsc{gadget-3} \citep{springel2005modelling}. With these numerical tools, we study the properties of Pop~III-seeded BHs, especially their GW signals, in a series of cosmological simulations, whose characteristics are summarised below (Sec.~\ref{s2.1}).

\subsection{Simulation setup}
\label{s2.1}

To explore both cosmic-average environments and overdense regions in the early Universe, our simulations are conducted in two simulation setups. 
The first setup (\texttt{zoom}) is the zoom-in region adopted in \citet{liu2019global}, which is defined around a halo of $\sim 10^{10}\ \mathrm{M}_{\odot}$ at $z\sim 10$, with a co-moving volume $V_{C}\sim 4\ h^{-3}\mathrm{Mpc}^{3}$. While the second setup (\texttt{box}) is a cubic box with co-moving side-length $l=4\ h^{-1}\mathrm{Mpc}$. The initial conditions for both setups in $\Lambda$CDM cosmology are generated with the \textsc{music} code \citep{hahn2011multi} at the initial redshift $z_{i}=99$ under the \textit{Planck} cosmological parameters \citep{planck}: $\Omega_{m}=0.315$, $\Omega_{b}=0.048$, $\sigma_{8}=0.829$, $n_{s}=0.966$, and $h=0.6774$. The chemical abundances are initialized with the results in \citet{galli2013dawn}, following \citet{liu2019global} (see their Table~1). To better appreciate the effects of stellar feedback on Pop~III-seeded BHs, in addition to the fiducial (\texttt{FD}) implementation of stellar feedback, we further explore an `extreme' case in the \texttt{zoom} setup (under the same resolution), \texttt{NSFDBK}, where photo-ionization heating and stellar winds from Pop~II stars are turned off\footnote{We never turn off the feedback from Pop~III stars and the LW radiation from Pop~II stars, as this leads to significant (a factor of $3-5$) overproduction of Pop~III stellar populations, and thus, BH seeds, relative to the \texttt{FD} case, so that the results will be of no comparison power. Note that the Pop~III stellar mass densities in our \texttt{FD} runs are consistent with observational constraints (see Sec. \ref{s3.1}).}. Besides, to evaluate the convergence of our methods, we conduct a higher-resolution (\texttt{HR}) simulation in the \texttt{zoom} setup, with the mass resolution for gas and DM particles increased by a factor of 8 compared with the fiducial runs. The basic information of the aforementioned simulations are summarised in Table~\ref{t1}. For illustration, Fig.~\ref{cswb} shows the thermal and DM structure in the center of the zoom-in region at $z=4$, from one of our fiducial runs, in terms of temperature distribution and projected DM density field. We use the \textsc{yt} \citep{turk2010yt} and \textsc{caesar} \citep{thompson2014pygadgetreader} 
software packages to analyse simulation results.

\subsection{Star formation and feedback}
\label{s2.2}
\begin{figure}
\includegraphics[width=1\columnwidth]{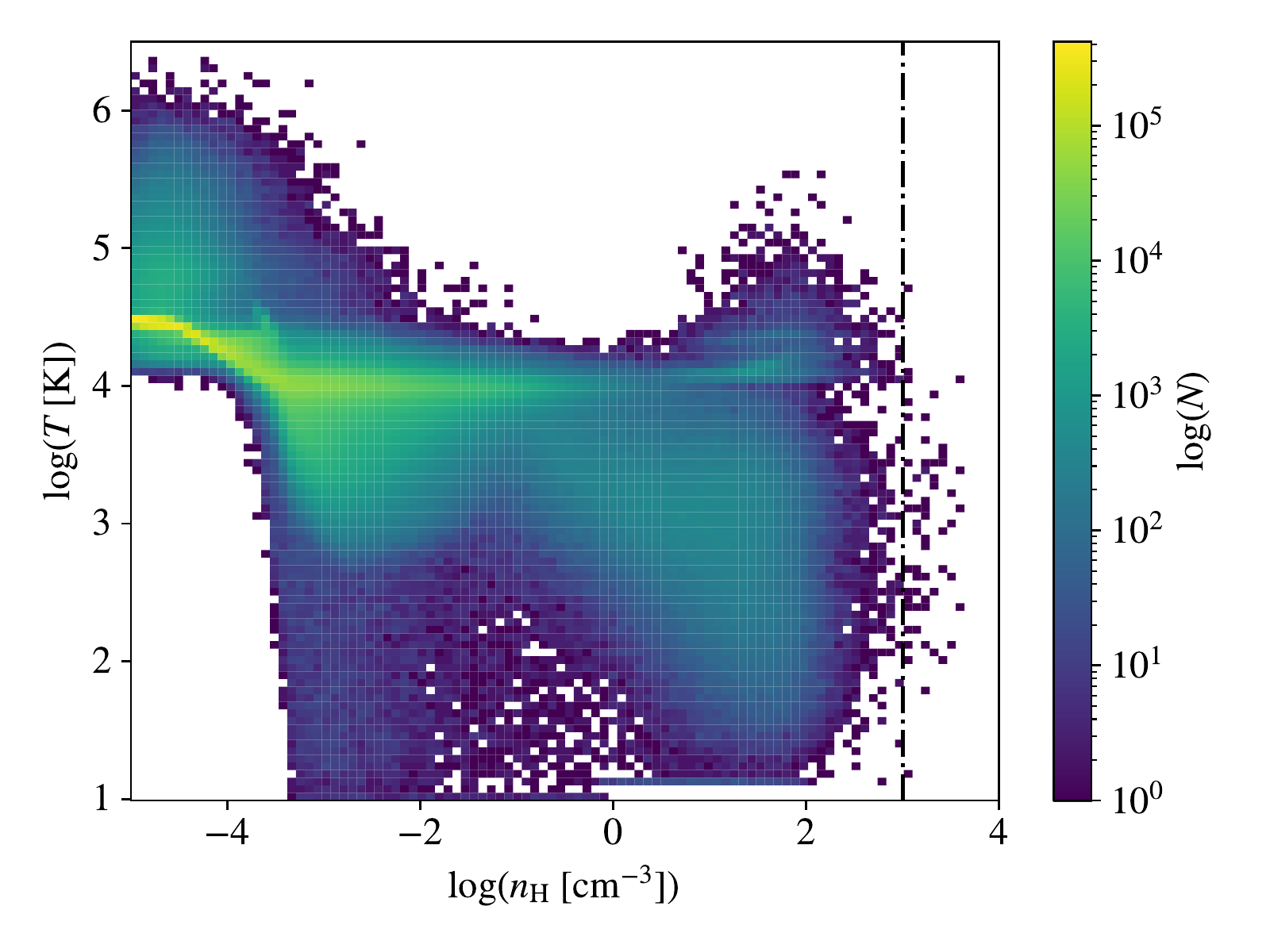}
\caption{Temperature-density phase diagram in \texttt{FDbox\_Lseed} (see Sec.~\ref{s2.3.1} for the meaning of \texttt{Lseed}) at $z=4$, where the vertical dashed-dotted line denotes the density threshold for photo-ionization heating $n_{\mathrm{th,\HII}}=10^{3}\ \mathrm{cm^{-3}}$ (see Sec.~\ref{s2.2.3} for details). Multiple phases of gas can be identified here, such as the hot diffuse ionized IGM ($n_{\uH}\lesssim 10^{-4}\ \mathrm{cm^{-3}}$, $T\sim 2\times 10^{4}\ \mathrm{K}$), cold dense star forming gas ($n_{\uH}\gtrsim 10^{2}\ \mathrm{cm^{-3}}$, $T\lesssim 10^{3}\ \mathrm{K}$) and hot ionized gas in \HII\ regions ($n_{\uH}\sim 10^{-3}-10^{3}\ \mathrm{cm^{-3}}$, $T\gtrsim 10^{4}\ \mathrm{K}$). The effect of SN feedback can also be seen from the reduction of the amount of low-density cold gas ($n_{\uH}\sim 10^{-4}-10^{-2}\ \mathrm{cm^{-3}}$, $T\lesssim 10^{3}\ \mathrm{K}$) and the presence of a very hot diffuse phase ($n_{\uH}\lesssim 10^{-3}\ \mathrm{cm^{-3}}$, $T\gtrsim 3\times 10^{4}\ \mathrm{K}$).}
\label{f2}
\end{figure}

Since individual stars cannot be resolved in our cosmological simulations, each stellar particle represents a stellar population whose member stars are sampled from the input initial mass function (IMF). We use the same Pop~III and Pop~II stellar population models as those used in \citet{jaacks2018baseline,jaacks2018legacy,liu2019global}. Pop~III stars are sampled on-the-fly from a top-heavy IMF $\Phi(M)\propto M^{-\alpha}\exp(-M^{2}_{\mathrm{cut}}/M^{2})$ with $\alpha=0.17$ and $M^{2}_{\mathrm{cut}}=20\ \mathrm{M}_{\odot}^{2}$ in the mass range $1-150\ \mathrm{M}_{\odot}$. While for Pop~II stellar populations, we pre-calculate all needed physical quantities (e.g. luminosity of ionizing photons) per unit stellar mass, by integrations of a Chabrier IMF over a mass range $0.08-100\ \mathrm{M}_{\odot}$ (see table~2 and equ.~(7) in \citealt{jaacks2018legacy} for details), and assume that all Pop~II stellar particles are identical. Again, following \citet{jaacks2018legacy}, a gas particle will be identified as a SF candidate when the number density of hydrogen exceeds $n_{\mathrm{th}}=100\ \mathrm{cm^{-3}}$, while the temperature $T$ remains below $T_{\mathrm{th}}=10^{3}$~K. However, in this work, in order to better simulate the interactions between BHs and stars, we do not turn SF candidates directly into stellar particles\footnote{In \citet{jaacks2018legacy}, a stellar particle represents not only the stellar population associated with it, but also the underlying interstellar medium (ISM) assumed to be coupled with the stellar population. In this work, we treat stars and their natal ISM separately to better model the dynamical friction of BHs by stars.}. Instead, we let SF candidates spawn stellar particles in a stochastic manner (see Sec.~\ref{s2.2.1}). A Pop~III stellar population is assigned to the newly-born stellar particle when its metallicity is below a critical value, $Z<Z_{\mathrm{crit}}=10^{-4}\ Z_{\odot}$ \citep{safranek2010, schneider2011}; otherwise, a Pop~II stellar population is assigned. Furthermore, we include stellar winds from Pop~II stars, using the methodology in \citet{springel2003wind} (see Sec.~\ref{s2.2.5}). In the following subsections, we briefly describe our implementations of SF and stellar feedback, focusing on the modifications with respect to the original model in \citet{jaacks2018baseline,jaacks2018legacy}. Equipped with these sub-grid models and primordial chemistry and cooling, our simulations can capture the multi-phase features of interstellar and intergalactic media (ISM and IGM) (see Fig.~\ref{f2} for an example of the temperature-density phase diagram in the \texttt{box} setup, in the post-reionization era). The resulting star formation and BH accretion histories are also consistent with observational constraints at high-$z$ (see Sec.~\ref{s3}).

\subsubsection{Stochastic star formation}
\label{s2.2.1}
For each SF candidate, we calculate the corresponding probability of SF as
\begin{align}
    p_{\mathrm{SF}}=\frac{m_{\mathrm{SF}}}{m_{\star}}\left[1-\exp(-\eta_{\star}\delta t/t_{\mathrm{ff},i})\right]\ ,\label{e1}
\end{align}
where $m_{\mathrm{SF}}$ and $m_{\star}$ are the masses of the SF candidate and stellar particle to be spawned, $\eta_{\star}$ is the star formation efficiency (SFE), $\delta t$ is the current simulation timestep, and $t_{\mathrm{ff},i}=\sqrt{3\pi/(32G\rho_{i})}$ is the free-fall timescale of the SF candidate with a gas density $\rho_{i}$. Here we set $\eta_{\star,\mathrm{PopIII}}=0.05$ for Pop~III stars and $\eta_{\star,\mathrm{PopII}}=0.1$ for Pop~II stars, consistent with \citet{jaacks2018legacy}. 
A random number $p$ following a uniform distribution in [0, 1] is generated, and a stellar particle will be spawned if $p<p_{\mathrm{SF}}$. We set $m_{\star}\simeq 600\ \mathrm{M}_{\odot}$, based on the results from high-resolution simulations of Pop~III star formation in individual minihaloes \citep{bromm2013,stacy2016building} and observational constraints from the global 21-cm absorption signal \citep{schauer2019constraining}, showing that the characteristic mass of Pop~III stellar populations is $500-1000\ \mathrm{M}_{\odot}$. For simplicity, we adopt the same $m_{\star}$ for both Pop~III and Pop~II stellar populations, having verified that the choice of $m_{\star}$ has little impact on processes involving Pop~II stars.

This stochastic implementation of SF is based on the assumption that the local SF rate density (SFRD) within gas of density $\rho_{\mathrm{g}}$ can be written as (e.g. \citealt{1992ApJ...391..502K,stinson2006star})
\begin{align}
    \frac{d\rho_{\star}}{dt}=\eta_{\star}\frac{\rho_{\mathrm{g}}}{t_{\mathrm{g}}}\ ,\label{e2}
\end{align}
where $t_{\mathrm{g}}$ is the characteristic timescale for gas inflow during the collapse of the star-forming cloud. This formalism is incorporated into our simulations with $\rho_{\mathrm{g}}=\rho_{i}$ and $t_{\mathrm{g}}=t_{\mathrm{ff},i}$. 

\subsubsection{Lyman-Werner radiation}
\label{s2.2.2}
Similar to \citet{jaacks2018baseline}, we calculate the contributions to the global uniform Lyman-Werner (LW) background from Pop~III and Pop~II stars with \citep{johnson2013formation}
\begin{align}
    J_{\mathrm{LW,bg}}(t)\simeq \frac{hc}{4\pi} \eta_{\uLW}\frac{\langle\dot{\rho}_{\star}(t)\rangle t_{\star}X}{m_{\mathrm{H}}}\ ,
\end{align}
where $t_{\star}$ is the typical lifetime of (massive stars in) the stellar population, $\langle\dot{\rho}_{\star}(t)\rangle$ is the global (physical) SFRD at time $t$, $\eta_{\uLW}$ is the number of LW photons produced per stellar baryon, and $X=0.76$ is the mass fraction of hydrogen in primordial gas. We adopt $\eta_{\uLW,\mathrm{PopIII}}=2\times 10^{4}$, $t_{\star,\mathrm{PopIII}}=3\ \mathrm{Myr}$ for Pop~III and $\eta_{\uLW,\mathrm{PopII}}=4\times 10^{3}$, $t_{\star,\mathrm{PopII}}=10\ \mathrm{Myr}$ for Pop~II stellar populations, consistent with our IMFs. To better simulate the formation of Pop~III stars, we also consider the local LW field under the optically thin assumption. That is to say, each newly-born stellar particle is labelled active for $t_{\star}$, during which it contributes to the local LW field with
\begin{align}
    J_{\uLW,\star}(r) = \frac{\langle L_{\nu}\rangle m_{\star}}{16\pi^{2}r^{2}}\ ,\quad \langle L_{\nu}\rangle\simeq \frac{h\eta_{\uLW}X}{m_{\uH}t_{\star}}\ ,
\end{align}
where $\langle L_{\nu}\rangle$ is the specific luminosity of LW radiation per unit stellar mass averaged across the stellar population, $m_{\star}$ is again the mass of the stellar particle, and $r$ are the distance to it. The total LW intensity at any position $\mathbf{x}$ in the simulation region is then estimated via
\begin{align}
    J_{\uLW}(t,\mathbf{x})=J_{\mathrm{LW,bg}}(t)+J_{\uLW,\mathrm{local}}(t,\mathbf{x})\ ,\\
    J_{\uLW,\mathrm{local}}(t,\mathbf{x})\equiv \sum_{i}\frac{\langle L_{\nu,i}\rangle m_{\star}}{16\pi^{2}|\mathbf{x}_{i}-\mathbf{x}|^{2}}\ .
\end{align}
In the second line, the summation extends over all active stellar particles at time $t$ that are within $200\ \mathrm{kpc}$ around $\mathbf{x}$ and have contributions to the local LW intensity above $10^{-3} J_{\mathrm{LW,bg}}(t)$. The choice of $200\ \mathrm{kpc}$ is based on the results from \citet{regan2019emergence}, which show that the LW radiation of star-forming galaxies has little effect (on the evolution of primordial gas) beyond $200\ \mathrm{kpc}$ (see also \citealt{johnson2007local}). 

The LW intensity distribution is then used to calculate the dissociation rates of the main molecular coolants $\uH_{2}$ and $\uH\mathrm{D}$ in our chemical network for each gas particle. The effect of self-shielding is approximated with dimensionless factors \citep{wolcott2011photodissociation,wolcott2011suppression}, based on the local $\uH_{2}$ column density $N_{\uH_{2}}\simeq n_{\uH_{2}}L_{\mathrm{J}}$, where $L_{\mathrm{J}}=\sqrt{15k_{B}T/(4\pi\rho Gm_{\uH})}$ is the local Jeans length (see equ.~(12) in \citealt{wolcott2011photodissociation}, as well as equ.~(12) and table~1 in \citealt{wolcott2011suppression} for details).

\subsubsection{Photo-ionization heating}
\label{s2.2.3}
Similar to \citealt{jaacks2018legacy}, globally, heating by the UV background is calculated with the redshift-dependent photo-ionization rate $\zeta (z)$ from \citet{faucher2009new}, taking into account self-shielding. Locally, photo-ionization (PI) heating is applied to the gas particles on-the-fly in the spherical region around each active stellar particle within the Str\"{o}mgren radius
\begin{align}
	R_{\mathrm{ion}}=\left(\frac{3\langle \dot{N}_{\mathrm{ion}}\rangle m_{\star}}{4\pi \hat{n}_{\uH}^{2}\alpha_{B}}\right)^{1/3}\ ,\label{e7}
\end{align}
where $\langle \dot{N}_{\mathrm{ion}}\rangle$ is the ionization luminosity per unit stellar mass, $\hat{n}_{\uH}$ is the typical average number density of hydrogen in the surrounding interstellar medium (ISM), and $\alpha_{B}=2.59\times 10^{-13}\ \mathrm{cm^{3}\ s^{-1}}$ is the case-B recombination coefficient. Based on our IMFs, we have $\langle \dot{N}_{\mathrm{ion,PopIII}}\rangle\sim 10^{48}\ \mathrm{s^{-1}\ M_{\odot}^{-1}}$ for Pop~III and $\langle \dot{N}_{\mathrm{ion,PopII}}\rangle\sim 10^{47}\ \mathrm{s^{-1}\ M_{\odot}^{-1}}$ for Pop~II. We set $\hat{n}_{\uH,\mathrm{PopIII}}=0.01\ \mathrm{cm^{-3}}$ for Pop~III while $\hat{n}_{\uH,\mathrm{PopII}}=1\ \mathrm{cm^{-3}}$ for Pop~II, considering the scenario that ionization fronts around Pop~III stars can break out of the host minihaloes and thus impact lower-density gas, compared with the case of Pop~II stars whose \HII\ regions are confined by the gravitational potentials of the host galaxies. The resulting radii are $R_{\mathrm{ion,PopIII}}\simeq 2\ \mathrm{kpc}\times [m_{\star}/(500\ \mathrm{M}_{\odot})]^{1/3}$ and $R_{\mathrm{ion,PopII}}\simeq 0.2\ \mathrm{kpc}\times [m_{\star}/(500\ \mathrm{M}_{\odot})]^{1/3}$. 

As pointed out by \citet{liu2019global}, since $\hat{n}_{\uH}$ is fixed regardless of the actual environments around stellar particles in the simulation, our model tends to over-predict the volume of dense \HII\ regions around Pop~II stars by neglecting the effect of self-shielding against ionizing photons in dense clumps. To avoid this problem, we restrict PI heating from Pop~II stars to gas particles with hydrogen number densities below $n_{\mathrm{th,\HII}}=10^{3}\ \mathrm{cm^{-3}}$ except in an inner sphere around each Pop~II stellar particle. This inner sphere captures the `realistic' dense \HII\ region, whose radius $R_{\mathrm{inner}}$ is estimated from Equation~(\ref{e7}) by replacing $\hat{n}_{\uH}$ with the density of the stellar particle's nearest gas neighbour. 



\subsubsection{Supernova legacy feedback}
\label{s2.2.4}

As in \citet{jaacks2018baseline,jaacks2018legacy}, our simulations cannot resolve the process of shell expansion in supernova (SN) explosions. We instead `paint' the chemical and thermal legacy of supernovae onto the affected simulation particles. When a stellar population dies (i.e. $t_{\star}$ after it was born), we calculate the final radius of shell expansion $r_{\mathrm{final}}$ based on the total SN energy $E_{\mathrm{tot}}$ with the fitting formula (see fig.~4 in \citealt{jaacks2018baseline} for details)
\begin{align}
	r_{\mathrm{final}}\simeq 0.8\ \mathrm{kpc}\ \left(\frac{E_{\mathrm{tot}}}{10^{52}\ \mathrm{erg}}\right)^{0.38}\ .
\end{align}
Metals produced by SNe are evenly (mass-weighted) distributed to the gas particles within $r_{\mathrm{final}}$ around the stellar particle. The total SN energy and ejected metal mass $M_{Z}$ are calculated on-the-fly by counting progenitors (i.e. core-collapse and pair-instability SNe) for each Pop~III stellar population (see Table~3 of \citealt{jaacks2018baseline}). Whereas for Pop~II stellar populations, we assign $E_{\mathrm{tot}}\simeq 10^{52}\ \mathrm{erg}\times m_{\star}/(10^{3}\ \mathrm{M}_{\odot})$ and $M_{Z}=0.016 m_{\star}$. 

\begin{figure*}
\includegraphics[width=2\columnwidth]{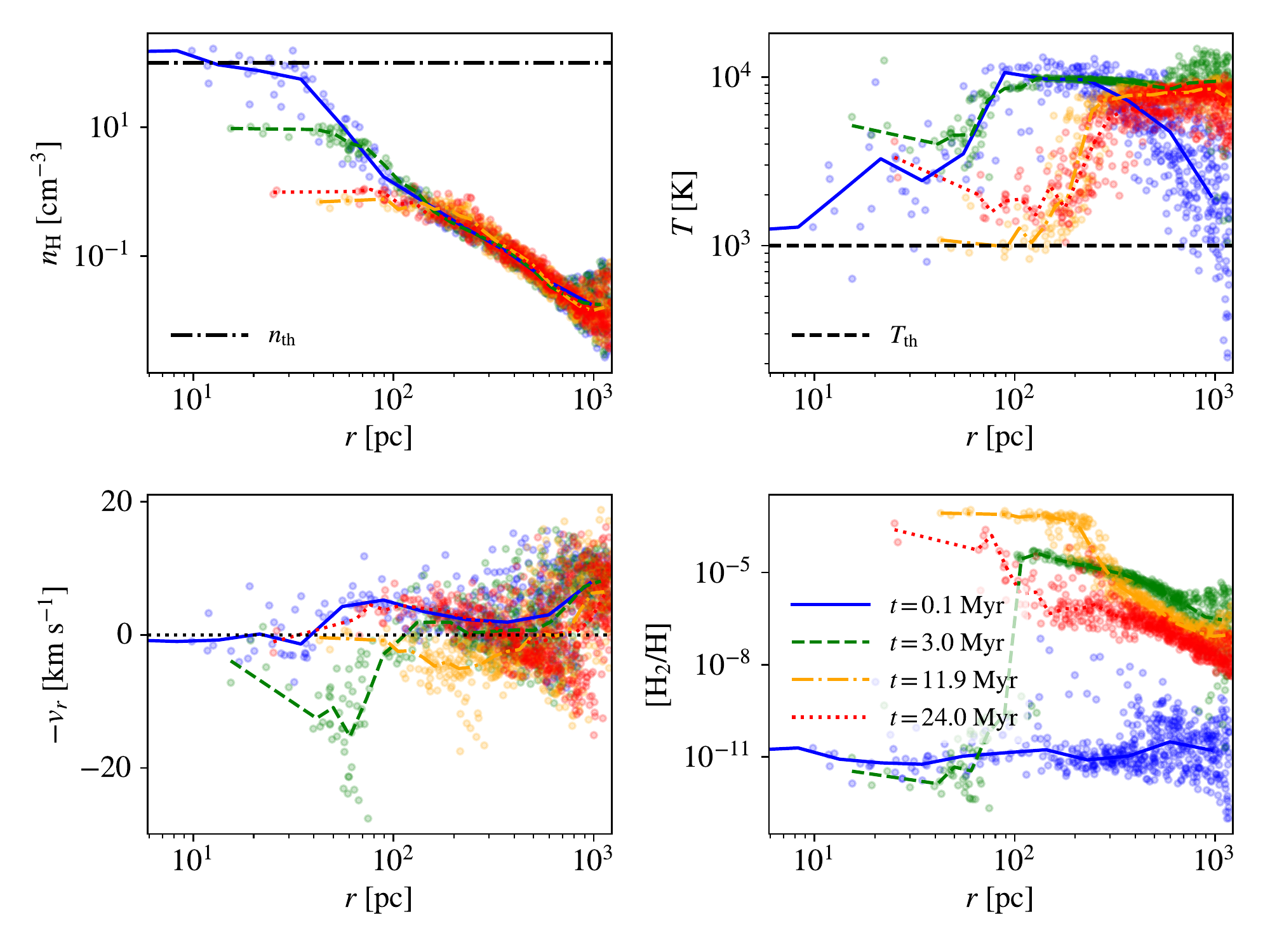}
\caption{Evolution of gas density, temperature, $\mathrm{H}_{2}$ abundance and radial velocity profiles (clockwise) around a Pop~III stellar particle formed at $z\sim 11.5$, from a test simulation in the \texttt{zoom} setup. Here $r$ is the physical distance to the Pop~III stellar particle. The median values at $t\simeq 0.1$, 3, 12 and 24~Myr are plotted with solid, dashed, dashed-dotted and dotted lines, on top of individual gas particles denoted by orange, blue, green and yellow dots, respectively, where $t$ is the time after the formation of the Pop~III stellar population. For $t\lesssim t_{\star}=3$~Myr, photo-ionization heating reduces the central density by a factor of 10, and generates inner ($r\lesssim 100$~pc) outflows with velocities up to $\sim 30~\mathrm{km\ s^{-1}}$. Then the \HII\ region keeps expanding for $\sim20$~Myr before re-collapse, during which process gas cools to $\sim 10^{3}$~K in the inner region with $n_{\uH}\gtrsim 0.1\ \mathrm{cm^{-3}}$, and the $\mathrm{H}_{2}$ abundance reaches up to $\sim 10^{-3}$ by enhanced $\mathrm{H}_{2}$ formation catalysed by free elections. The decrease in $\mathrm{H}_{2}$ abundance from $t\simeq 12$~Myr to $t=24$~Myr is caused by external LW radiation. Note that, in our case, $J_{\mathrm{LW,bg}}\sim 10^{-21}\ \mathrm{erg\ s^{-1}\ cm^{-2}\ Hz^{-1}\ sr^{-1}}$ at $z\sim 11.5$.}
\label{popIII}
\end{figure*}

In addition to metal enrichment, we also model the thermal legacy of Pop~III SNe by applying thermal energy injection (corresponding to a temperature increase of $T_{c}\simeq 2\times 10^{4}$~K) and instantaneous ionization of hydrogen to the gas particles within $R_{\mathrm{ion}}$ around each Pop~III stellar particle at the end of its lifetime. This thermal legacy, combined with photo-ionization heating, is sufficient to capture the effect of Pop~III feedback on surrounding gas, as exemplified in Fig.~\ref{popIII}, consistent with the results in high-resolution simulations (e.g. \citealt{johnson2007local,ritter2012confined,ritter2015metal}). 

\subsubsection{SN-driven winds}
\label{s2.2.5}
Above, we present our implementations of thermal and radiation feedback. However, mechanical (or kinematic) feedback is also crucial for simulating self-regulated SF and the state of the ISM, especially for Pop~II stars. In light of this, we further include SN-driven winds for Pop~II stars based on \citet{springel2003wind}. For each Pop~II SF candidate (gas particle) with a mass $m_{\mathrm{SF}}$ about to spawn a stellar population, we calculate a probability
\begin{align}
p_{w}=1-\exp(-\eta_{w,\mathrm{SF}}\frac{m_{\star}}{m_{\mathrm{SF}}})\label{e9}
\end{align} 
for the gas particle to be launched as a wind particle, where $\eta_{w,\mathrm{SF}}$ is the wind-loading factor. Similar to our stochastic SF model (Sec.~\ref{s2.2.1}), a random number is then generated to determine whether to launch a wind particle based on $p_{w}$. Once launched, the gas particle receives a kick of $v_{w,\mathrm{SF}}\simeq 170\ \mathrm{km\ s^{-1}}$ in a random direction. Here $v_{w,\mathrm{SF}}$ is calculated as \citep{springel2003wind}: 
\begin{align}
\begin{split}
v_{w,\mathrm{SF}}&=\sqrt{\frac{2\beta\chi u_{\mathrm{SN}}}{(1-\beta)\eta_{w,\mathrm{SF}}}}\ ,\\
u_{\mathrm{SN}}&=\frac{1}{\mu(\gamma-1)}\frac{k_{B}T_{\mathrm{SN}}}{m_{\uH}}\ ,
\end{split}
\end{align}
where we adopt $\beta=0.16$, $\chi=0.05$ as the efficiencies for SN-hot-gas and hot-gas-wind energy couplings. Furthermore, $u_{\mathrm{SN}}$ is the specific SN explosion energy, written in terms of the initial SN blast wave temperature $T_{\mathrm{SN}}\simeq 1.5\times 10^{8}$~K, the adiabatic index $\gamma=5/3$, and the mean molecular weight $\mu$ ($\simeq 0.63$ for ionized primordial gas)\footnote{The values of $\beta$ and $T_{\mathrm{SN}}$ are consistent with our IMF for which the total energy and mass of SN are $E_{\mathrm{tot}}\sim 10^{52}\ \mathrm{erg}\times m_{\star}/(10^{3}\ \mathrm{M}_{\odot})$ and $M_{\mathrm{SN}}=0.16 m_{\star}$, such that $u_{\mathrm{SN}}=E_{\mathrm{tot}}/M_{\mathrm{SN}}\simeq 3\times 10^{16}\ \mathrm{cm^{2}\ s^{-2}}$.}. The wind particle is ineligible to become a SF or DCBH candidate, unless a time interval of $t_{w}$ has past or its (hydrogen number) density is below $n_{w}$. Here we set $t_{w}=0.1H^{-1}(t)$ and $n_{w}=0.1 n_{\mathrm{th}}=10\ \mathrm{cm^{-3}}$ for wind-ISM recoupling, where $H^{-1}(t)$ is the local Hubble time. To be conservative, we only apply this sub-grid wind model to Pop~II stars, as the the effect of SN blast waves from Pop~III stars has already been captured by the photo-ionization and thermal energy injection (see Sec.~\ref{s2.2.3} and \ref{s2.2.4}). We set $\eta_{w,\mathrm{SF}}=2$ and $\chi=0.05$ to reproduce the observed SFRD at $z\lesssim 10$ (see Sec.~\ref{s3} for details).


\subsection{Black hole model}
\label{s2.3}
In this section, we briefly describe our numerical methods of simulating Pop~III-seeded BHs at high redshifts, which are mostly based on existing models designed for SMBHs \citep{tremmel2015off,tremmel2017romulus,negri2017black}. 
Following the BH model in \textsc{gadget-3} \citep{springel2005modelling}, we calculate physical quantities reflecting the environment of each BH particle based on the simulation particles within a specific search radius $h_{\mathrm{BH}}$. Here $h_{\mathrm{BH}}$ is adjusted on-the-fly such that $\sim 64=2 N_{\mathrm{ngb}}$ gas particles are enclosed, but cannot exceed the upper limit $h_{\max}=5\epsilon_{\mathrm{gas}}=1\ h^{-1}\mathrm{kpc}$ (in co-moving coordinates). This upper limit is adopted to avoid over-predicting the BH accretion rate in low-density regions\footnote{Without this upper limit, the median value of $h_{\mathrm{BH}}$ is $\sim 0.5\ h^{-1}\mathrm{kpc}$ in presence of stellar feedback, so that the simulation results are insensitive to $h_{\max}$ for $h_{\max}\gtrsim 0.5\ h^{-1}\mathrm{kpc}$. }. 
For conciseness, below we denote the physical (i.e. non-comoving) gravitational softening length of BH particles as $\epsilon_{\mathrm{g}}\equiv 2.8a\epsilon_{\mathrm{BH}}$, beyond which scale the force is completely Newtonian, where $a$ is the cosmic scale factor.

\subsubsection{Black hole seeding}
\label{s2.3.1}

\begin{figure}
\includegraphics[width=1\columnwidth]{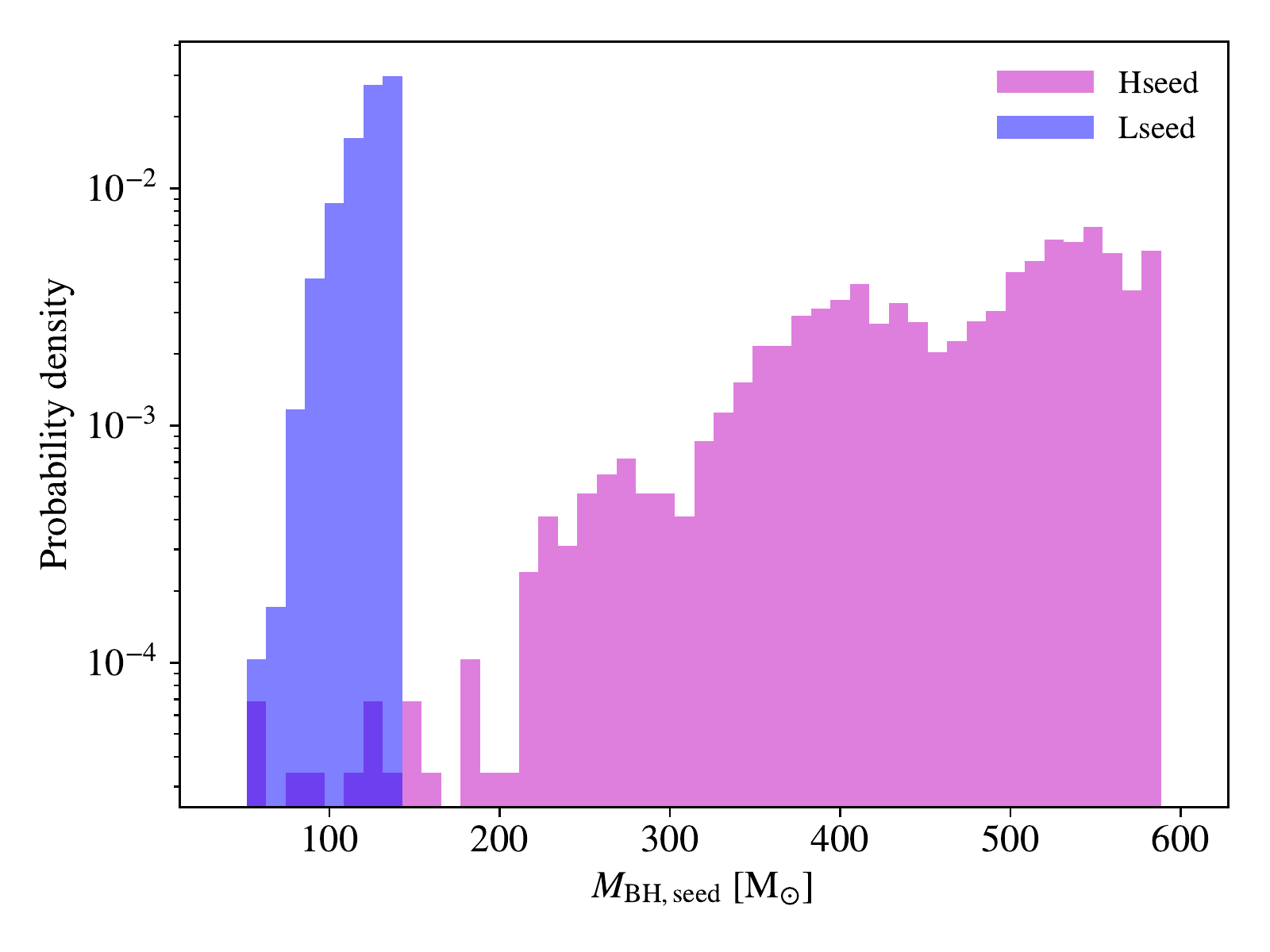}
\caption{Distributions of Pop~III BH seed masses for \texttt{Hseed} (purple) and \texttt{Lseed} (blue), measured with the Pop~III particles in \texttt{FDzoom\_Hseed} at $z=4$.}
\label{f4}
\end{figure}

As mentioned in Section~\ref{s2.2}, each Pop~III population is sampled from a top-heavy IMF, where we identify stars in the mass range $40\le M/\mathrm{M}_{\odot} \le 140$ as BH progenitors, consistent with our feedback model. Stars with $M>140\ \mathrm{M}_{\odot}$ will evolve into pair-instability SNe, leaving no remnant behind, while the majority of stars with masses below $40\ \mathrm{M}_{\odot}$ will not end up in black holes. Besides, even for the stars that become core-collapse SNe at last and give birth to BHs with some fractions of their stellar masses (i.e. for $8\ \mathrm{M}_{\odot}<M<40\ \mathrm{M_{\odot}}$), the resulting BHs are low-mass in nature, and the SN natal kicks (with typical velocities above $100\ \mathrm{km\ s^{-1}}$, e.g. \citealt{repetto2012investigating}) can eject them out of the host minihaloes, which will significantly suppress their participation in mass growth and mergers \citep{whalen2012formation}. 
We assume that Pop~III stars in the range $40\le M/\mathrm{M}_{\odot} \le 140$ will convert all their stellar mass into BH mass, and that natal kicks are negligible. Note that our stellar population models do not explicitly include binary stars, as we here focus on the GW signals from \textit{ex-situ} BH binaries (formed by dynamical capture) with members from different stellar populations. 

Given the small number of Pop~III BH progenitors with $40\le M/\mathrm{M}_{\odot} \le 140$ from IMF sampling, there are still uncertainties in the resulting mass spectrum of BH seeds, because of small-scale dynamical interactions between stars, beyond our resolution. It remains unknown whether/how these progenitors will merge to form more massive BHs or eject some of them out of the host halo, since simulations resolving individual stars are computationally expensive and only able to track their evolution for up to $\sim 10^{5}\ \mathrm{yr}$ (e.g. \citealt{stacy2013constraining,susa2014mass,machida2015accretion,stacy2016building,hirano2017formation,hosokawa2020}). 
For simplicity and generality, we consider two extreme cases with relatively heavy (\texttt{Hseed}) and light (\texttt{Lseed}) BHs as the final products. In \texttt{Hseed}, we assume that all BH progenitors will coalesce to form one BH in the end, resulting in seed masses ($\sim 300-590\ \mathrm{M}_{\odot}$, see Fig.~\ref{f4}) that are on average $\sim 80$\% of the total initial stellar mass $m_{\star}$. Such a coalescence process can be driven by runaway stellar collisions or gas inflows (e.g. \citealt{lupi2014constraining,giersz2015mocca,sakurai2017imbh}). In \texttt{Lseed}, we only keep track of the most massive BH progenitor (for accretion and feedback), while the rest is assumed to be inactive in terms of accretion, feedback and dynamical capture\footnote{The untracked BH progenitors can either remain within or be ejected from the original SF disc/host halo.}. 
The typical seed mass ($\sim 80-140\ \mathrm{M}_{\odot}$, see Fig.~\ref{f4}) is then only $\sim 20$\% of $m_{\star}$. When a Pop~III stellar population reaches the end of its life, we turn it into a BH particle with a mass assigned with the methods above. For \texttt{Hseed}, the dynamical mass is unchanged when the stellar particle is turned into a BH particle, under the assumption that the remaining $\sim 20\%$ of mass is locked-up in low-mass stars (and/or their remnants) bound to the BH. While for \texttt{Lseed}, the dynamical mass is set to $f_{\mathrm{rm}}\simeq 0.6$ of the original stellar mass $m_{\star}$, according to the finding that on average $\sim 0.4$ of Pop~III protostars are ejected out of the SF disc \citep{stacy2013constraining}.

For DCBH candidates, we design a set of criteria to capture the hot, dense and metal-poor phase in which the collapse is almost isothermal, leading to formation of BH seeds with $M_{\mathrm{BH}}\gtrsim 10^{4}\ \mathrm{M_{\odot}}$. We turn a gas particle into a DCBH candidate, when its (hydrogen number) density reaches $n_{\mathrm{DCBH}}=2\times 10^{3}\ \mathrm{cm^{-3}}$ with a temperature $7000<T/ \mathrm{K}<10^{4}$, a metallicity $Z<Z_{\mathrm{DCBH}}=2\times 10^{-4}\ \mathrm{Z}_{\odot}$ and a $\mathrm{H}_{2}$ abundance $x_{\mathrm{H_{2}}}<1\times 10^{-6}$. The criteria here are based on the predictions from our one-zone model with the same cooling functions and chemical network as adopted in the simulations \citep{jaacks2018legacy,lithium}\footnote{The metallicity threshold $Z_{\mathrm{DCBH}}=2\times 10^{-4}\ \mathrm{Z}_{\odot}$ is consistent with the results in the early semi-analytical study by \citet{omukai2008can} and the empirical choice in the \textsc{romulus} simulations \citep{tremmel2017romulus}. }, in which the critical density $n_{\mathrm{DCBH}}$ is a bifurcation point, such that gas in the metal-poor, $\mathrm{H}_{2}$-poor and hot phase described by the above criteria remains hot ($T\sim 7500$~K), even when as dense as $n_{\uH}\sim 10^{10}\ \mathrm{cm^{-3}}$; otherwise, gas can cool to $T<10^{3}$~K and fragment during collapse, resulting in normal star formation. In reality, this metal-poor, $\mathrm{H}_{2}$-poor and hot phase occurs under strong gas inflows and/or LW fields\footnote{Such situations are rare even in over-dense regions. For instance, there are typically a few DCBH candidates while $\sim 2000$ stellar BH particles formed at $z>4$ in \texttt{FDbox} runs.}. 
Similar to \citet{tremmel2017romulus}, our criteria are used to select the regions where BH seeds can form and grow quickly to large masses, regardless of the specific formation pathways at unresolved scales, such as supermassive stars (e.g. \citealt{schleicher2013massive,sakurai2016supermassive}). As the initial mass growth (and feedback) cannot be resolved in our simulations, we set the initial (candidate) DCBH mass to the mass of the parent gas particle for simplicity ($M_{\mathrm{BH}}\simeq m_{\mathrm{gas}}\simeq 10^{4}\ \mathrm{M}_{\odot}$ under fiducial resolution), and use sub-grid models to keep track of the subsequent evolution (see below).

\subsubsection{Black hole capture}
\label{s2.3.2}
Once two black hole particles (with dynamical masses $m_{1}$ and $m_{2}$) get closer than two softening lengths in relative distance, i.e. $\Delta r<2\epsilon_{\mathrm{g}}$, and meanwhile remain gravitationally bound to each other (when the relative velocity $\Delta v$ is smaller than the escape speed $v_{\mathrm{esc}}=\sqrt{2G(m_{1}+m_{2})/\Delta r}$), they are combined into one \textit{simulation} BH particle. The resulting BH particle then represents a BH binary or multiple system. Here the combination of two \textit{simulation} particles represents the formation of a gravitationally bound system by dynamical capture, whose subsequent internal evolution is beyond our resolution, and it is by no means equivalent to the final coalescence of two or multiple BHs that would result in GW emission. In reality, it can take several billion years for the BH binary to harden and eventually produce GW signals (e.g. \citealt{sesana2015scattering, conselice2019gravitational}). 

Note that the binaries identified in this way may not be hard binaries, while only hard binaries will be hardened to become BH-BH mergers. Therefore, the resulting GW rates should be regarded as optimistic estimations. We discuss the difference between this optimistic model with a more `pessimistic/realistic' model that only considers hard binaries in Section~\ref{s4.3}. Since we cannot resolve the hardening processes of BH binaries in our cosmological simulations, we estimate the lifetime of each BH binary analytically from observed/simulated scaling relations (see Sec.~\ref{s4.1} for details). To do so, we record the masses of the primary ($M_{1}$) and secondary ($M_{2}$) members, as well as the initial (orbital) eccentricity $e$ for each newly-formed binary/multiple BH system.

\subsubsection{Dynamical friction}
\label{s2.3.3}
In our simulations, the masses of Pop~III BH particles are usually comparable to that of stellar particles ($m_{\mathrm{BH}}\sim m_{\star}$), and much smaller than the masses of gas and DM particles ($m_{\mathrm{BH}}\sim 0.1 m_{\mathrm{gas}}\sim 0.02 m_{\mathrm{DM}}$). Under this condition, the corresponding gravitational softening lengths of background particles must be large enough to avoid spurious collisionality, which meanwhile will suppress dynamical friction (DF) by preventing close encounters with BH particles. Thus, DF of BHs by background objects is not naturally simulated with the gravity solver on scales smaller than the background gravitational softening length, which is a common problem in cosmological simulations with limited mass/spatial resolution. Since dynamical interactions are crucial for BH mergers, we adopt the sub-grid model from \citet{tremmel2015off} to better simulate DF of BHs by background stars\footnote{We only apply the sub-grid DF model to stars because they are the dominant source for DF, and our resolution for gas and DM particles is too low for the sub-grid model to work.}. 

For each BH particle, the additional acceleration from the sub-grid DF model is \citep{tremmel2015off}
\begin{align}
	\mathbf{a}_{\mathrm{DF}}=-4\pi G^{2}M_{\mathrm{BH}}\rho_{\star}(<v_{\mathrm{BH}})\ln\Lambda\frac{\mathbf{v}_{\mathrm{BH}}}{v_{\mathrm{BH}}^{3}}\ ,
\end{align}
where $\mathbf{v}_{\mathrm{BH}}$ is the velocity of the BH relative to the local background center of mass (COM)\footnote{The local COM velocity is defined with all stellar particles around the BH within $h_{\mathrm{BH}}$.}, $\rho_{\star}(<v_{\mathrm{BH}})$ is the mass density of stars with velocities relative to the COM smaller than $v_{\mathrm{BH}}$, and $\ln\Lambda$ is the Coulomb logarithm. In our case, $\rho_{\star}(<v_{\mathrm{BH}})$ is estimated with
\begin{align}
	\rho_{\star}(<v_{\mathrm{BH}})=\frac{M_{\star}(<v_{\mathrm{BH}})}{[4\pi r_{\mathrm{DF}}^{3}/3]}\ ,
\end{align}
where $M_{\star}(<v_{\mathrm{BH}})$ is the total mass of stellar particles within $r_{\mathrm{DF}}$ (physical) around the BH, whose velocities relative to the COM are smaller than $v_{\mathrm{BH}}$. We set $r_{\mathrm{DF}}=ah_{\mathrm{BH}}$. The Coulomb logarithm is
\begin{align}
\begin{split}
	\ln\Lambda &= \ln(1+b_{\max}/b_{\min})\ ,\\
	b_{\max}&=\epsilon_{\mathrm{g},\star}\ ,\quad b_{\min}=\frac{GM_{\mathrm{BH}}}{v_{\mathrm{BH}}^{2}}\ .
\end{split}
\end{align}
Here we use $b_{\max}=\epsilon_{\mathrm{g},\star}\equiv 2.8a\epsilon_{\star}$ and multiply the acceleration $\mathbf{a}_{\mathrm{DF}}$ by a factor $1/[1+M_{\mathrm{BH}}/(5m_{\star})]$ to avoid double counting the frictional forces on resolved (larger) scales\footnote{The factor $1/[1+M_{\mathrm{BH}}/(5m_{\star})]$ is only significant for DCBH candidate particles with $M_{\mathrm{BH}}\simeq m_{\mathrm{gas}}\gtrsim 10 m_{\star}$. The Coulomb logarithm $\ln\Lambda$ is around 8 in our simulations, as $M_{\mathrm{BH}}\sim 10^{2}-10^{3}\ \mathrm{M}_{\odot}$, $v_{\mathrm{BH}}\sim 30\ \mathrm{km\ s^{-1}}$ and $\epsilon_{\mathrm{g},\star}\sim 8$~pc, typically.}.

\subsubsection{Black hole accretion}
\label{s2.3.4}
We adopt a modified Bondi-Hoyle formalism developed by \citet{tremmel2017romulus} to calculate the BH accretion rate $\dot{M}_{\mathrm{acc}}$, which takes into account the angular momentum of gas. For each BH particle, the characteristic rotational velocity of surrounding gas $\epsilon_{\mathrm{g}}$ away from the BH is estimated as $v_{\theta}=j/\epsilon_{\mathrm{g}}$, where $j$ is the specific angular momentum of gas particles in the radius range $3h_{\mathrm{acc}}-4h_{\mathrm{acc}}$, and $h_{\mathrm{acc}}=\min({\epsilon_{\mathrm{g}}},h_{\mathrm{BH}}/4)$. Then $v_{\theta}$ is compared with the characteristic bulk motion velocity $v_{\mathrm{bulk}}$, approximated by the smallest relative velocity of gas particles within $h_{\mathrm{BH}}$. When $v_{\theta}\le v_{\mathrm{bulk}}$, the effect of angular momentum is negligible, so that the original Bondi-Hoyle accretion formula is used:
\begin{align}
	\dot{M}_{\mathrm{acc}}=\frac{4\pi (G M_{\mathrm{BH}})^{2}\rho_{\mathrm{g}}}{(c_{s}^{2}+v_{\mathrm{g}}^{2})^{3/2}}\ ,\label{e14}
\end{align}
where $\rho_{\mathrm{g}}$ is the gas density computed from the hydro kernel at the position of the BH, $c_{s}$ is the sound speed and $v_{\mathrm{g}}$ the mass-weighted average relative velocity of gas with respect to the BH. Here $c_{s}$ is calculated with the mass-weighted average temperature of surrounding gas. While for $v_{\theta}> v_{\mathrm{bulk}}$, a rotation-based formula is used \citep{tremmel2017romulus}:
\begin{align}
	\dot{M}_{\mathrm{acc}}=\frac{4\pi (G M_{\mathrm{BH}})^{2}\rho_{\mathrm{g}}c_{s}}{\left(c_{s}^{2}+v_{\theta}^{2}\right)^{2}}\ ,\label{e15}
\end{align}
We also place an upper limit on $\dot{M}_{\mathrm{acc}}$ as 10 times the Eddington accretion rate (for primordial gas with $\mu\simeq 1.22$)\footnote{Throughout our simulations, accretion is always highly sub-Eddington, especially for Pop~III-seeded BHs. We include this upper limit for completeness, although it is never invoked here.}
\begin{align}
	\dot{M}_{\mathrm{Edd}}=2.7\times 10^{-6}\ \mathrm{M_{\odot}\ yr^{-1}}\ \left(\frac{M_{\mathrm{BH}}}{1000\ \mathrm{M_{\odot}}}\right)\left(\frac{\epsilon_{0}}{0.1}\right)^{-1}\ ,\label{e16}
\end{align}
where $\epsilon_{0}$ is the default radiative efficiency. Here we set $\epsilon_{0}=0.125$.

Given $\dot{M}_{\mathrm{acc}}$, we update the BH mass at each timestep via $\delta M_{\mathrm{BH}}=\dot{M}_{\mathrm{acc}}\delta t$. The dynamical masses of BH particles are also updated smoothly\footnote{This implies that mass conservation is not explicitly enforced in our simulations. Nevertheless, the effect is negligible since $M_{\mathrm{BH,res}}\sim m_{\mathrm{gas}}$, and the total number of BH particles is much smaller than the total number of gas particles. Besides, the fraction of accreted mass in BH mass is typically less than one percent in our simulations.} with $\delta m_{\mathrm{BH}}=\delta M_{\mathrm{BH}}$ for $M_{\mathrm{BH}}<M_{\mathrm{BH,res}}=10^{4}\ \mathrm{M}_{\odot}\sim m_{\mathrm{gas}}$. While for $M_{\mathrm{BH}}\ge M_{\mathrm{BH,res}}$, we adopt the algorithm from \citealt{springel2005modelling} (see their equ.~(35)), in which BH particles swallow nearby gas particles stochastically, and the dynamical masses are only updated when gas particles are swallowed instead of smoothly at each timestep. As in \citet{springel2005modelling}, here we also apply drag forces from accretion to BH particles according to momentum conservation.

\subsubsection{Black hole feedback}
\label{s2.3.5}
We include both thermal and mechanical feedback from BH accretion in our simulations. Following \citet{springel2005modelling}, the thermal feedback is implemented as internal energy injection into gas particles within $h_{\mathrm{BH}}$ for each BH particle. The total amount of energy to be distributed among gas particles within the hydro kernel of size $h_{\mathrm{BH}}$, over a timestep $\delta t$, is $\delta E=\epsilon_{r}L_{\mathrm{BH}}\delta t$, where $\epsilon_{r}$ is the efficiency of radiation-thermal coupling, and $L_{\mathrm{BH}}$ is the luminosity from BH accretion
\begin{align}
	L_{\mathrm{BH}}=\epsilon_{\mathrm{EM}}\dot{M}_{\mathrm{acc}}c^{2}\ .
\end{align}
Instead of using a fixed radiation efficiency $\epsilon_{\mathrm{EM}}$, we here use the method in \citet{negri2017black} to calculate $\epsilon_{\mathrm{EM}}$ as
\begin{align}
\epsilon_{\mathrm{EM}}=\frac{\epsilon_{0}A\eta}{1+A\eta}\ ,\quad \eta\equiv \dot{M}_{\mathrm{acc}}/\dot{M}_{\mathrm{Edd}}\ ,
\end{align}
where $A=100$ and $\epsilon_{0}=0.125$. This approach takes into account the transition from optically thick and geometrically thin, radiatively efficient accretion discs, to optically thin, geometrically thick, radiatively inefficient advection dominated accretion flows \citep{negri2017black}. We set $\epsilon_{r}=0.02$ as a conservative choice, based on the calibration with SMBHs in \citet{tremmel2017romulus}.

The mechanical feedback in terms of broad absorption line (BAL) winds is only applied to BHs with $M_{\mathrm{BH}}\ge M_{\mathrm{BH,res}}=10^{4}\ \mathrm{M}_{\odot}$. Once turned on, the probability of swallowing gas particles is boosted by a factor of $1/f_{\mathrm{acc}}$. For each swallowed gas particle with mass $\delta m$ above (below) the accretion plane\footnote{The accretion plane is orthogonal to the (specific) momentum of surrounding gas.}, a wind gas particle with mass $m_{w}=(1-f_{\mathrm{acc}})\delta m$ is launched with a kick velocity of $v_{w}$ relative to the swallowed particle, along (opposite to) the direction of the angular momentum of surrounding gas. Here we assume that $v_{w}=200\ \mathrm{km\ s^{-1}}\ (M_{\mathrm{BH}}/M_{\mathrm{BH,res}})^{1/2}$, and calculate $f_{\mathrm{acc}}$ with \citep{negri2017black}
\begin{align}
	f_{\mathrm{acc}}=\frac{1}{1+\eta_{w}}\  ,\quad \eta_{w}=2\epsilon_{w}\frac{c^{2}}{v_{w}^{2}}\  ,\quad\epsilon_{w}=\frac{\epsilon_{w0}A_{w}\eta}{1+A_{w}\eta}\ ,
\end{align}
where $\epsilon_{w0}=10^{-4}$ and $A_{w}=1000$. Note that since accretion is insignificant for the dominant Pop~III-seeded BHs under stellar and BH feedback, we always have $M_{\mathrm{BH}}\lesssim M_{\mathrm{BH,res}}$, except for rare DCBH candidates, and BAL winds have little effect in our simulations.


\section{Simulation results}
\label{s3}
In this section we summarize the main features of our simulations in the EM window, in terms of star formation histories, BH growth and halo-stellar-BH mass scaling relations, in comparison with observational constraints and previous theoretical predictions. For conciseness, we here only show the results for the fiducial stellar feedback model and resolution under the \texttt{Lseed} BH seeding scenario. The results for \texttt{Hseed} are similar. The effects of stellar feedback and resolution are discussed in appendices~\ref{a1} and \ref{a2}. We will discuss the GW signals in the next section.

\subsection{Star formation history}
\label{s3.1}

\begin{figure*}
\hspace{-5pt}
\centering
\subfloat[Total]{\includegraphics[width= 1.065\columnwidth]{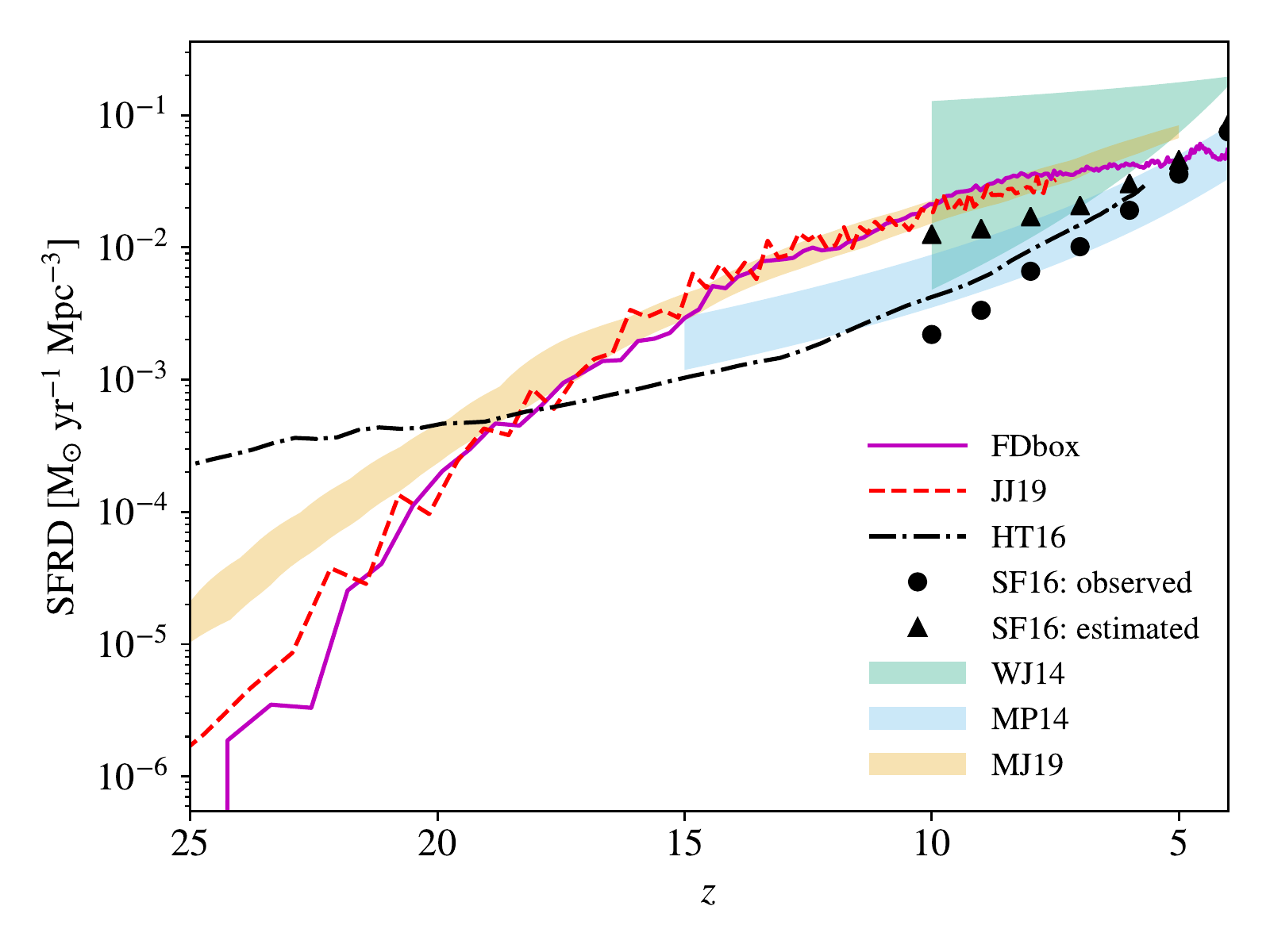}}
\subfloat[Pop~III]{\includegraphics[width= 1.065\columnwidth]{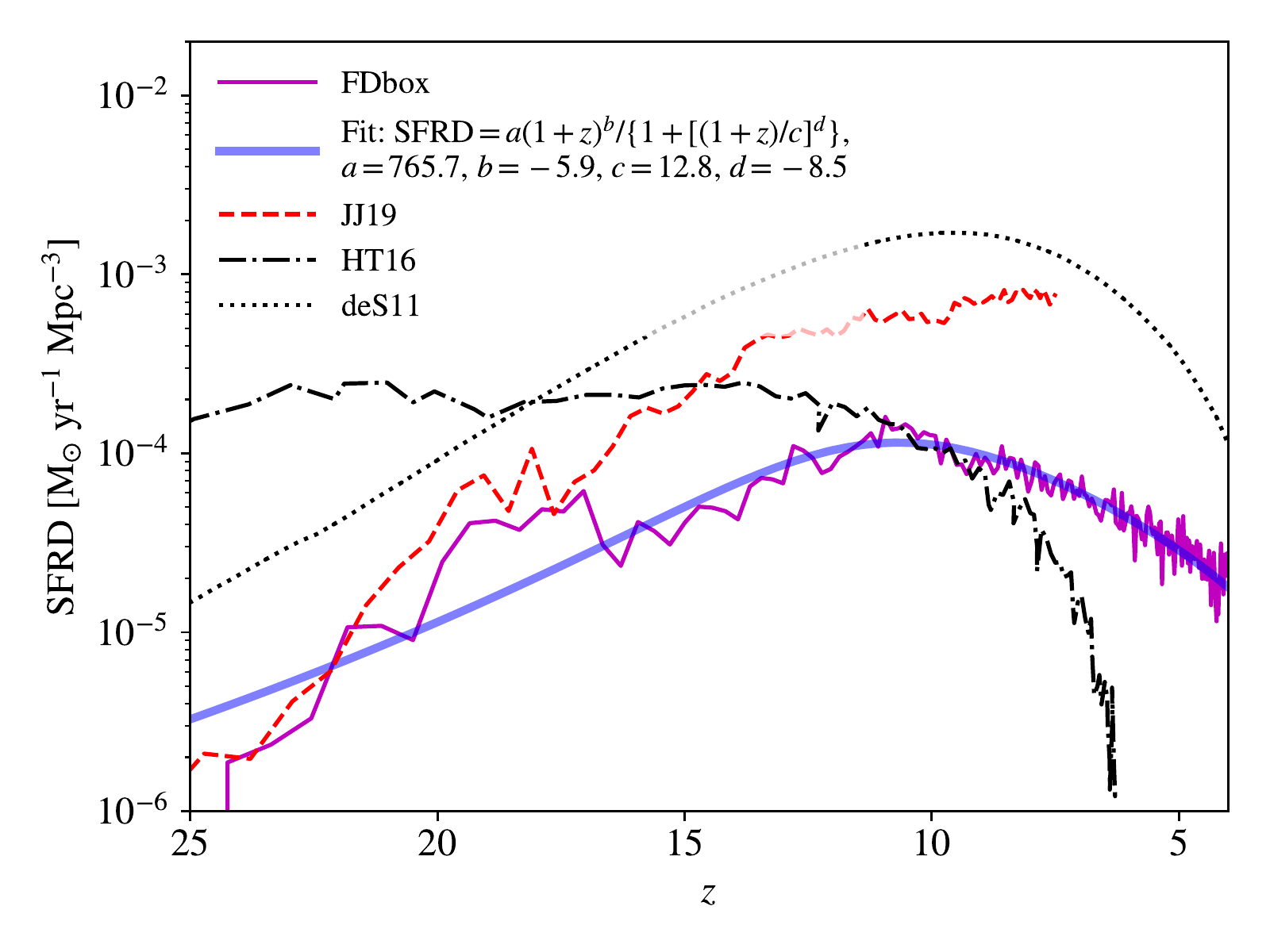}}
\vspace{-10pt}
\caption{Star formation histories from \texttt{FDbox\_Lseed}. \textit{Left panel}: total (co-moving) SFRD, in comparison with the simulation results from \citealt{jaacks2018legacy} (JJ19), semi-analytical models in \citealt{hartwig2016} (HT16) and \citealt{Mirocha2018} (MJ19, based on the observed 21-cm absorption signal), and (extrapolated) observations from \citealt{wei2014cosmological} (WJ14), \citealt{madau2014araa} (MP14) and \citealt{finkelstein2016observational} (SF16). \textit{Right panel}: Pop~III SFRD, in comparison with the JJ19 simulation, as well as semi-analytical models from HT16 and \citealt{deS2011} (deS11). For deS11, we have rescaled their most optimistic SFRD by a factor of 0.5 to be consistent with the recent \textit{Planck} measurement of the optical depth to electron scattering \citep{inayoshi2016gravitational}. Our simulated SFRD is fitted to the form $\dot{\rho}_{\star,\mathrm{PopIII}}(z)\mathrm{\ [M_{\odot}\ yr^{-1}\ Mpc^{-3}]}=a(1+z)^{b}/\{1+[(1+z)/c]^{d}\}$. For \texttt{FDbox\_Lseed}, the best-fitting parameters are $a=765.7$, $b=-5.92$, $c=12.83$ and $d=-8.55$ (thick blue curve). The total density of Pop~III stars formed across cosmic time inferred from this best-fitting formula is $\simeq 7\times 10^{4}\ \mathrm{M_{\odot}\ Mpc^{-3}}$, consistent with the constraints in \citet{visbal2015} set by \textit{Planck} data.}
\label{fsfh}
\end{figure*}

Fig.~\ref{fsfh} shows the (co-moving) SFRDs for all stars (left panel, including both Pop~III and Pop~II) and Pop~III stars (right panel) measured in \texttt{FDbox\_Lseed}. We compare our simulated SFRDs with the results from \citet{jaacks2018legacy}, semi-analytical models in \citet{hartwig2016}, \citet{deS2011} and \citealt{Mirocha2018} (based on the observed 21-cm absorption signal), and (extrapolated) observational results from \citet{wei2014cosmological}, \citet{madau2014araa} and \citet{finkelstein2016observational}. In the \texttt{box} setup, the total SFRD is dominated by Pop~II stars at $z\lesssim 18$. Despite the different implementations of SF and stellar feedback, our total SFRD agrees well with that of \citet{jaacks2018legacy}, whose simulation setup is identical to our \texttt{box} run. It is also within the range of observational constraints at $z\lesssim 10$, with an upper limit from \citet{wei2014cosmological} based on \textit{Swift} long gamma-ray bursts and a lower limit from \citet{madau2014araa,finkelstein2016observational} based on UV and IR galaxy surveys. However, it is lower than the semi-analytical models from \citet{Mirocha2018} and \citet{hartwig2016} (by up to 2 orders of magnitude) in the Pop~III-dominated regime at $z\gtrsim 18$. One explanation is that our simulations have more realistic/stricter conditions for Pop~III star formation, while semi-analytical models use idealized criteria (based on efficient $\mathrm{H_{2}}$ cooling) to identify star-forming minihaloes, which may overproduce the abundance of Pop~III hosts and thus, the resulting Pop~III SFRD. We do not expect these uncertainties at the highest redshifts to play an important role in inferring the GW signals from Pop~III-seeded BBHs, for which the SFRD at lower redshifts is more relevant.

For Pop~III stars, the simulated SFRD reaches a peak at $z\sim 11$ and declines rapidly thereafter. We fit our Pop~III SFRD to the form $\dot{\rho}_{\star,\mathrm{PopIII}}(z)\ \mathrm{[M_{\odot}\ yr^{-1}\ Mpc^{-3}]}=a(1+z)^{b}/\{1+[(1+z)/c]^{d}\}$ and find the best-fitting parameters $a=765.7$, $b=-5.92$, $c=12.83$ and $d=-8.55$ for \texttt{FDbox\_Lseed}. By integrating this best-fitting SFRD to $z=4$, which is assumed as the end of metal-free SF\footnote{Our simulations do not self-consistently model the global reionization process, which can suppress Pop~III star formation by Jeans mass filtering. For simplicity, we assume that there is no significant Pop~III star formation after $z=4$.}, we obtain a total density of Pop~III stars formed across cosmic time as $\rho_{\star,\mathrm{PopIII}}=\int_{4}^{\infty} dz\left[\dot{\rho}_{\star,\mathrm{PopIII}}(z) dt/dz\right] \simeq 7\times 10^{4}\ \mathrm{M_{\odot}\ Mpc^{-3}}$, which is consistent with the constraints in \citet{visbal2015} set by \textit{Planck} data. However, the discrepancies among different models for Pop~III SFRD are significant. For instance, our result is lower than that of \citet{jaacks2018legacy} by up to one order of magnitude at $z\lesssim 17$. The reason is that we have implemented stronger stellar feedback such as local LW fields, suppressing $\mathrm{H}_{2}$ formation, and SN-driven winds (which enhance metal enrichment in the IGM). When compared with the semi-analytical Pop~III SFRDs in \citet{deS2011} and \citet{hartwig2016}, which are widely used to derive the GW signals of \textit{in-situ} Pop~III BBHs, disagreements of up to a factor of 10 occur (at $z\lesssim 7$ and $z\gtrsim 15$ for \citealt{hartwig2016} and $z>4$ for \citealt{deS2011}). Since direct measurement of the Pop~III SFRD remains challenging even in the \textit{JWST} era, we regard the agreement with the indirect constraint on $\rho_{\star,\mathrm{PopIII}}(z\ge 0)\simeq 10^{5}\ \mathrm{M_{\odot}\ Mpc^{-3}}$ from \textit{Planck} data \citep{visbal2015} as a consistency check for our SFRD. 
Note that the indirect constraints from the \textit{Planck} optical depth are sensitive to assumptions on the escape fraction of ionizing photons $f_{\rm esc}$, and the Pop~III ionization efficiency $\eta_{\rm ion}$. The constraint in \citet{visbal2015} of $\sim 10^{4-5}\ \mathrm{M_{\odot}\ Mpc^{-3}}$ is on the strong side. Weaker constraints are also possible, e.g. $\simeq 10^{6}\ \mathrm{M_{\odot}\ Mpc^{-3}}$ for $f_{\rm esc}=0.1$ and $\eta_{\rm ion}=5\times 10^{4}$ \citep{inayoshi2016gravitational}. 
To be self-consistent, when comparing the GW signals of our simulated ex-situ BBHs with those of in-situ BBHs in the literature, we rescale the latter results to meet the same integral condition. The BH seeding models have little effect on star formation in our simulations, as BH accretion is always unimportant and the relevant feedback is also too weak to make any difference.

\subsection{Black hole growth}
\label{s3.2}

\begin{figure}
\includegraphics[width=1\columnwidth]{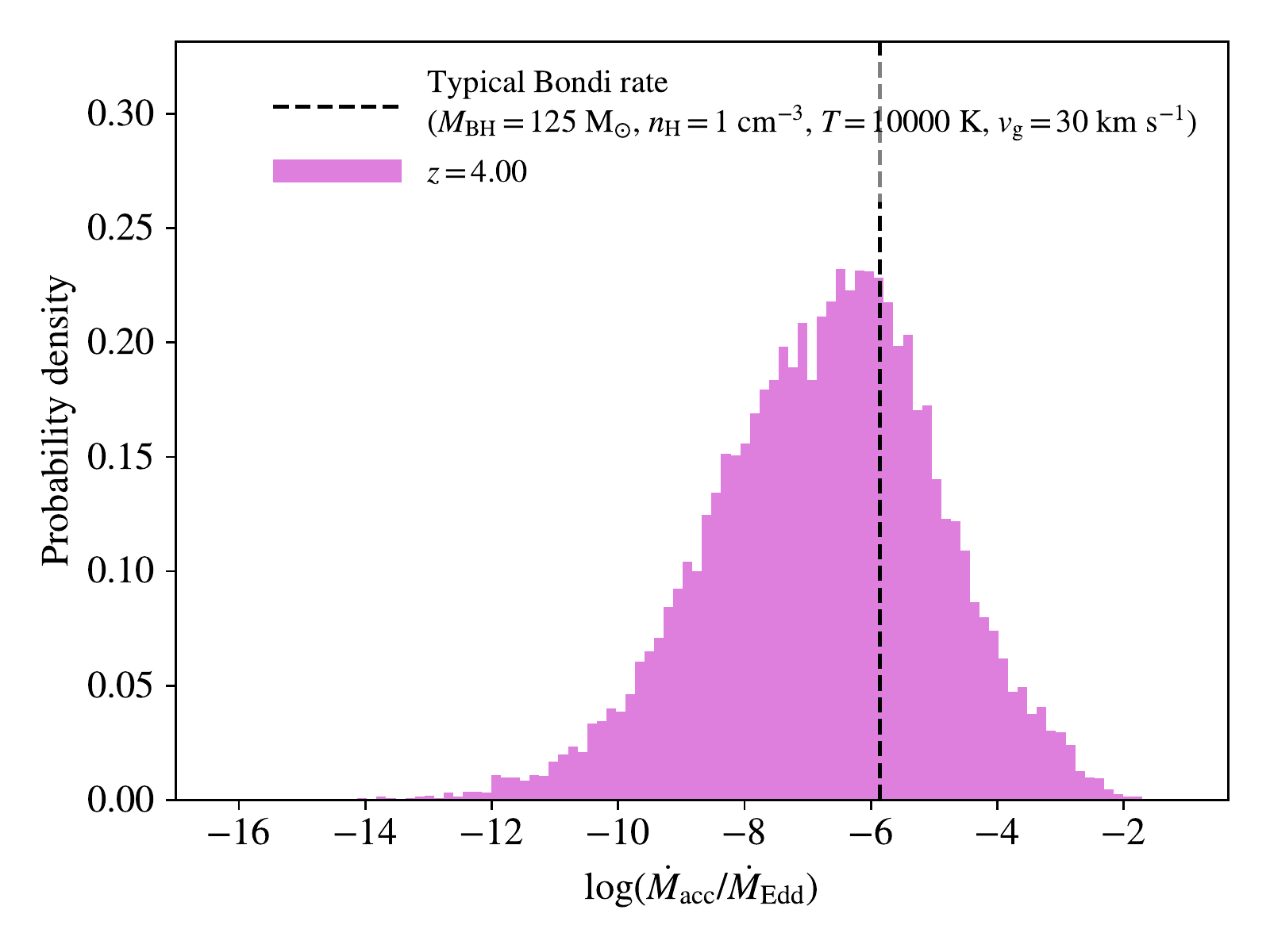}
\caption{Distribution of Eddington ratio $\dot{M}_{\mathrm{acc}}/\dot{M}_{\mathrm{Edd}}$ in \texttt{FDbox\_Lseed} at $z=4$. This log-normal-like distribution peaks at $\dot{M}_{\mathrm{acc}}/\dot{M}_{\mathrm{Edd}}\sim 10^{-6}$, consistent with the results in high-resolution simulations with detailed BH feedback (e.g. \citealt{alvarez2009accretion}). This also corresponds to the typical Bondi accretion rate (dashed vertical line, from Equ.~\ref{e14}) for Pop~III-seeded BHs embedded in the ISM of typical haloes ($M_{\mathrm{halo}}\sim 10^{8}-10^{9}\ \mathrm{M_{\odot}}$) at $z=4$, given $M_{\mathrm{BH}}=125\ \mathrm{M_{\odot}}$, $n_{\mathrm{H}}=1\ \mathrm{cm^{-3}}$, $T=10^{4}\ \mathrm{K}$ and $v_{\mathrm{g}}=30\ \mathrm{km\ s^{-1}}$.}
\label{f6}
\end{figure}

\begin{figure}
\includegraphics[width=1\columnwidth]{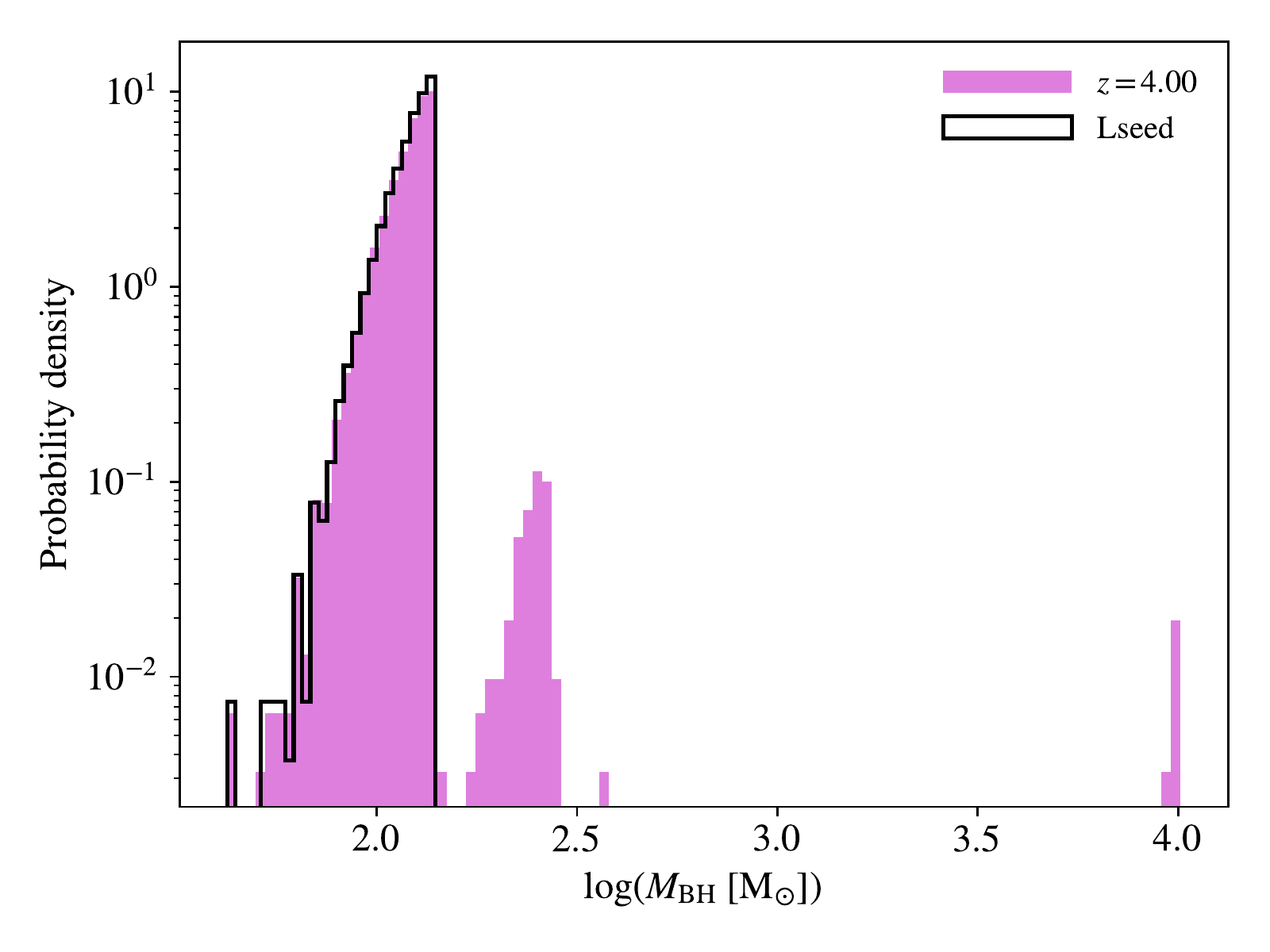}
\caption{Distribution of BH particle masses in \texttt{FDbox\_Lseed} at $z=4$ (purple histograms). The majority of BH particles follows the seed mass distribution (black contour), and a group of more massive BH particles with $M_{\mathrm{BH}}\sim 200-500\ \mathrm{M}_{\odot}$ (which denote binary/multiple BH systems) have been formed from combinations of BH particles. A tiny fraction ($\sim2\times 10^{-4}$) of BH particles with $M_{\rm{BH}}\simeq 10^{4}\ \rm{M}_{\odot}$ represent DCBH candidates.}
\label{f7}
\end{figure}

Throughout our simulations, accretion is mostly sub-Eddington for Pop~III-seeded BHs in the presence of stellar and BH feedback. For instance, Fig.~\ref{f6} shows the distribution of Eddington ratio $\dot{M}_{\mathrm{acc}}/\dot{M}_{\mathrm{Edd}}$ in \texttt{FDbox\_Lseed} at $z=4$, which has a log-normal-like shape and peaks at $\dot{M}_{\mathrm{acc}}/\dot{M}_{\mathrm{Edd}}\sim 10^{-6}$. This is consistent with the results in high-resolution 3D simulations with detailed BH feedback (e.g. \citealt{alvarez2009accretion}). By the end of the \texttt{FD} simulations in the \texttt{zoom} setup ($z=4$), the average (median) values of the accreted mass ratio $M_{\mathrm{acc}}/M_{\mathrm{BH}}$ are $1.8\ (1.3)\times 10^{-3}$ for \texttt{Lseed} and $5.2\ (4.1)\times 10^{-3}$ for \texttt{Hseed}. Even without Pop~II PI heating and SN-driven winds (\texttt{NSFDBK}), the average/median value of $M_{\mathrm{acc}}/M_{\mathrm{BH}}$ is still less than 0.01, and the maximum accreted ratio is $\sim 0.04$ across all simulation snapshots, indicating that accretion is usually unimportant. Our results agree with those in previous studies based on 3D cosmological simulations, showing that Pop~III seeds can hardly grow via accretion at high redshifts (e.g. \citealt{johnson2007aftermath,alvarez2009accretion,hirano2014one,smith2018growth}). However, semi-analytical models and simulations have found that it is possible for Pop~III seeds to grow sufficiently by super-Eddington accretion under peculiar conditions (e.g. \citealt{madau2014super,volonteri2015case,pezzulli2016super,
inayoshi2016hyper,toyouchi2019super}). Our simulations did not capture this scenario, possibly due to different treatments of BH and stellar feedback and limited resolution. 
The mass spectrum of BH particles at $z=4$ from \texttt{FDbox\_Lseed} is shown in Fig.~\ref{f7}, which reflects the initial distribution of Pop~III seed masses (see Fig.~\ref{f4}) and combinations of BH particles into binary/multiple systems.

\begin{figure}
\includegraphics[width=1\columnwidth]{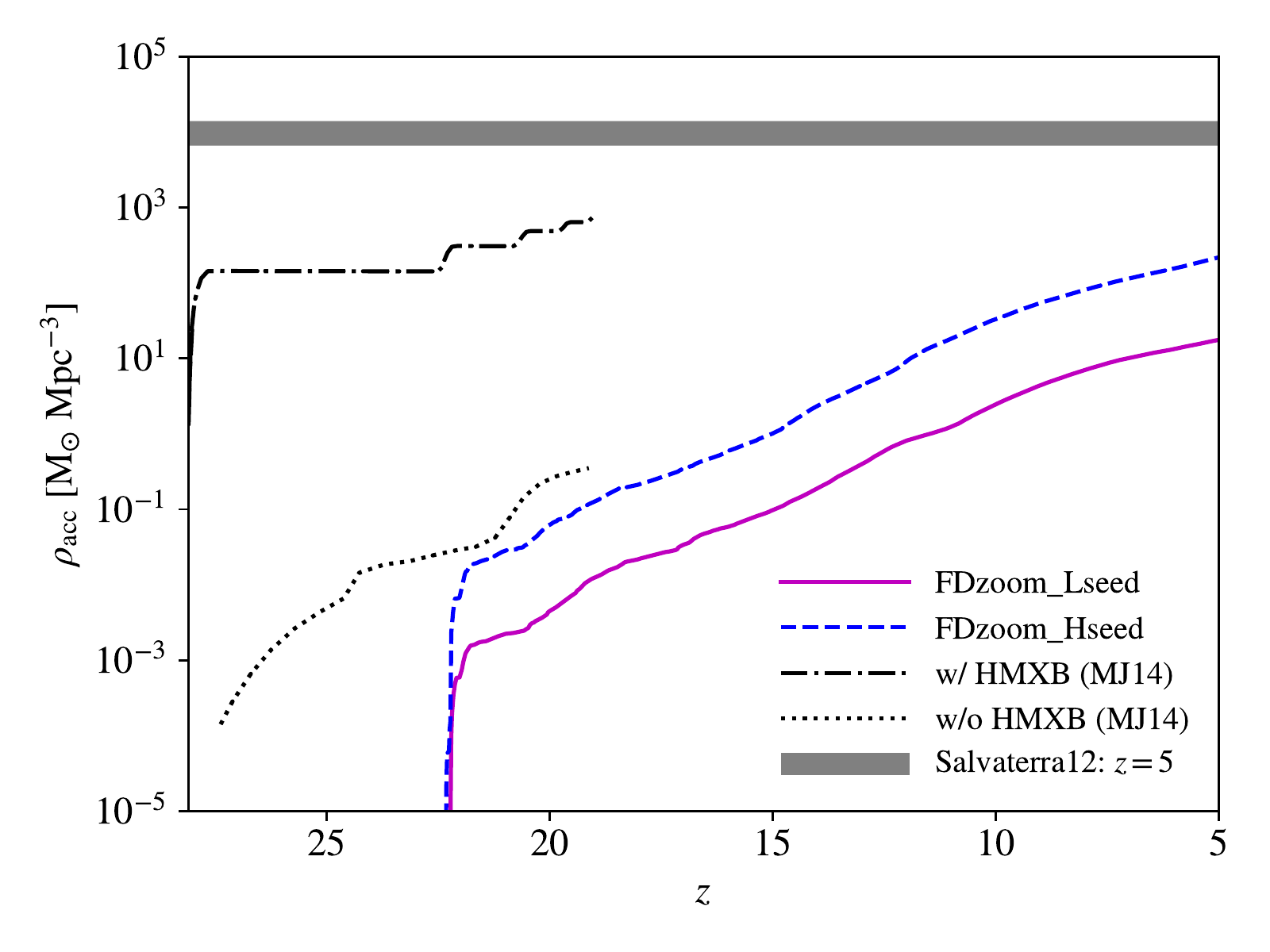}
\caption{Total (co-moving) accreted BH mass density vs. redshift in \texttt{FDzoom\_Lseed} (solid) and \texttt{FDzoom\_Hseed} (dashed). The shaded region shows the upper limit on $\rho_{\mathrm{acc}}$ at $z=5$, placed by the unresolved cosmic x-ray background, from \citealt{salvaterra2012limits}. Our simulations results are always below this limit (by at least a factor of 50), implying that under stellar feedback, BH accretion is dominated by more massive haloes ($M_{\mathrm{halo}}\gtrsim 10^{11}\ \mathrm{M}_{\odot}$) and/or HMXBs not modelled in our simulations with limited volumes and resolution. For comparison, we also show the results of the zoom-in simulation in \citealt{jeon2014radiative} (MJ14) with (dashed-dotted) and without (dotted) HMXBs, which indicate that including HMXBs can enhance BH accretion by two orders of magnitude. Our simulations do not capture the early BH accretion in MJ14 at $z\gtrsim 22.5$ due to limited resolution.}
\label{rhoacc}
\end{figure}

Regarding global properties, Fig.~\ref{rhoacc} shows the redshift evolution of the total (co-moving) accreted mass density $\rho_{\mathrm{acc}}$ for BHs in \texttt{FDzoom\_Lseed} and \texttt{FDzoom\_Hseed}. 
By the end of the simulation ($z=4$), the total accreted mass only makes up a small fraction ($1.6-5.5 \times 10^{-3}$) of the total BH mass. Besides, we have $\rho_{\mathrm{acc}}(z=4)\simeq 20\ (200)\ \mathrm{M_{\odot}\ Mpc^{-3}}$ for \texttt{Lseed} (\texttt{Hseed}), much lower than the upper limit $\sim 10^{4}\ \mathrm{M_{\odot}\ Mpc^{-3}}$ placed by the unresolved cosmic X-ray background \citep{salvaterra2012limits}. 
Actually, the values of $\rho_{\mathrm{acc}}$ in our simulations 
are always below the observational upper limit by at least one order of magnitude, even in the \texttt{zoom} setup for \texttt{Hseed} without Pop~II stellar feedback\footnote{Interestingly, BH accretion remains almost unchanged when Pop~II stellar feedback is turned off, as shown in Appendix~\ref{a1}. The reason is that without Pop~II winds and PI heating, all cold gas is efficiently turned into stars, unable to enhance BH accretion.}. This implies that 
BH accretion is dominated by more massive haloes ($M_{\mathrm{halo}}\gtrsim 10^{11}\ \mathrm{M}_{\odot}$), missing in our simulations with limited volumes, especially in the Pop~II-dominated regime ($z\lesssim 18$). Note that our simulations do not include high-mass X-ray binaries (HMXBs), consisting of Pop~III stars and BHs, which can significantly boost BH accretion (e.g. \citealt{jeon2014radiative}). Furthermore, our sub-grid model for BH accretion may underestimate rates due to insufficient resolution, as small ($\lesssim 10^{5}\ \mathrm{M_{\odot}}$), cold gas clumps in the ISM are not resolved.



\subsection{Mass scaling relations}
\label{s3.3}
To evaluate the halo-stellar-BH mass scaling relations in our simulations, we identify DM haloes and their (central/satellite) galaxies using the standard friends-of-friends (FOF) method within the \textsc{caesar} code \citep{thompson2014pygadgetreader}.  
The linkage parameter is $b=0.2$ for DM and gas particles, and $b=0.02$ for stellar particles. For simplicity, we define the central BH mass as the total mass of all BHs within 0.5$R_{1/2}$ from the galaxy center, where $R_{1/2}$ is the galaxy's half mass radius for baryons. 

\begin{figure}
\includegraphics[width=1\columnwidth]{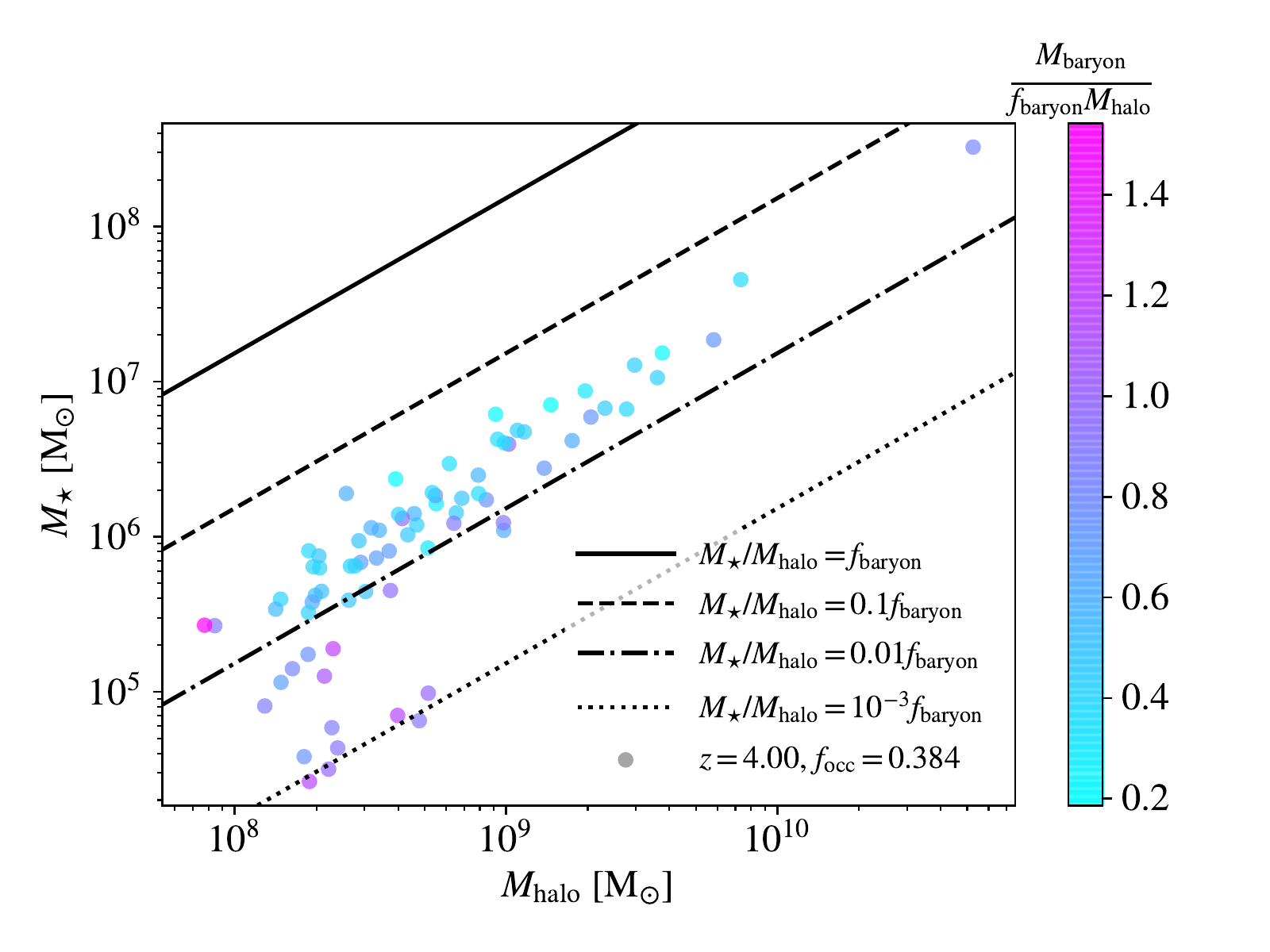}
\caption{Stellar-halo mass relation for haloes hosting resolved galaxies (with $M_{\star}\gtrsim 2\times 10^{4}\ \mathrm{M}_{\odot}$) at $z=4$ in \texttt{FDzoom\_Lseed}. The baryon fraction $M_{\mathrm{baryon}}/M_{\mathrm{halo}}$ with respect to the cosmic average $f_{\mathrm{baryon}}=\Omega_{b}/\Omega_{m}$ is color coded. $f_{\mathrm{occ}}$ is the fraction of atomic cooling haloes that host resolved galaxies. The median stellar mass fraction $M_{\star}/M_{\mathrm{baryon}}$ is $\sim 3\%$, and the maximum is $\sim 20\%$, consistent with abundance matching results for low-mass galaxies where stellar feedback is efficient. Galaxies with higher stellar mass fractions tend to have lower gas fractions, which is a signature of outflows driven by strong stellar feedback. }
\label{fMs_Mh}
\end{figure}

Fig.~\ref{fMs_Mh} shows the stellar-halo mass relation for atomic-cooling haloes, with $M_{\mathrm{halo}}>2.5\times 10^{7}\ \mathrm{M}_{\odot}[(1+z)/10]^{-3/2}$ (\citealt{yoshida2003simulations,trenti2009formation}), hosting resolved galaxies (with $M_{\star}> 32m_{\star}\simeq 2\times 10^{4}\ \mathrm{M}_{\odot}$) at $z=4$ in \texttt{FDzoom\_Lseed}, where the baryon fraction $M_{\mathrm{baryon}}/(f_{\mathrm{baryon}}M_{\mathrm{halo}})$ with respect to the cosmic mean $f_{\mathrm{baryon}}=\Omega_{b}/\Omega_{m}$ is color coded. We find that the median stellar mass fraction $M_{\star}/M_{\mathrm{baryon}}$ is $\sim 3\%$, and the maximum is $\sim 20\%$, consistent with abundance matching results for low-mass galaxies where stellar feedback is efficient. It is also evident that galaxies with higher stellar mass fractions tend to have lower gas fractions, which is a signature of outflows driven by strong stellar feedback. Similar trends exist for \texttt{Hseed}, under fiducial stellar feedback.

\begin{figure}
\includegraphics[width=1\columnwidth]{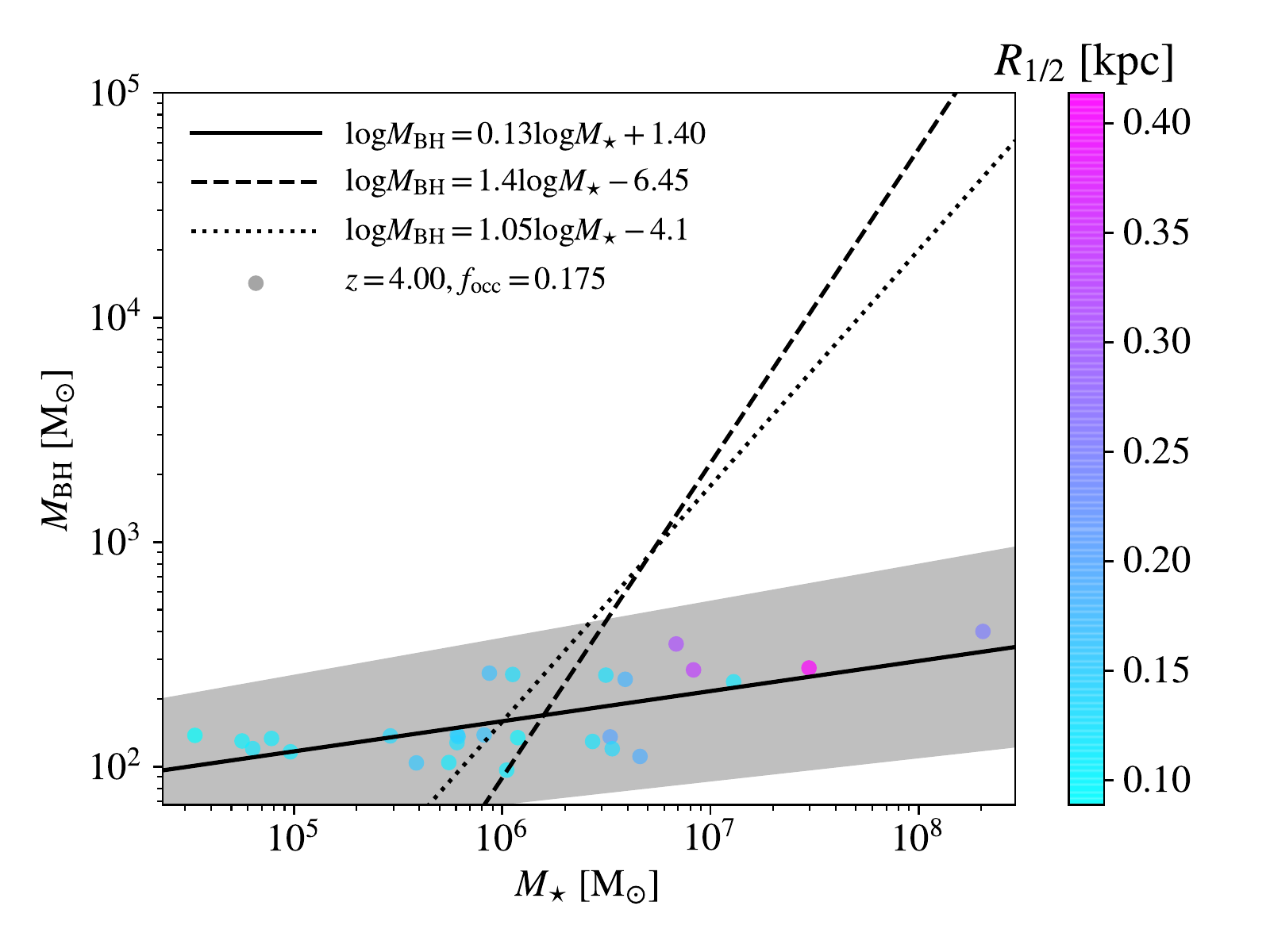}
\caption{BH-stellar mass relation at $z=4$ in \texttt{FDzoom\_Lseed}. The central BH mass is defined as the total mass of all BHs within 0.5$R_{1/2}$ from the galaxy center, where $R_{1/2}$ (color coded) is the half mass radius for baryons. $f_{\mathrm{occ}}$ is the fraction of galaxies (with $M_{\star}\gtrsim 2\times 10^{4}\ \mathrm{M}_{\odot}$) that host central BHs. For comparison, we plot the extrapolated relations derived from local observations for high-stellar mass ellipticals and bulges (dashed), as well as for moderate-luminosity AGNs in low-mass haloes (dotted), from \citealt{reines2015relations} (with $M_{\star}\gtrsim 3\times 10^{8}\ \mathrm{M}_{\odot}$ and $z<0.055$). The best-fitting relation from our simulation is shown with the solid line and the shaded region (for 1$\sigma$ uncertainties).}
\label{fMbh_Ms}
\end{figure}

Fig.~\ref{fMbh_Ms} shows the BH-stellar mass relation at $z=4$ in \texttt{FDzoom\_Lseed} for all resolved galaxies (including satellites), in comparison with the extrapolated relations derived from local observations for high-stellar mass ellipticals and bulges, $\log M_{\mathrm{BH}}=1.4\log M_{\star}-6.45$, as well as for moderate-luminosity AGNs in low-mass haloes, $\log M_{\mathrm{BH}}=1.05\log M_{\star}-4.1$ (with $M_{\star}\gtrsim 3\times 10^{8}\ \mathrm{M}_{\odot}$ and $z<0.055$; \citealt{reines2015relations}), where all masses are in $\mathrm{M_{\odot}}$. It turns out that our simulations tend to overpredict the BH mass in the low-mass regime ($M_{\star}\lesssim 10^{6}\ \mathrm{M_{\odot}}$), while underestimating the BH mass at the high-mass end ($M_{\star}\gtrsim 10^{7}\ \mathrm{M_{\odot}}$). The former trend will be particularly obvious for galaxies hosting DCBHs, in which the BH mass may even exceed the stellar mass (e.g. \citealt{agarwal2013unravelling}). 

The best-fitting power-law relation for \texttt{FDzoom\_Lseed} at $z=4$ is $\log M_{\mathrm{BH}}=0.13\log M_{\star}+1.4$. This indicates that the correlation between BH and stellar masses is much weaker in the simulated low-mass systems ($M_{\star}\lesssim 10^{8}\ \mathrm{M}_{\odot}$) at high redshifts ($z\gtrsim 4$) than observed in the nearby Universe. 
Interestingly, we find that the correlation strength (i.e. the power-law index) slightly decreases with redshift, or in general increases when BH accretion and halo mergers proceed. 
This outcome implies that sufficient BH accretion and structure formation at $z\lesssim 4$ are necessary to explain the observed BH-stellar mass scaling relation, relevant only for massive haloes ($M_{\mathrm{halo}}\gtrsim 10^{11}\ \mathrm{M}_{\odot}$) at low-redshifts ($z\lesssim 4$). Actually, the recent work by \citet{delvecchio2019galaxy} predicts that a tight super-linear stellar-BH mass relation only applies at $z\lesssim 2$ for galaxies with $M_{\star}\gtrsim 10^{9}\ \mathrm{M}_{\odot}$, based on the star-forming `main-sequence' and stellar mass dependent ratio between BH accretion rate and star formation rate. Compared with the representative case of \texttt{FD} and \texttt{Lseed} discussed here, the BH-stellar mass scaling relations for \texttt{NSFDBK} and/or \texttt{Hseed} runs are only different in terms of normalization, while the evolution of the slope (i.e. correlation strength) is almost identical, so that the above trends also apply.

\section{Gravitational wave signals}
\label{s4}
\subsection{BH binary evolution}
\label{s4.1}
As mentioned in Section~\ref{s2.3.2}, combinations of BH particles in our simulations represent formation of \textit{ex-situ} BH binaries or multiple systems (by dynamical capture), whose subsequent evolution is beyond our resolution. Actually, the dynamics of coalescence is rather complex, involving various astrophysical aspects such as a clumpy ISM, DM distribution, and the presence of nuclear star clusters \citep{rovskar2015orbital,tamfal2018formation,ogiya2019}, many of which are not taken into account in our simulations. As an exploratory approach, we employ semi-analytical models to calculate the lifetimes of BH binaries under simple assumptions and parameterizations, depending on their large-scale environments. In this way, we can investigate a broad range of parameters to estimate the lower and upper bounds for the GW signals of our simulated BBHs.

\subsubsection{Orbital evolution}
\label{s4.1.1}
Assuming that the orbital evolution of BBHs is driven mostly by surrounding stars (in `dry' galaxy mergers)\footnote{It is a reasonable approximation to ignore the effect of gas on binary evolution, since Pop~III stars are typically massive with strong PI and SN feedback which leads to low gas density around the newly-born BHs. Besides, considering the uncertainties in the density profiles of stars around BBHs, the additional friction by gas can be captured with steep stellar density profiles.} 
and GW emission, the evolution of the semimajor axis takes the form \citep{sesana2015scattering} 
\begin{align}
\frac{da}{dt}=\left.\frac{da}{dt}\right\vert_{\mathrm{3b}}+\left.\frac{da}{dt}\right\vert_{\mathrm{GW}}=-Aa^{2}-\frac{B}{a^{3}}\ .\label{e20}
\end{align}
The first term denotes the effect of interactions with surrounding stars (i.e. three-body hardening), while the second term corresponds to the energy and angular momentum loss via GWs. Here  
\begin{align}
\begin{split}
A&=\frac{GH\rho_{\mathrm{inf}}}{\sigma_{\mathrm{inf}}}\ ,\\
B&=\beta F(e)\ ,\quad \beta=\frac{64G^{3}M_{1}M_{2}M}{5 c^{2}}\ ,\label{e21}
\end{split}
\end{align} 
where $M_{1}$ and $M_{2}$ are the masses of the primary and secondary BHs, $M=M_{1}+M_{2}$, $\sigma_{\mathrm{inf}}$ and $\rho_{\mathrm{inf}}$ are the velocity dispersion and stellar density at the radius of influence $r_{\mathrm{inf}}$ of the BH binary, $F(e)=(1-e^{2})^{-7/2}[1+(73/24)e^{2}+(37/96)e^{4}]$ given the eccentricity $e$, and $H\sim 15-20$ is a dimensionless parameter.

The binary system spends most of its lifetime in a phase with a characteristic semimajor axis that can be estimated by imposing $(da/dt)|_{\mathrm{3b}}=(da/dt)|_{\mathrm{GW}}$:
\begin{align}
a_{\star/\mathrm{GW}}=\left(\frac{B}{A}\right)^{1/5}=\left[\frac{64G^{2}\sigma_{\mathrm{inf}}M_{1}M_{2}MF(e)}{5 c^{2} H\rho_{\mathrm{inf}}}\right]^{1/5}\ .
\end{align}
The binary hardening time, i.e. the time taken for $a$ to reach $a_{\star/\mathrm{GW}}$, can be estimated with
\begin{align}
t_{\mathrm{HD}}\simeq \frac{1}{Aa_{\star/\mathrm{GW}}}=\left(\frac{1}{A^{4}B}\right)^{1/5}\ .
\end{align}
For typical parameters $e=0.99$, $M_{1}=M_{2}\sim 100\ \mathrm{M_{\odot}}$, $\sigma_{\mathrm{inf}}\sim 10\ \mathrm{km\ s^{-1}}$, $\rho_{\mathrm{inf}}\sim 10^{3}\ \mathrm{M_{\odot}\ pc^{-3}}$ and $H=17.5$ (see the next subsection for details), we have $a_{\star/\mathrm{GW}}\sim 5\times 10^{-5}$~pc and $t_{\mathrm{HD}}\sim 2.4\ \mathrm{Gyr}$. Given such small $a_{\star/\mathrm{GW}}$ and large $t_{\mathrm{HD}}$, it is very challenging to fully simulate the binary hardening process in cosmological simulations.

We assume that the eccentricity is fixed to the initial value in the three-body hardening stage. The contribution from GW emission to hardening is comparable to that from the surrounding stars when $a$ is close to $a_{\star/\mathrm{GW}}$. In other words, the system only emits strong GW signals at $a\lesssim a_{\star/\mathrm{GW}}$, when actual coalescence is about to happen. The time spent in the subsequent GW dominated phase ($a<a_{\star/\mathrm{GW}}$) can be estimated with\footnote{\url{http://www.physics.usu.edu/Wheeler/GenRel2013/Notes/GravitationalWaves.pdf}}
\begin{align}
\begin{split}
t_{\mathrm{col}}&=\frac{12}{19}\frac{c_{0}^{4}}{\beta}\int_{0}^{e}dx\frac{x^{29/19}[1+(121/304)x^{2}]^{1181/2299}}{(1-x^{2})^{3/2}}\ ,\\
c_{0}&=\frac{a_{\star/\mathrm{GW}}(1-e^{2})}{e^{12/19}}\left[1+\frac{121}{304}e^{2}\right]^{-870/2299}\ ,
\end{split}
\end{align}
assuming that the effect of surrounding stars is negligible. In our model, $t_{\mathrm{col}}$ and $t_{\mathrm{HD}}$ are comparable, and the ratio $t_{\mathrm{col}}/t_{\mathrm{HD}}$ increases with $e$, covering the range $1/4-2$. 

For simplicity, we set $H=17.5$, and the only unknowns in the above formulas are $\sigma_{\mathrm{inf}}$ and $\rho_{\mathrm{inf}}$. However, for Pop~III-seeded BHs with $M_{\mathrm{BH}}\lesssim 10^{3}\ \mathrm{M}_{\odot}$, the influence radii are typically sub-parsec, far beyond our spatial resolution. Besides, $\sigma_{\mathrm{inf}}$ and $\rho_{\mathrm{inf}}$ are defined for the long-term (Gyr-scale) quasi-steady state of the background stellar system, unavailable from our simulations designed for high redshifts ($z\gtrsim 4$). In light of this, as in \citet{sesana2015scattering}, we estimate $\sigma_{\mathrm{inf}}$ and $\rho_{\mathrm{inf}}$ with BH-bulge scaling relations, as described below.

\subsubsection{Environments around BBHs}
\label{s4.1.2}
We assume that the stars surrounding a BBH follow the Dehnen density profile \citep{dehnen1993family}
\begin{align}
\rho_{\star}(r)=\frac{(3-\gamma)M_{\star}}{4\pi}\frac{r_{0}}{r^{\gamma}(r+r_{0})^{4-\gamma}}\ ,\label{e25}
\end{align} 
where $\gamma$ is the inner slope, $M_{\star}$ is the bulge mass, and $r_{0}$ is the core size, which is related to the bulge effective radius $R_{\mathrm{eff}}$, as well as $r_{\mathrm{inf}}$:
\begin{align}
R_{\mathrm{eff}}&=\frac{3r_{0}}{4[2^{1/(3-\gamma)}-1]}\ ,\label{e26}\\
r_{\mathrm{inf}}&=\frac{r_{0}}{[M_{\star}/(2M)]^{1/(3-\gamma)}-1}\ .\label{e27}
\end{align}
To estimate the bulge mass $M_{\star}$, we consider two models. The first model (\texttt{obs-based}) extrapolates the observed scaling relation $M_{\star}=10^{11}\ \mathrm{M}_{\odot}\left(M_{9}/0.49\right)^{1/1.16}$ \citep{kormendy2013coevolution} to our low-mass regime, where $M_{9}\equiv M/(10^{9}\ \mathrm{M}_{\odot})$. In the second model (\texttt{sim-based}), we estimate the bulge mass with the best-fitting BH-stellar mass relation measured at the last snapshot in the corresponding simulation ($z=4$, see Fig.~\ref{fMbh_Ms} for an example): $M_{\star}=f_{\mathrm{bulge}}M_{\star,\mathrm{sim}}(M)$, where $M_{\star,\mathrm{sim}}(M)$ is the best-fitting formula for \textit{total} stellar mass versus BH mass, and $f_{\mathrm{bulge}}$ is the bulge mass fraction, which we treat as an adjustable parameter in the range $f_{\mathrm{bulge}}\sim 0.2-1$. In principle, the scaling relation evolves with redshift such that the simulation results should converge to the observed results at $z\sim 0$. However, limited by computational resources, we can only run our simulations down to $z=4$, which is only 11\% of the age of the Universe, so that we cannot take into account the entire cosmic evolution of the scaling relations self-consistently. The \texttt{obs-based} and \texttt{sim-based} models adopted here with parameters $\gamma$ and $f_{\mathrm{bulge}}$ are meant to explore the range of GW rates under uncertainties in the environments of BBHs embodied by the scaling relations.

Once $M_{\star}$ is known, the bulge effective radius is given by \citep{dabringhausen2008star}
\begin{align}
R_{\mathrm{eff}}=\max(2.95M_{\star,6}^{0.596},34.8M_{\star,6}^{0.399})\ \mathrm{pc}\ ,\label{e28}
\end{align}
where $M_{\star,6}\equiv M_{\star}/(10^{6}\ \mathrm{M}_{\odot})$. The first term in the bracket is for ultracompact objects (such as ultracompact dwarfs, globular and nuclear star clusters), whereas the second is for regular elliptical galaxies. In our case with $M_{\mathrm{BH}}\sim 100-1000\ \mathrm{M_{\odot}}$, we have $M_{\star}\sim 10^{5}-10^{6}\ \mathrm{M_{\odot}}\ll 10^{9}\ \mathrm{M_{\odot}}$, such that the second term always applies, i.e. $R_{\mathrm{eff}}=34.8M_{\star,6}^{0.399}\ \mathrm{pc}$, leading to conservative estimations of $\rho_{\mathrm{inf}}$. Finally, we obtain $\rho_{\mathrm{inf}}$ by evaluating the density profile~(\ref{e25}) at $r_{\mathrm{inf}}$, which is derived from equations~(\ref{e26})-(\ref{e28}) for the estimated $M_{\star}$. Given $M\sim 100-1000\ \mathrm{M_{\odot}}$, $r_{\mathrm{inf}}\sim 0.01-1\ \mathrm{pc}$, and $\rho_{\mathrm{inf}}\sim 10^{2}-10^{4}\ \mathrm{M_{\odot}\ pc^{-3}}$ for $\gamma\sim 0-1.5$. However, if the BBHs reside in ultracompact systems such that $R_{\mathrm{eff}}=2.95M_{\star,6}^{0.596}$, we have $r_{\mathrm{inf}}\sim 0.001-0.1\ \mathrm{pc}$ and $\rho_{\mathrm{inf}}\sim 10^{6}-10^{9}\ \mathrm{M_{\odot}\ pc^{-3}}$ for $\gamma\sim 0-1.5$. We assume that such ultra-dense environments are rare, and defer investigating their GW signals to future work.

For $\sigma_{\mathrm{inf}}$, we also have two schemes. In the \texttt{obs-based} model, the observed scaling relation \citep{kormendy2013coevolution}
\begin{align}
\sigma_{\mathrm{inf}}&=200\ \mathrm{km\ s^{-1}} \left(\frac{M_{9}}{0.309}\right)^{1/4.38}\ \label{e29}
\end{align}
is used. While in the \texttt{sim-based} model, we estimate $\sigma_{\mathrm{inf}}$ with the velocity dispersion of surrounding stellar particles around the primary member of the BH binary at the moment of binary formation. It turns out that these two models are consistent with each other within a factor of 3, and typically $\sigma_{\mathrm{inf}}\sim 10\ \mathrm{km\ s^{-1}}$.

\begin{figure}
\includegraphics[width=1\columnwidth]{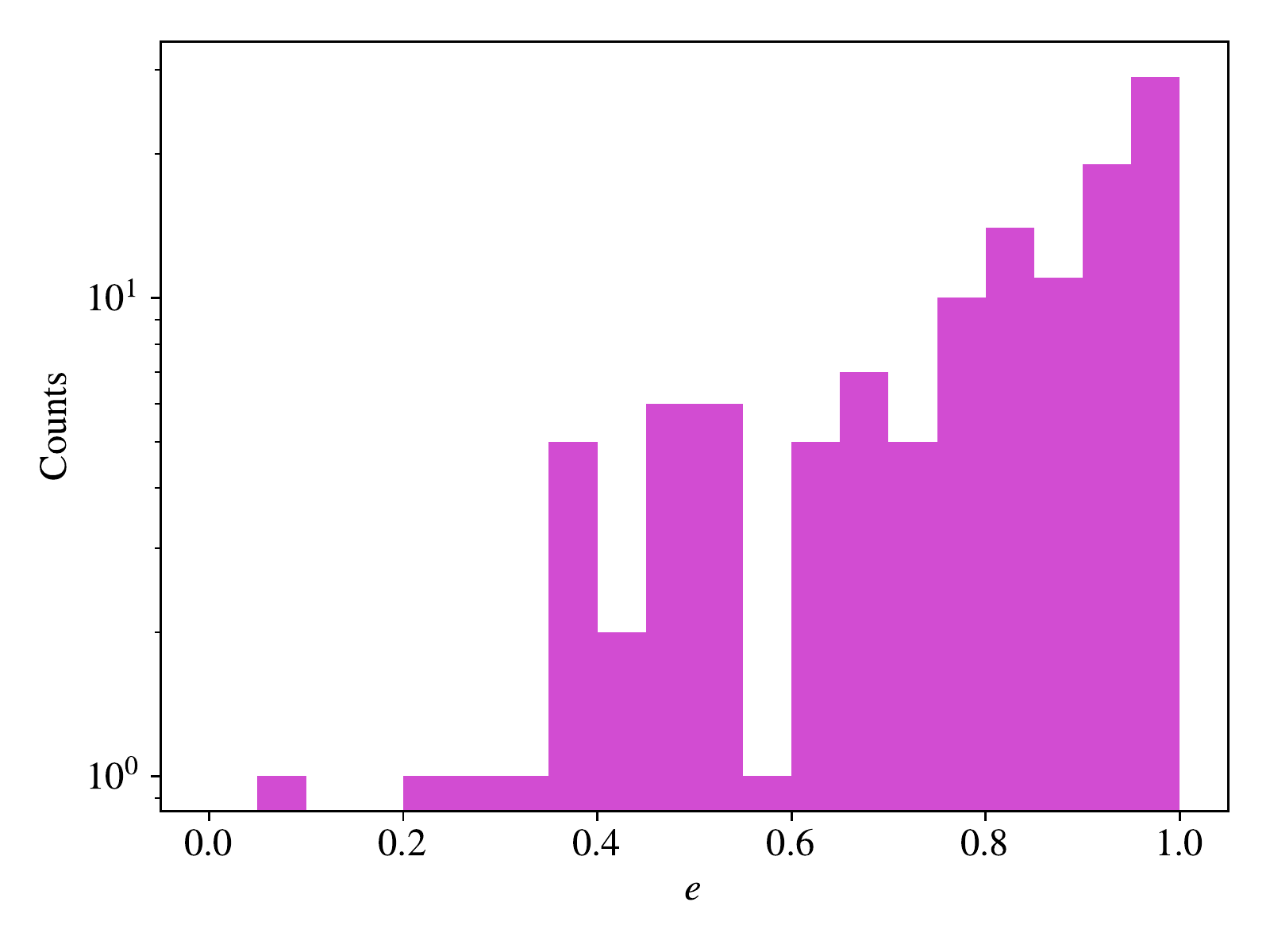}
\caption{Distribution of initial eccentricity for ex-situ BBHs formed at $z_{\mathrm{BBH}}> 4$ in \texttt{FDbox\_Lseed}.} 
\label{f9}
\end{figure}

We treat $\gamma$ as a parameter characterizing the inner structures of high-$z$ (dwarf) galaxies, and explore the range $0\le \gamma\le 1.5$. For a BH binary formed at $t_{\mathrm{BBH}}$ in our simulation, we expect the corresponding GW signals to be emitted at $t=t_{\mathrm{BBH}}+t_{\mathrm{GW}}$ (GW time, henceforth), where $t_{\mathrm{GW}}=t_{\mathrm{HD}}+t_{\mathrm{col}}$. For most cases we have $t_{\mathrm{GW}}\sim 1\ -10\ \mathrm{Gyr}\gtrsim t_{\mathrm{BBH}}$, and $t_{\mathrm{GW}}$ is highly sensitive to eccentricity. The simulated BBHs tend to have very high initial eccentricities (by the nature of dynamical capture), as shown in Fig.~\ref{f9} as a sample distribution of initial eccentricity for BBHs formed at $z_{\mathrm{BBH}}> 4$ in \texttt{FDbox\_Lseed}. 

Note that our formalism treats multiple systems (with more than two BHs) in a hierarchical manner, such that the primary coalesces with other members in turn according to their individual GW times. In reality, BHs may not merge hierarchically in a multiple system which itself can be transient, as close encounters of BBHs with another single or binary BH system can have diverse outcomes (e.g. ejection, exchange and GW recoil) beyond our resolution. Since multiple systems (with up to 3 BHs) only count for less than 1\% of the dynamical capture events in our simulations, we expect them to have negligible impact on the GW rates.

\subsection{Intrinsic rate density of GW events}
\label{s4.2}

Here and in the following subsection, we discuss the GW signals from the simulated ex-situ BBHs, mainly showing results for the most representative (fiducial) case of \texttt{FDbox\_Lseed}. The results for \texttt{Hseed} are similar. The effects of stellar feedback and resolution are explored in Appendices~\ref{a1} and \ref{a2}. We focus on the intrinsic and detection rates of GW events. It is also interesting to explore the statistics of detectable sources, such as distributions of masses, spins and eccentricity. However, given the uncertainties in the simulated GW events from our idealized treatment of BBH evolution, we defer such discussions to future work.

\begin{figure}
\includegraphics[width=1.0\columnwidth]{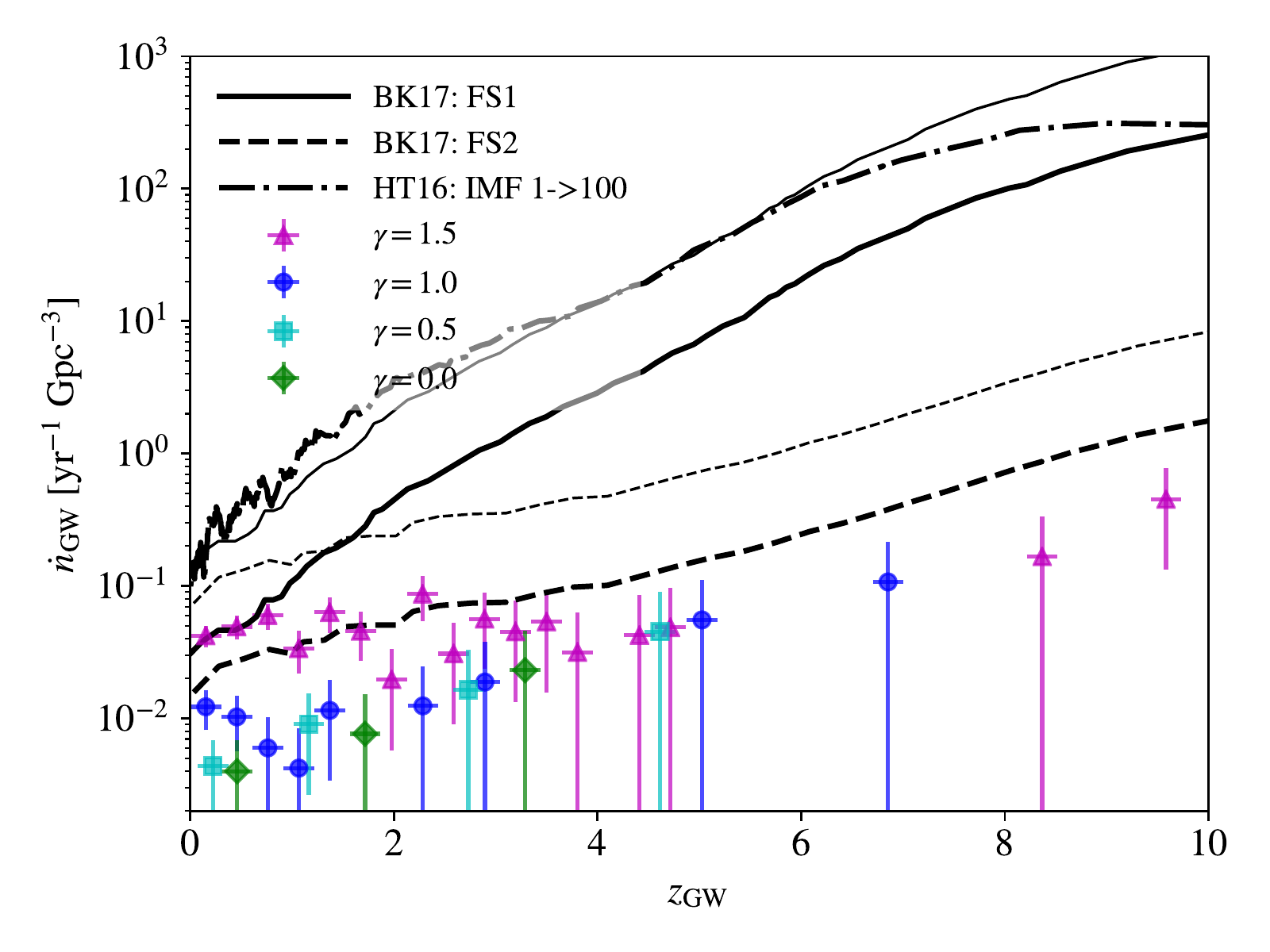}
\caption{Co-moving rest-frame rate densities of GW events from ex-situ BBHs formed at $z_{\mathrm{BBH}}> 4$ in \texttt{FDbox\_Lseed}, for $\gamma=1.5$ (triangle), 1 (circle) 0.5 (square) and 0 (diamond), measured with a bin size $\Delta z_{\mathrm{GW}}=0.3$, where the \texttt{obs-based} scaling relations are used. Error bars denote 1$\sigma$ uncertainties assuming that the number of GW events in each redshift bin follows a Poisson distribution such that $\sigma_{i}=\dot{n}_{\mathrm{GW},i}/\sqrt{N_{i}}$ for bin $i$ which contains $N_{i}$ events. 
For comparison, we show the original results for in-situ BBHs from \citealt{belczynski2017likelihood} (BK17, for two different scenarios of Pop~III star-forming clouds: FS1 and FS2, see their Fig.~12) and \citealt{hartwig2016} (HT16, with a IMF mass range $1-100\ \mathrm{M_{\odot}}$, see their Fig.~2). We also rescale the results of BK17, calibrated to our total Pop~III stellar mass density $\simeq 7\times 10^{4}\ \mathrm{M_{\odot}\ Mpc^{-3}}$, as shown with the thin curves.}
\label{fgw1}
\end{figure}

\begin{figure}
\includegraphics[width=1\columnwidth]{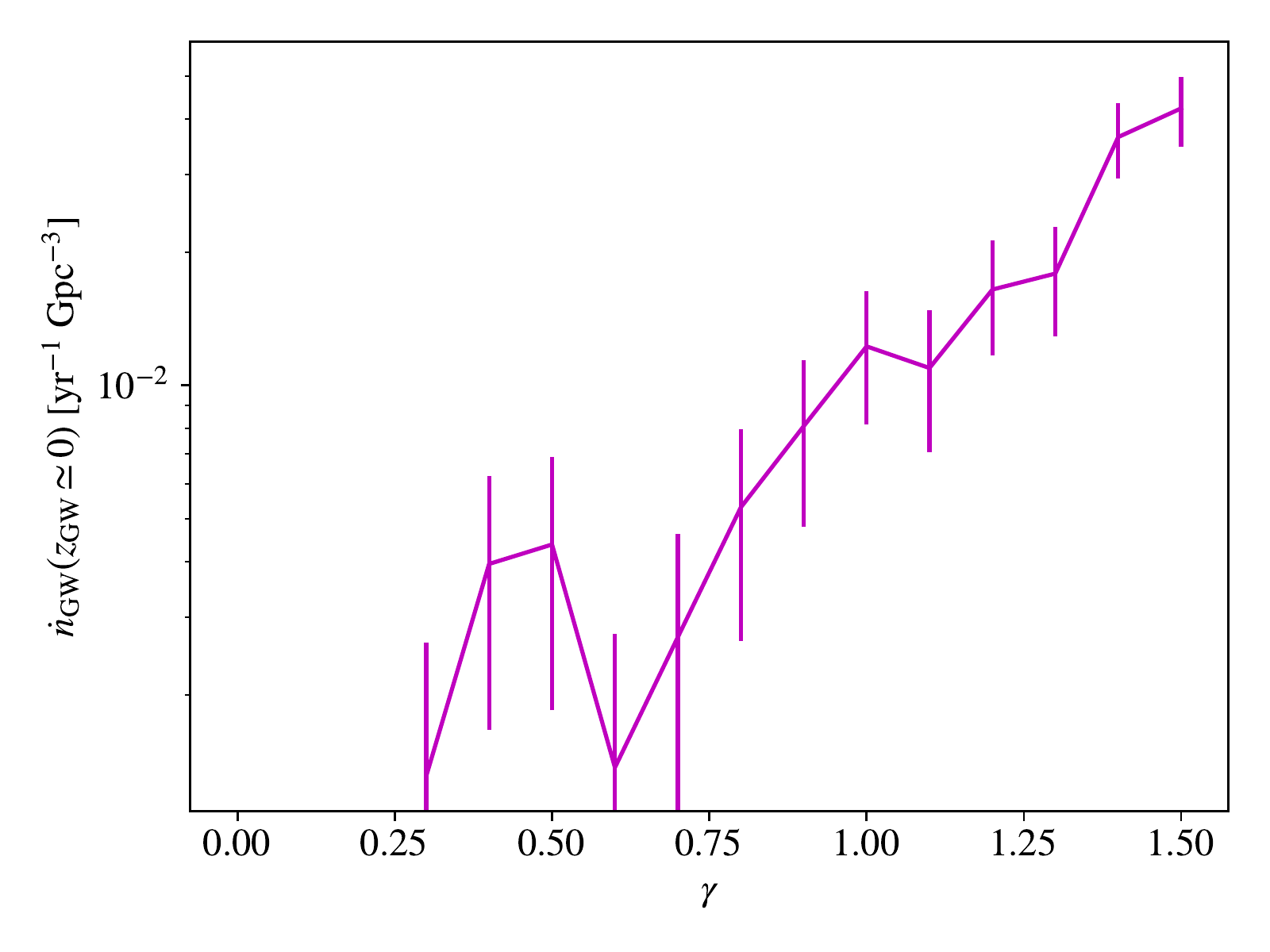}
\caption{Local ($z_{\mathrm{GW}}\lesssim 0.3$) rate density of GW events from ex-situ BBHs formed at $z_{\mathrm{BBH}}> 4$ in \texttt{FDbox\_Lseed}, with the \texttt{obs-based} scaling relations, as a function of the inner slope $\gamma$.}
\label{fgwrd_gamma}
\end{figure}

Fig.~\ref{fgw1} shows the intrinsic GW rate densities of ex-situ BH-BH merger events at $z_{\mathrm{BBH}}> 4$ in \texttt{FDbox\_Lseed}, vs. $z_{\mathrm{GW}}$ for $\gamma=1.5,\ 1.0,\ 0.5$ and 0, with the \texttt{obs-based} scaling relations. The results for \texttt{sim-based} scaling relations are similar, and thus not shown. The main difference between the \texttt{obs-based} and \texttt{sim-based} models is that GW events tend to occur at higher redshifts in the \texttt{sim-based} model compared with the case of \texttt{obs-based} scaling relations\footnote{For $\gamma>1.25$ in the \texttt{sim-based} model, almost all BBHs merge at $z_{\rm GW}\gtrsim 0.3$, leading to a zero local rate density of GW events.}. For the \texttt{sim-based} model, the local rate density is insensitive to $f_{\mathrm{bulge}}$ for $f_{\mathrm{bulge}}\sim 0.2-1$. In both models, the local ($z_{\mathrm{GW}}\lesssim 0.3$) rate density generally increases with $\gamma$ (see Fig.~\ref{fgwrd_gamma}), and resides in the range $\sim 10^{-3}-0.04\ \mathrm{yr^{-1}\ Gyr^{-3}}$ for $\gamma\sim 0-1.5$. Actually, the intrinsic GW rate density is a convolution of the distribution of $t_{\mathrm{GW}}$ and the formation rate density of BBHs. The former will be a `redshifted' version of the latter when the delay time $t_{\mathrm{GW}}$ is non-negligible (compared with $t_{\mathrm{BBH}}$)\footnote{However, if all the BBHs reside in ultracompact systems with $\rho_{\mathrm{inf}}\gtrsim 10^{6}\ \mathrm{M_{\odot}\ pc^{-3}}$, we have $t_{\mathrm{GW}}\lesssim 10\ \mathrm{Myr}\ll t_{\mathrm{BBH}}$, so that the GW rate will closely trace the BBH formation rate. }. In our case, $t_{\mathrm{GW}}\gtrsim 1\ \mathrm{Gyr}$ typically for $\gamma<1.5$, such that GW events only occur at $z_{\mathrm{GW}}\lesssim 10$, even though BBH formation starts at $z_{\mathrm{BBH}}\sim 18$ and has a rate density of $\sim 0.1-1\ \mathrm{yr^{-1}\ Gpc^{-3}}$ for $z_{\mathrm{BBH}}\sim 4-18$. 

Interestingly, the rate density of ex-situ Pop~III BBHs is comparable to the predictions for in-situ Pop~III BBHs in some previous studies at the level of $\sim 0.01-0.1\ \mathrm{yr^{-1}\ Gpc^{-3}}$ \citep{belczynski2017likelihood,hartwig2016}, especially when their results are rescaled to our simulated total density of Pop~III stars formed across cosmic time $\simeq 7\times 10^{4}\ \mathrm{M_{\odot}\ Mpc^{-3}}$. The efficiency of BH-BH mergers $N_{\mathrm{merger}}/M_{\star,\rm{PopIII}}$ is $\sim 10^{-6}-10^{-5}\ \rm{M_{\odot}^{-1}}$, also comparable to the in-situ values (e.g. see Table 4 of \citealt{belczynski2017likelihood}), where $N_{\mathrm{merger}}$ is the total number of Pop~III BBHs that merge within the age of the Universe and $M_{\star,\rm{PopIII}}$ is the total mass of Pop~III stars from which they were born. However, the in-situ rate of Pop~III BH-BH mergers is still in debate, and the literature results shown in Fig.~\ref{fgw1} should be regarded as conservative estimations. Optimistic in-situ rates can be as high as $\sim 1-10\ \mathrm{yr^{-1}\ Gpc^{-3}}$ \citep{kinugawa2014possible}, even if the Pop~III SFRD is constrained by reionization, or calibrated to simulations \citep{inayoshi2016gravitational}. Such discrepancies in the literature for the in-situ channel arise from uncertainties in the initial binary parameters and Pop~III binary stellar evolution models. The (typical) metallicity (threshold) adopted for Pop~III stars also varies in different studies, such that cautions are required to compare their results. We defer more comprehensive comparison between the GW signals from the in-situ and ex-situ channels to future work. 

The local rate density of ex-situ Pop~III BBHs only counts for a tiny fraction ($\sim 10^{-5}-0.005$) of the total local rate density $9-240\ \mathrm{yr^{-1}\ Gpc^{-3}}$ measured by LIGO \citep{abbott2019gwtc}, which is dominated by mergers of SBHs from more metal-enriched progenitors (Pop~II and Pop~I stars). Note that, the highest local rate density $\simeq 0.04\ \mathrm{yr^{-1}\ Gpc^{-3}}$ achieved in \texttt{FDbox\_Lseed} with $\gamma\sim 1.5$ is lower than the upper limit $0.36\ \mathrm{yr^{-1}\ Gpc^{-3}}$ for IMBH binaries of $M=210\ \mathrm{M_{\odot}}$ (similar to the mass scale of Pop~III-seeded BBHs in the \texttt{Lseed} scenario) inferred from the first advanced LIGO observing run \citep{2020arXiv200210666C}.

\subsection{Effects of binary identification}
\label{s4.3}
As described in Section~\ref{s2.3.2}, we form BBHs as bound systems at our force resolution limit of $\epsilon_{\mathrm{g}}\sim 8\ \mathrm{pc}$, which may not be small enough to ensure that the resulting BBHs are hard binaries, especially for BBHs with highly eccentric orbits which can be easily disrupted at apocenters. Since only hard binaries will be hardened to emit GWs, the GW rates may be overestimated by our optimistic binary identification scheme. In this subsection, we evaluate the relevant effects with semi-analytical post-processing and test simulations.

For each BBH in the fiducial run \texttt{FDbox\_Lseed}, we estimate the unresolved true initial semi-major axis $a_{\rm true}$ by a random draw from a logarithmically flat distribution, which is observed in binary-stars \citep{abt1983normal}. Similar distributions (dominated by close binaries) are also obtained in N-body simulations of Pop~III star clusters (see fig.~8 of \citealt{belczynski2017likelihood}), where close binaries dominate. Therefore, we expect it to be a good approximation to the case of dynamically formed BBHs. Actually, as long as the distribution of $a_{\rm true}$ is dominated by close binaries, the fraction of hard binaries is more sensitive to the lower bound $a_{\min}$ than the detailed shape of the distribution. In our case, we set $a_{\min}=0.1\ \mathrm{pc}$, which is the typical size of Pop~III star clusters (e.g. \citealt{susa2014mass,sugimura2020birth}). Beyond this scale, the two Pop~III clusters from which the two BHs originate should be regarded as one initially, and governed by the in-situ channel. We adopt $2\epsilon_{\rm g}/(1-e)$ as the upper bound, since $(1-e)a_{\rm true}<2\epsilon_{\rm g}$ is required for the binary to be identified in the simulation (at the pericenter). Given $a_{\rm true}$, only hard binaries are preserved for GW calculations, which satisfy $K_{\rm ap}>(1/2)\langle\sigma_{\star}^{2}M_{\star}\rangle$, where $K_{\rm ap}=[(1-e)/(1+e)]E_{\rm b}=[(1-e)/(1+e)]Gm_{\rm BH}^{2}/(2a_{\rm true})$ is the kinetic energy at the apocenter, and $\langle\sigma_{\star}^{2}M_{\star}\rangle=(10\ \mathrm{km\ s^{-1}})^{2}\ \mathrm{M_{\odot}}$ is assumed for typical surrounding (Pop~II/I) stars. In this way, the efficiency of BH-BH mergers drops by a factor of $\sim 4$, for the \texttt{obs-based} scaling relations given $\gamma=1.5$, with removal of most highly eccentric binaries ($e\gtrsim 0.8$). The resulting all-sky detection rates are reduced by a factor of $\sim 4-17$ (see the next subsection for details)\footnote{The detection rates are suppressed more for GW detectors more sensitive to high-$z$ events ($z\gtrsim 2$) such as ET and DO when the analysis is restricted to hard binaries. The reason is that high-$z$ events are dominated by highly eccentric BBHs with shorter delay times, whose number is significantly reduced when the hard binary criterion is imposed.}. Note that if the distribution of $a_{\rm true}$ is instead dominated by wide binaries, the reduction of GW rates will be larger.

To verify the above results, we run test simulations under the same setup and feedback model as those for \texttt{FDzoom} runs. In the test runs, $\epsilon_{\mathrm{g}}\simeq 3~\mathrm{pc}$, and only hard BH binaries are considered with $[(1-e)/(1+e)]E_{\rm b}>(1/2)\langle\sigma_{\star}^{2}M_{\star}\rangle$, where the binding energy $E_{\rm b}$ is measured on-the-fly. We find similar trends as those seen in analytical calculations: When the simulation only identifies hard binaries, the efficiency of BH-BH mergers is reduced by a factor of $\sim3$ (4), compared with the optimistic case of \texttt{FDzoom\_Hseed} (\texttt{FDzoom\_Lseed}), with up to a factor of $\sim 22$ (16) suppression in detection rates, for the \texttt{obs-based} scaling relations with $\gamma=1.5$.

\subsection{Detectability}
\label{s4.4}

To evaluate the detectability for our simulated GW sources, we adopt the phenomenological model `PhenomC' from \citet{santamaria2010matching} to calculate the waveform of the GW signal in the frequency domain $|\tilde{h}(f|M_{1},M_{2},D_{L},z_{\mathrm{GW}})|$, under the simplifying assumption that the two BHs before merger are non-spinning in circular orbits, and where $D_{L}$ is the luminosity distance to the source. This assumption of zero spin and circular orbit, as well as the intrinsic waveform uncertainties may lead to uncertainties in the resulting signal-to-noise ratio (SNR) of a few tens of percent. We consider several GW detectors covering the frequency range $10^{-4}-10^{3}\ \rm{Hz}$: LIGO O2 (\citealt{abbott2019gwtc}, for the instrument at Livingston as an example), advanced LIGO by design (AdLIGO; \citealt{martynov2016sensitivity}), ET under xylophone configuration (ETxylophone; \citealt{hild2009xylophone}), DO with optimal performance (DOoptimal; \citealt{dechihertz}) and LISA \citep{robson2019construction}. For each source, given $|\tilde{h}(f)|$, the SNR for an instrument with a noise power spectral density (PSD) function $S(f)$ is obtained by integrating in the \textit{observer-frame} frequency domain:
\begin{align}
\mathrm{SNR}^{2}=4\int_{f_{\min}}^{f_{\max}}\frac{|\tilde{h}(f)|^{2}}{S(f)}df\ ,
\end{align}
assuming that the noise is stationary and Gaussian with zero mean. Here we set\footnote{Note that a GW event only resides at $(1+z_{\mathrm{GW}})\times f=f_{\mathrm{rest}}>f_{\mathrm{GW},0}=\pi^{-1}\sqrt{GM}(1+e)^{1.1954}/[ a_{\star/\mathrm{GW}}(1-e^{2})]^{1.5}$ in the frequency domain, where $f_{\mathrm{GW},0}$ is the peak frequency at the onset of the GW-driven inspiral. In our case, $f_{\mathrm{GW},0}/(1+z_{\mathrm{GW}})\lesssim 10^{-5}\ \mathrm{Hz}$, such that $f_{\min}=10^{-4}\ \mathrm{Hz}>f_{\mathrm{GW},0}/(1+z_{\mathrm{GW}})$ always holds.} $f_{\min}=10^{-4}\ \mathrm{Hz}$ and $f_{\max}=2f_{\mathrm{RD}}/(1+z_{\mathrm{GW}})$, where $f_{\mathrm{RD}}$ marks the beginning of the ringdown phase (see equation (5.5) in \citealt{santamaria2010matching}). 

\begin{figure}
\includegraphics[width=1\columnwidth]{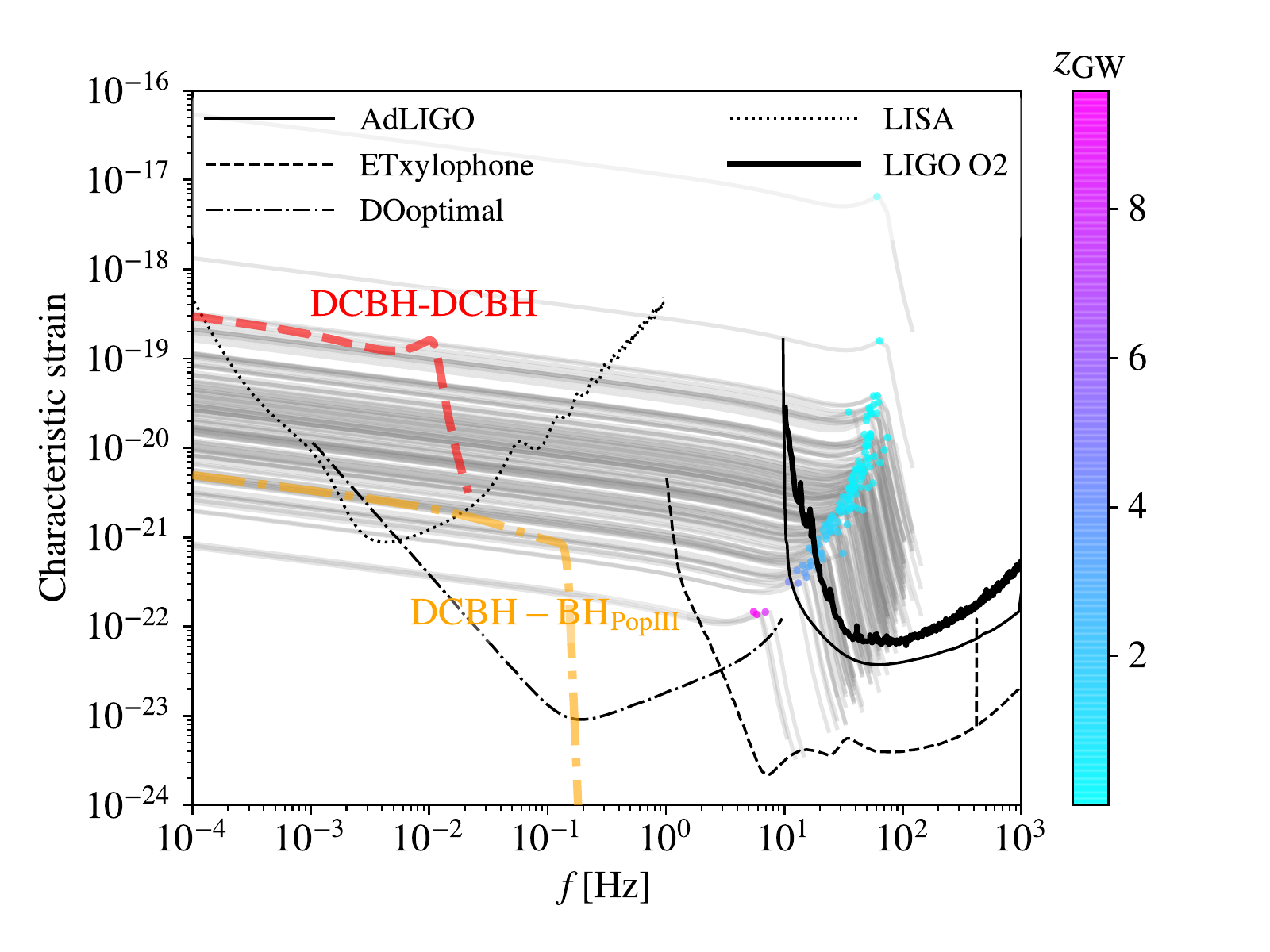}
\caption{Waveforms ($2f|\tilde{h}(f)|$) of all simulated GW events with the \texttt{obs-based} scaling relations under $\gamma=1.5$ from \texttt{FDbox\_Lseed}, together with sensitivity curves ($\sqrt{f\times S(f)}$) for selected GW detectors. 
The simulated BH mergers involving two Pop~III seeds are plotted with faint gray curves. 
For illustration, we also label the transition point from inspiral to ringdown for each event. The labels are color coded by $z_{\rm{GW}}$. 
The sensitivity curves for LIGO O2, AdLIGO (by design), ETxylophone, DOoptimal and LISA are shown in thick and normal solid, dashed, dashed-dotted and dotted, respectively (see the text of Sec.~\ref{s4.4} for descriptions of the detectors). 
For illustration, we also plot the typical waveforms of DCBH-DCBH mergers ($M_{1}=M_{2}\sim 10^{5}\ \mathrm{M_{\odot}}$) and DCBH-BH$_{\mathrm{PopIII}}$ mergers ($M_{1}\sim 10^{4}\ \mathrm{M_{\odot}}$, $M_{2}\sim 100\ \mathrm{M_{\odot}}$) at $z_{\mathrm{GW}}\sim 7$ with the red dashed and orange dashed-dotted curves. 
}
\label{wf}
\end{figure}

\begin{figure}
\includegraphics[width=1\columnwidth]{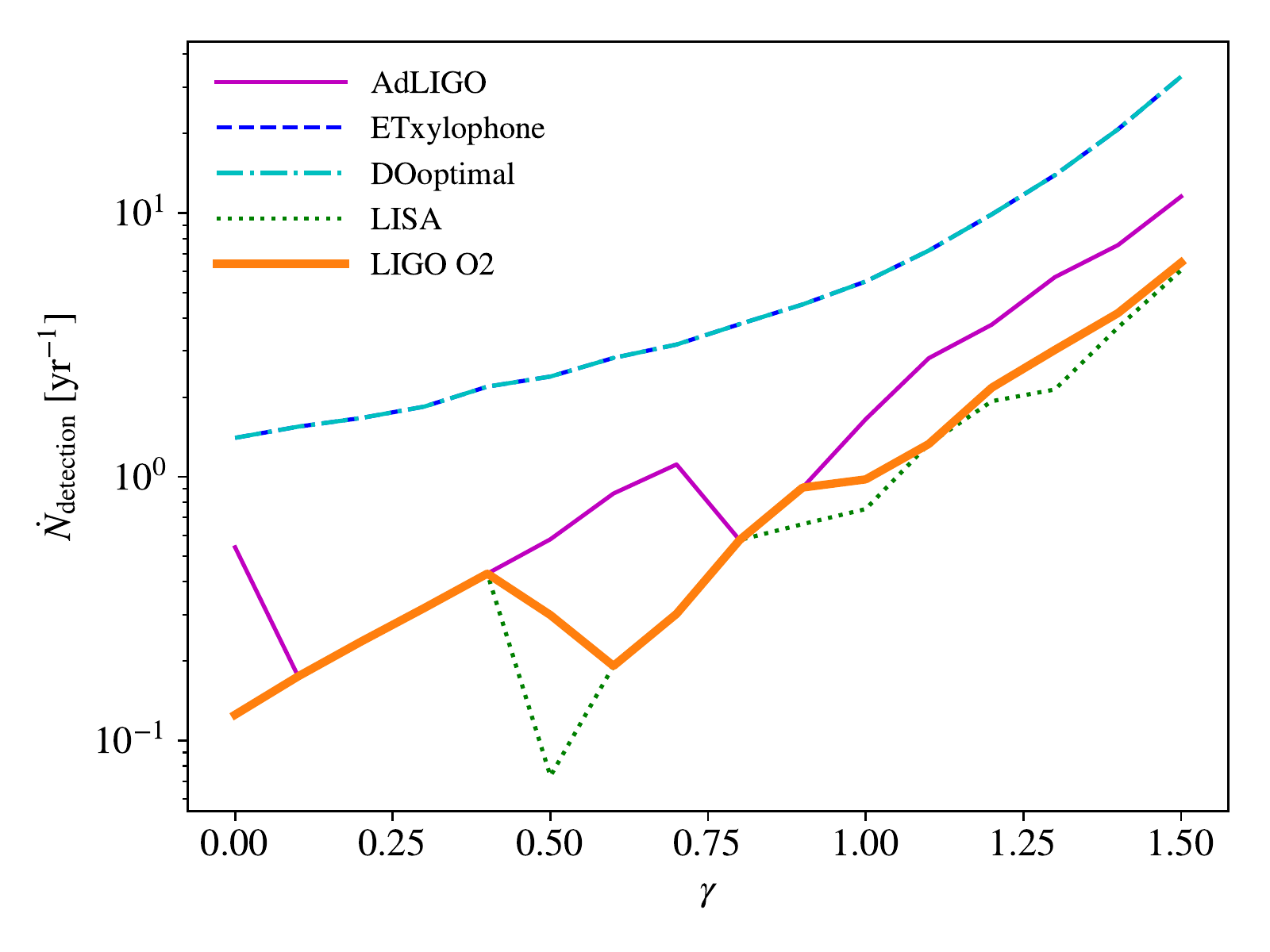}
\caption{All-sky detection rates (with $\mathrm{SNR}>10$) of GW events from ex-situ BBHs formed at $z_{\mathrm{BBH}}>4$ in \texttt{FDbox\_Lseed}, for AdLIGO by design (solid), ET's xylophone configuration (dashed), the optimal performance of DO (dashed-dotted) and LISA (dotted), as functions of $\gamma$, where the stellar environments are characterized with the \texttt{obs-based} scaling relations.}
\label{fdrate}
\end{figure}

The waveforms of all simulated GW events (at $z_{\rm{GW}}\ge 0$) with the \texttt{obs-based} scaling relations under $\gamma=1.5$ from \texttt{FDbox\_Lseed} are shown in Fig.~\ref{wf} (in terms of the frequency evolution of the characteristic strain $2f|\tilde{h}(f)|$), on top of the sensitivity curves of the detectors considered here. Due to the rareness of DCBH candidates, we only found events involving Pop~III-seeded BHs ($\mathrm{BH_{PopIII}}$s). Under \texttt{Lseed}, the median total mass of detectable sources is $\simeq 250\ \mathrm{M_{\odot}}$ for all the instruments considered here, while the median redshifts are $\simeq0.55$ for AdLIGO (by design), 0.69 for ETxylophone and DOoptimal (whose typical waveforms overlap), and 0.45 for LISA and LIGO O2. The integrated all-sky detection rates (with $\mathrm{SNR}>10$) as functions of $\gamma$ are shown in Fig.~\ref{fdrate}, under the \texttt{obs-based} scaling relations. The results in the \texttt{sim-based} model is similar, where the detection rates are insensitive to $f_{\rm{bulge}}$ (within a factor of 2 variations for $f_{\mathrm{bulge}}\sim 0.2-1$). 
The detection rate generally increases with $\gamma$ in both models, especially for ETxylophone and DOoptimal which can reach high-$z$ sources. The reason is that with denser stellar environments (embodied by higher values of $\gamma$) the delay times are on-average smaller such that more BH binaries can merge within the age of the Universe. The simulated events are almost always detectable (with $\mathrm{SNR}>10$) by ETxylophone and DOoptimal up to $z_{\rm{GW}}\sim 10$, leading to detection rates of $\sim 1-30\ \mathrm{yr^{-1}}$. Among them, the nearby sources are also detectable by AdLIGO (according to the design sensitivity) for $z_{\mathrm{GW}}\lesssim 2$ in the ringdown phase, and by LISA for $z_{\mathrm{GW}}\lesssim 1$ during the inspiral phase.

Interestingly, the sources at $z\lesssim 1$ are detectable by LIGO during O2, and the predicted detection rate can be as high as $6.5\ \mathrm{yr^{-1}}$, if dense environments around BBHs are assumed with $\gamma=1.5$ (i.e. $\rho_{\mathrm{inf}}\sim 10^{4}\ \mathrm{M_{\odot}\ pc^{-3}}$). This means that it is possible for ex-situ BBHs originating from Pop~III stars to account for a fraction (up to $\sim 50\%$) of currently detected events\footnote{The BH masses involved in our simulated GW events are typically $\gtrsim 100\ \mathrm{M_{\odot}}$, higher than the estimated BH masses from almost all detected sources. However, this could be an artifact of our BH seeding models, as we only track the most massive BHs in each Pop~III population even in \texttt{Lseed}. In principle, low-mass Pop~III-seeded BHs of $\sim 40-50\ \mathrm{M_{\odot}}$ can also form ex-situ BH binaries and merge at $z\lesssim 1$ to be detected by LIGO O2.}, i.e. the 10 highly-significant BH-BH mergers in GWTC-1 during O1 and O2 of LIGO for 8.5 months (i.e. 0.7~yr) of data (\citealt{abbott2019gwtc}, corresponding to a detection rate of $\sim 14\ \mathrm{yr^{-1}}$), especially for massive systems such as GW170729 and GW170823. However, when less dense environments are assumed (e.g. $\gamma\lesssim 1.0$, $\rho_{\mathrm{inf}}\sim 10^{2}\ \mathrm{M_{\odot}\ pc^{-3}}$), the detection rate drops to $\lesssim 1\ \mathrm{yr^{-1}}$, making detection by past LIGO observations unlikely. Note that the definition of confident events by LIGO during O1 and O2 may be more strict than our criterion for detection (i.e. $\mathrm{SNR}>10$ for the O2 Livingston sensitivity curve). 

Besides, it is challenging to distinguish the GW events from ex-situ Pop~III BBHs, if detected, with those from the in-situ channel and more metal-enriched progenitors (Pop~II and I stars), which most likely dominate the local Universe \citep{belczynski2017likelihood}: First, it is possible to form BHs as massive as $\sim 50\ \mathrm{M_{\odot}}$ (the maximum primary mass in GWTC-1) with a metallicity $Z\sim 0.01\ \mathrm{Z_{\odot}}$ (e.g. \citealt{belczynski2010maximum}), much higher than the Pop~III threshold $Z_{\mathrm{crit}}=10^{-4}\ \mathrm{Z_{\odot}}$. Different progenitors cannot be easily distinguished from the mass distribution of BH-BH mergers at $z\sim 0$. The evolution of mass distribution with redshift may be more useful, which could be measured with third-generation detectors. Particularly, high-$z$ sources ($z_{\mathrm{GW}}\gtrsim 15$) tend to be dominated by Pop~III progenitors, especially for $M_{\mathrm{BH}}\gtrsim 50\ \mathrm{M_{\odot}}$ (see \citealt{belczynski2017likelihood}). Comparing the mass distribution of BBHs observed via GWs to BH pairs drawn randomly from certain BH mass functions, may also reveal the main formation channel and progenitors of BBHs. 

Second, if Pop~III-seeded BHs grow significantly by accretion across cosmic time, they could be distinguished from BHs formed at lower redshifts in more metal-enriched environments by higher spins. However, mass growth by accretion has been found typically insignificant for Pop~III seeds, at least for $z\gtrsim 4$ (see Fig.~\ref{f6}-\ref{rhoacc} and \citealt{johnson2007aftermath,alvarez2009accretion,hirano2014one,smith2018growth}). If there are detectable spins, the relative direction of spins is expected to be isotropic (uncorrelated) for the ex-situ channel, while spin alignment is expected for the in-situ channel \citep{farr2017distinguishing}. Therefore, the two channels can de distinguished with the distribution of effective spins (e.g. \citealt{safarzadeh2020branching}. 
Last but not least, one salient feature of ex-situ BBHs formed by dynamical capture, is their high initial orbital eccentricities (see Fig.~\ref{f9}). However, as GW emissions circularize binary orbits, such that even with the highest initial eccentricity found in our simulations ($1-e\sim 10^{-5}$), the eccentricity drops to $\lesssim 0.1$ for $f_{\mathrm{rest}}\gtrsim 10^{-2}\ \mathrm{Hz}$, making it difficult to measure with ground-based GW detectors (see Fig.~\ref{wf}), given the parameter-estimation degeneracy between BH spins and orbital eccentricity \citep{PhysRevD.97.024031}. Nevertheless, it is possible to make robust eccentricity measurements for nearby sources with future planned space-based instruments sensitive at low frequencies, such as LISA and DOs.

\begin{table}
\caption{Detection rates per year and percentages (in brackets) of simulated GW events with $\mathrm{SNR}>10$ from ex-situ BBHs formed at $z_{\mathrm{BBH}}> 4$ in \texttt{FDzoom\_Lseed} and \texttt{FDzoom\_Hseed}, under the \texttt{obs-based} scaling relations with $\gamma=1.5$ and \texttt{sim-based} scaling relations with $\gamma=1.5$ and $f_{\mathrm{bulge}}=1.0$. The detection percentage reflects the ratio of the number of sources with $\mathrm{SNR}>10$ and the total number of sources (at $z_{\mathrm{GW}}\ge 0$). The first column is the flag \texttt{SD} for BH seeding model with \texttt{L} for \texttt{Lseed} and \texttt{H} for \texttt{Hseed}. See the text of Sec.~\ref{s4.4} for descriptions of the detectors considered here.
}
\begin{tabular}{ccccc}
\hline
& \texttt{obs-based} & $\gamma=1.5$\\
\texttt{SD} & AdLIGO & ETxylophone & DOoptimal & LISA\\
\texttt{L} & 18 (90\%) & 30 (100\%) & 30 (100\%) & 5.4 (70\%) \\
\texttt{H} & 15 (60\%) & 53 (90\%) & 98 (100\%) & 53 (90\%)\\
\hline
& \texttt{sim-based} & $\gamma=1.5$ & $f_{\mathrm{bulge}}=1.0$\\
\texttt{SD} & AdLIGO & ETxylophone & DOoptimal & LISA\\
\texttt{L} & 0 & 200 (100\%) & 200 (100\%) & 0\\
\texttt{H} & 0  & 202 (84\%) & 289 (100\%) & 41 (32\%) \\
\hline
\end{tabular}
\label{t1_}
\end{table}

To demonstrate the dependence of detection rates on BH seeding models, we show the results for selected detectors from \texttt{FDzoom\_Lseed} and \texttt{FDzoom\_Hseed} in Table~\ref{t1_}, under the optimistic conditions of \texttt{obs-based} scaling relations with $\gamma=1.5$ and \texttt{sim-based} scalings with $\gamma=1.5$ and $f_{\mathrm{bulge}}=1.0$. Generally speaking, the detection rates for AdLIGO (by design), ETxylophone and DOoptimal are not very sensitive to our choice of the BH seeding models such that the differences are generally within a factor of 3 for \texttt{Lseed} and \texttt{Hseed}. While for LISA, the detection rate is significantly higher (by up to 10 times) for \texttt{Hseed} than for \texttt{Lseed}. The reason is that LISA is especially sensitive to massive BBHs. Since accretion is unimportant for our Pop~III seeds, the typical masses of BBHs are mainly determined by the seeding schemes (Sec.~\ref{s2.3.1}), such that $M=M_{1}+M_{2}\sim 300\ \mathrm{M}_{\odot}$ for \texttt{Lseed} and $M=M_{1}+M_{2}\sim 1000\ \mathrm{M}_{\odot}$ for \texttt{Hseed}. This overall increase in $M$ moves the waveforms of BBHs to the top-left in Fig.~\ref{wf} into the region where LISA can reach.


\section{Summary and discussions}
\label{s5}
We use meso-scale (box sizes of a few Mpc) cosmological hydrodynamic simulations to study the gravitational wave (GW) signals of the \textit{ex-situ} binary black holes (BBHs) formed by dynamical capture, from the remnants of the first stars. Our results in the electromagnetic window are consistent with theoretical and observational constraints on star formation and BH accretion histories, as well as halo-stellar-BH mass relations, at $z>4$. 
For BBHs originating from Pop~III stars, we for the first time predict the intrinsic and detection rates of GW events from the \textit{ex-situ} channel of BBH formation, complementing previous results for the \textit{in-situ} channel. In our fiducial run (\texttt{FDbox\_Lseed}), terminated at $z=4$, we found a local intrinsic GW event rate density of $\sim 10^{-3}-0.04\ \mathrm{yr^-1\ Gpc^{-3}}$, comparable to the conservative estimations for \textit{in-situ} BBHs $\sim 0.01-0.1\ \mathrm{yr^-1\ Gpc^{-3}}$, but much lower than the optimistic predictions $\sim 1-10\ \mathrm{yr^-1\ Gpc^{-3}}$ (e.g. 
\citealt{kinugawa2014possible,kinugawa2015detection,dvorkin2016metallicity,
hartwig2016,inayoshi2016gravitational,
belczynski2017likelihood,mapelli2019properties}). 

We also found promising all-sky detection rates for selected GW detectors covering the frequency range $10^{-4}-10^{3}\ \rm{Hz}$: $<6.5\ \mathrm{yr^{-1}}$ for LIGO O2, $<17.7\ \mathrm{yr^{-1}}$ for the advanced LIGO by design \citep{martynov2016sensitivity}, $1.4-202\ \mathrm{yr^{-1}}$ for the Einstein Telescope under xylophone configuration \citep{hild2009xylophone}, $1.4-289\ \mathrm{yr^{-1}}$ for the Deciherz Observatory with optimal performance \citep{dechihertz}, and $<52.9\ \mathrm{yr^{-1}}$ for LISA \citep{robson2019construction}. Our results indicate that the \textit{ex-situ} channel of BBH formation can be as important as the \textit{in-situ} channel for Pop~III-seeded BHs and deserves further investigation. However, given the large uncertainties (up to 2 orders of magnitude) for both the in-situ and ex-situ pathways (see below), the (relative) contributions to Pop-III BH-BH merger events from the two channels are still uncertain, to be revealed by future observations and more advanced theoretical models. 

Since cosmological hydrodynamic simulations are always limited in resolution, volume and redshift range, we cannot self-consistently model the small-scale (sub-parsec) physics of BH seeding, as well as formation and evolution of BBHs. Instead, we use idealized sub-grid models whose parameter spaces are explored to estimate the range of GW rates. There are several caveats in our methodology, which may significantly impact the results:
\begin{itemize}
\item On the one hand, we only keep track of at most one BH from one Pop~III stellar population that typically includes a few BHs, leading to possible underestimation of the BBH formation rate. Actually, the average number of BHs is 6 according to our IMF, such that the BBH formation rate can be boosted by up to a factor of 36 (for \texttt{Lseed}) if multiple BHs are considered from each Pop~III stellar population. 
\item On the other hand, we form BBHs as bound systems at our force resolution limit $\epsilon_{\mathrm{g}}\sim 8\ \mathrm{pc}$, which may not be small enough to ensure that the resulting BBHs are hard binaries, especially for BBHs with highly eccentric orbits which can be easily disrupted at apocenters. Since only hard binaries will be hardened to emit GWs, we may have overestimated the GW rates. With analytical calculations and test simulations, we estimate the overestimation to be up to a factor of $\sim4$ and $\sim 20$ in the efficiency of BH-BH mergers and detection rates of GW events, respectively, under the optimal condition of BBH evolution (see Sec.~\ref{s4.3}).
\item We also rely on semi-analytical post-processing to calculate BBH evolution, based on parameterized BH-stellar mass scaling relations and stellar density profile. We found that the predicted GW rates are highly sensitive to the environments around high-$z$ BBHs, which are captured by the inner slope $\gamma$ of stellar density profile in our model (see Fig.~\ref{fgw1} and \ref{fdrate}). Generally speaking, denser environments (i.e. larger $\gamma$) lead to more rapid binary hardening by three-body encounters, such that more BBHs can merge within the Hubble time, resulting in higher GW rates. The realistic long-term environments around Pop~III-seeded BBHs are still unknown, which can be more complex than the cases covered by our simple parameterized model. 
\end{itemize}

To arrive at more robust predictions for the GWs from high-$z$ \textit{ex-situ} BBHs, one must design more realistic sub-grid models addressing the following aspects:
\begin{itemize}
\item The mass and phase space distributions of BH systems (including single BHs, \textit{in-situ} BBHs and X-ray binaries) as the end products of Pop~III star-forming clouds, which, involving the initial mass function, star cluster evolution (over at least a few Myr) and supernova explosion/direct collapse mechanisms, may not be universal.
\item The (erratic) dynamics of Pop~III-seeded BHs ($M_{\mathrm{BH}}\sim 100-1000\ \mathrm{{M_{\odot}}}$) in galaxies, which is crucial for dynamical capture (and BH accretion), but challenging for cosmological simulations with limited mass resolution.
\item The stellar and gas environments around BBHs that drive binary evolution across cosmic time, which is non-trivial as their host systems undergo mergers and accretion.
\end{itemize}
Actually, such considerations are general for the \textit{ex-situ} channel of BBH formation by dynamical capture, from any generations of progenitor stars in the Universe. Moreover, it is generally interesting and crucial to evaluate the relative contributions of the \textit{ex-situ} and \textit{in-situ} channels, if we were to derive useful information (on the evolution of single and binary stars, BH seeding, dynamics and growth, cosmic structure formation) from the GW signals of coalescing compact objects. %

In future work, we will use small-scale simulations (for individual minihaloes/dwarf galaxies) to develop more physically-justified sub-grid models for BH seeding, dynamics, as well as binary formation and evolution. For instance, we can derive the fractions of ejected and remaining BHs from N-body simulations of Pop~III stellar groups (e.g. \citealt{ryu2015formation}), and allow Pop~III stellar particles to spawn multiple BH particles. In this way, we can also identify X-ray binaries among BH seeds whose accretion and feedback will be included \citep{jeon2014radiative}. 
It is also interesting to study the GW signals from Pop~III seeded BBHs in high-$z$ dense star clusters (dominated by Pop~II stars), such as nuclear star clusters and globular clusters, by combining models of their formation and evolution (e.g. \citealt{devecchi2009formation,devecchi2010high,devecchi2012high,lupi2014constraining,kim2018formation,el2019formation}) with direct N-body and Monte Carlo simulations of dense star clusters (e.g. \citealt{o2016dynamical,rodriguez2016binary,rodriguez2018post,hoang2018black}). 

The ultimate goal is to predict the distributions of GW events involving Pop~III-seeded BHs from both the ex-situ and in-situ channels in the GW parameter space (mass, redshift, eccentricity and spins), in comparison with the results of the events from other origins (e.g. Pop~II and I stars). In this way, we may be able to identify a region in the parameter space dominated by Pop~III-seeded BBHs, that can provide guidance for future GW instruments and insights on extracting information of early structure formation from GW observations. In the next decades, this region can be populated by hundreds to thousands of GW events, from which new constraints can be derived, e.g. for the Pop~III binary statistics and stellar evolution models (for the in-situ channel), as well as typical environments of BBH evolution and BH mass function (for the ex-situ channel). 

It is only by considering both the \textit{ex-situ} and \textit{in-situ} channels of binary formation that the power of the GW window can be fully realized in the era of GW astrophysics. With more advanced GW detectors coming into operation over the next decades, it is timely for the fields of star and galaxy formation to address the theoretical challenges involved in understanding binary remnants from the first stars. 

\section*{Acknowledgements}
This work was supported by National Science Foundation (NSF) grant AST-1413501. The authors acknowledge the Texas Advanced Computing Center (TACC) for providing HPC resources under XSEDE allocation TG-AST120024.

\bibliographystyle{mnras}
\bibliography{ref} 

\begin{thebibliography}{}
\makeatletter
\relax
\def\mn@urlcharsother{\let\do\@makeother \do\$\do\&\do\#\do\^\do\_\do\%\do\~}
\def\mn@doi{\begingroup\mn@urlcharsother \@ifnextchar [ {\mn@doi@}
  {\mn@doi@[]}}
\def\mn@doi@[#1]#2{\def\@tempa{#1}\ifx\@tempa\@empty \href
  {http://dx.doi.org/#2} {doi:#2}\else \href {http://dx.doi.org/#2} {#1}\fi
  \endgroup}
\def\mn@eprint#1#2{\mn@eprint@#1:#2::\@nil}
\def\mn@eprint@arXiv#1{\href {http://arxiv.org/abs/#1} {{\tt arXiv:#1}}}
\def\mn@eprint@dblp#1{\href {http://dblp.uni-trier.de/rec/bibtex/#1.xml}
  {dblp:#1}}
\def\mn@eprint@#1:#2:#3:#4\@nil{\def\@tempa {#1}\def\@tempb {#2}\def\@tempc
  {#3}\ifx \@tempc \@empty \let \@tempc \@tempb \let \@tempb \@tempa \fi \ifx
  \@tempb \@empty \def\@tempb {arXiv}\fi \@ifundefined
  {mn@eprint@\@tempb}{\@tempb:\@tempc}{\expandafter \expandafter \csname
  mn@eprint@\@tempb\endcsname \expandafter{\@tempc}}}

\bibitem[\protect\citeauthoryear{Abbott et~al.,}{Abbott
  et~al.}{2017}]{abbott2017exploring}
Abbott B.~P.,  et~al., 2017, Classical and Quantum Gravity, 34, 044001

\bibitem[\protect\citeauthoryear{Abbott et~al.,}{Abbott
  et~al.}{2018}]{abbott2018prospects}
Abbott B.~P.,  et~al., 2018, Living Reviews in Relativity, 21, 3

\bibitem[\protect\citeauthoryear{Abbott et~al.,}{Abbott
  et~al.}{2019a}]{abbott2019gwtc}
Abbott B.,  et~al., 2019a, Phys. Rev. X, 9, 031040

\bibitem[\protect\citeauthoryear{Abbott et~al.,}{Abbott
  et~al.}{2019b}]{abbott2019search}
Abbott B.,  et~al., 2019b, \prd, 100, 064064

\bibitem[\protect\citeauthoryear{Abt}{Abt}{1983}]{abt1983normal}
Abt H.~A.,  1983, \araa, 21, 343

\bibitem[\protect\citeauthoryear{Adhikari, Fishbach, Holz, Wechsler  \&
  Fang}{Adhikari et~al.}{2020}]{adhikari2020binary}
Adhikari S.,  Fishbach M.,  Holz D.~E.,  Wechsler R.~H.,   Fang Z.,  2020,
  arXiv preprint arXiv:2001.01025

\bibitem[\protect\citeauthoryear{Agarwal, Davis, Khochfar, Natarajan  \&
  Dunlop}{Agarwal et~al.}{2013}]{agarwal2013unravelling}
Agarwal B.,  Davis A.~J.,  Khochfar S.,  Natarajan P.,   Dunlop J.~S.,  2013,
  mnras, 432, 3438

\bibitem[\protect\citeauthoryear{Alvarez, Wise  \& Abel}{Alvarez
  et~al.}{2009}]{alvarez2009accretion}
Alvarez M.~A.,  Wise J.~H.,   Abel T.,  2009, \apjl, 701, L133

\bibitem[\protect\citeauthoryear{{Arca Sedda} et~al.,}{{Arca Sedda}
  et~al.}{2019}]{dechihertz}
{Arca Sedda} M.,  et~al., 2019, arXiv e-prints, \href
  {https://ui.adsabs.harvard.edu/abs/2019arXiv190811375A} {p. arXiv:1908.11375}

\bibitem[\protect\citeauthoryear{Barack et~al.,}{Barack
  et~al.}{2019}]{barack2019black}
Barack L.,  et~al., 2019, Classical and quantum gravity, 36, 143001

\bibitem[\protect\citeauthoryear{{Basu} \& {Das}}{{Basu} \&
  {Das}}{2019}]{SMBHmf}
{Basu} S.,  {Das} A.,  2019, \mn@doi [\apjl] {10.3847/2041-8213/ab2646}, \href
  {https://ui.adsabs.harvard.edu/abs/2019ApJ...879L...3B} {879, L3}

\bibitem[\protect\citeauthoryear{{Becerra}, {Marinacci}, {Bromm}  \&
  {Hernquist}}{{Becerra} et~al.}{2018}]{becerra2018assembly}
{Becerra} F.,  {Marinacci} F.,  {Bromm} V.,   {Hernquist} L.~E.,  2018, \mn@doi
  [\mnras] {10.1093/mnras/sty2210}, \href
  {https://ui.adsabs.harvard.edu/abs/2018MNRAS.480.5029B} {480, 5029}

\bibitem[\protect\citeauthoryear{Belczynski, Bulik, Fryer, Ruiter, Valsecchi,
  Vink  \& Hurley}{Belczynski et~al.}{2010}]{belczynski2010maximum}
Belczynski K.,  Bulik T.,  Fryer C.~L.,  Ruiter A.,  Valsecchi F.,  Vink J.~S.,
    Hurley J.~R.,  2010, \apj, 714, 1217

\bibitem[\protect\citeauthoryear{Belczynski, Holz, Bulik  \&
  O’Shaughnessy}{Belczynski et~al.}{2016}]{belczynski2016first}
Belczynski K.,  Holz D.~E.,  Bulik T.,   O’Shaughnessy R.,  2016, \nat, 534,
  512

\bibitem[\protect\citeauthoryear{Belczynski, Ryu, Perna, Berti, Tanaka  \&
  Bulik}{Belczynski et~al.}{2017}]{belczynski2017likelihood}
Belczynski K.,  Ryu T.,  Perna R.,  Berti E.,  Tanaka T.~L.,   Bulik T.,  2017,
  \mnras, 471, 4702

\bibitem[\protect\citeauthoryear{Bond, Arnett  \& Carr}{Bond
  et~al.}{1984}]{bond1984evolution}
Bond J.,  Arnett W.,   Carr B.~J.,  1984, \apj, 280, 825

\bibitem[\protect\citeauthoryear{{Bromm}}{{Bromm}}{2013}]{bromm2013}
{Bromm} V.,  2013, \mn@doi [Reports on Progress in Physics]
  {10.1088/0034-4885/76/11/112901}, \href
  {https://ui.adsabs.harvard.edu/abs/2013RPPh...76k2901B} {76, 112901}

\bibitem[\protect\citeauthoryear{{Bromm} \& {Loeb}}{{Bromm} \&
  {Loeb}}{2003}]{BrommLoeb2003}
{Bromm} V.,  {Loeb} A.,  2003, \mn@doi [\apj] {10.1086/377529}, \href
  {https://ui.adsabs.harvard.edu/abs/2003ApJ...596...34B} {596, 34}

\bibitem[\protect\citeauthoryear{Bromm \& Yoshida}{Bromm \&
  Yoshida}{2011}]{bromm2011first}
Bromm V.,  Yoshida N.,  2011, \araa, 49, 373

\bibitem[\protect\citeauthoryear{{Chandra}, {Gayathri}, {Calderon Bustillo}  \&
  {Pai}}{{Chandra} et~al.}{2020}]{2020arXiv200210666C}
{Chandra} K.,  {Gayathri} V.,  {Calderon Bustillo} J.,   {Pai} A.,  2020, arXiv
  e-prints, \href {https://ui.adsabs.harvard.edu/abs/2020arXiv200210666C} {p.
  arXiv:2002.10666}

\bibitem[\protect\citeauthoryear{Conselice, Bhatawdekar, Palmese  \&
  Hartley}{Conselice et~al.}{2019}]{conselice2019gravitational}
Conselice C.~J.,  Bhatawdekar R.,  Palmese A.,   Hartley W.~G.,  2019, arXiv
  preprint arXiv:1907.05361

\bibitem[\protect\citeauthoryear{Dabringhausen, Hilker  \&
  Kroupa}{Dabringhausen et~al.}{2008}]{dabringhausen2008star}
Dabringhausen J.,  Hilker M.,   Kroupa P.,  2008, \mnras, 386, 864

\bibitem[\protect\citeauthoryear{Dayal, Rossi, Shiralilou, Piana, Choudhury  \&
  Volonteri}{Dayal et~al.}{2019}]{dayal2019hierarchical}
Dayal P.,  Rossi E.~M.,  Shiralilou B.,  Piana O.,  Choudhury T.~R.,
  Volonteri M.,  2019, \mnras, 486, 2336

\bibitem[\protect\citeauthoryear{Dehnen}{Dehnen}{1993}]{dehnen1993family}
Dehnen W.,  1993, \mnras, 265, 250

\bibitem[\protect\citeauthoryear{Delvecchio et~al.,}{Delvecchio
  et~al.}{2019}]{delvecchio2019galaxy}
Delvecchio I.,  et~al., 2019, arXiv preprint arXiv:1910.08114

\bibitem[\protect\citeauthoryear{Devecchi \& Volonteri}{Devecchi \&
  Volonteri}{2009}]{devecchi2009formation}
Devecchi B.,  Volonteri M.,  2009, \apj, 694, 302

\bibitem[\protect\citeauthoryear{Devecchi, Volonteri, Colpi  \&
  Haardt}{Devecchi et~al.}{2010}]{devecchi2010high}
Devecchi B.,  Volonteri M.,  Colpi M.,   Haardt F.,  2010, \mnras, 409, 1057

\bibitem[\protect\citeauthoryear{Devecchi, Volonteri, Rossi, Colpi  \&
  Portegies~Zwart}{Devecchi et~al.}{2012}]{devecchi2012high}
Devecchi B.,  Volonteri M.,  Rossi E.,  Colpi M.,   Portegies~Zwart S.,  2012,
  \mnras, 421, 1465

\bibitem[\protect\citeauthoryear{{Di Carlo}, {Giacobbo}, {Mapelli}, {Pasquato},
  {Spera}, {Wang}  \& {Haardt}}{{Di Carlo} et~al.}{2019}]{dicarlo2019}
{Di Carlo} U.~N.,  {Giacobbo} N.,  {Mapelli} M.,  {Pasquato} M.,  {Spera} M.,
  {Wang} L.,   {Haardt} F.,  2019, \mn@doi [\mnras] {10.1093/mnras/stz1453},
  \href {https://ui.adsabs.harvard.edu/abs/2019MNRAS.487.2947D} {487, 2947}

\bibitem[\protect\citeauthoryear{{Dunn}, {Bellovary}, {Holley-Bockelmann},
  {Christensen}  \& {Quinn}}{{Dunn} et~al.}{2018}]{dunn2018dcbh}
{Dunn} G.,  {Bellovary} J.,  {Holley-Bockelmann} K.,  {Christensen} C.,
  {Quinn} T.,  2018, \mn@doi [\apj] {10.3847/1538-4357/aac7c2}, \href
  {https://ui.adsabs.harvard.edu/abs/2018ApJ...861...39D} {861, 39}

\bibitem[\protect\citeauthoryear{Dvorkin, Vangioni, Silk, Uzan  \&
  Olive}{Dvorkin et~al.}{2016}]{dvorkin2016metallicity}
Dvorkin I.,  Vangioni E.,  Silk J.,  Uzan J.-P.,   Olive K.~A.,  2016, \mnras,
  461, 3877

\bibitem[\protect\citeauthoryear{El-Badry, Quataert, Weisz, Choksi  \&
  Boylan-Kolchin}{El-Badry et~al.}{2019}]{el2019formation}
El-Badry K.,  Quataert E.,  Weisz D.~R.,  Choksi N.,   Boylan-Kolchin M.,
  2019, \mnras, 482, 4528

\bibitem[\protect\citeauthoryear{Farr, Stevenson, Miller, Mandel, Farr  \&
  Vecchio}{Farr et~al.}{2017}]{farr2017distinguishing}
Farr W.~M.,  Stevenson S.,  Miller M.~C.,  Mandel I.,  Farr B.,   Vecchio A.,
  2017, \nat, 548, 426

\bibitem[\protect\citeauthoryear{Farr, Fishbach, Ye  \& Holz}{Farr
  et~al.}{2019}]{farr2019future}
Farr W.~M.,  Fishbach M.,  Ye J.,   Holz D.~E.,  2019, \apj, 883, L42

\bibitem[\protect\citeauthoryear{Faucher-Giguere, Lidz, Zaldarriaga  \&
  Hernquist}{Faucher-Giguere et~al.}{2009}]{faucher2009new}
Faucher-Giguere C.-A.,  Lidz A.,  Zaldarriaga M.,   Hernquist L.,  2009, \apj,
  703, 1416

\bibitem[\protect\citeauthoryear{{Feng}, {Wang}, {Hu}, {Hu}  \& {Wang}}{{Feng}
  et~al.}{2019}]{tianqin}
{Feng} W.-F.,  {Wang} H.-T.,  {Hu} X.-C.,  {Hu} Y.-M.,   {Wang} Y.,  2019,
  \mn@doi [\prd] {10.1103/PhysRevD.99.123002}, \href
  {https://ui.adsabs.harvard.edu/abs/2019PhRvD..99l3002F} {99, 123002}

\bibitem[\protect\citeauthoryear{Finkelstein}{Finkelstein}{2016}]{finkelstein2016observational}
Finkelstein S.~L.,  2016, \pasa, 33

\bibitem[\protect\citeauthoryear{Fishbach, Holz  \& Farr}{Fishbach
  et~al.}{2018}]{fishbach2018does}
Fishbach M.,  Holz D.~E.,   Farr W.~M.,  2018, \apj, 863, L41

\bibitem[\protect\citeauthoryear{{Fragione}, {Ginsburg}  \&
  {Kocsis}}{{Fragione} et~al.}{2018a}]{Fragione2018gw}
{Fragione} G.,  {Ginsburg} I.,   {Kocsis} B.,  2018a, \mn@doi [\apj]
  {10.3847/1538-4357/aab368}, \href
  {https://ui.adsabs.harvard.edu/abs/2018ApJ...856...92F} {856, 92}

\bibitem[\protect\citeauthoryear{{Fragione}, {Leigh}, {Ginsburg}  \&
  {Kocsis}}{{Fragione} et~al.}{2018b}]{Fragione2018tidal}
{Fragione} G.,  {Leigh} N. W.~C.,  {Ginsburg} I.,   {Kocsis} B.,  2018b,
  \mn@doi [\apj] {10.3847/1538-4357/aae486}, \href
  {https://ui.adsabs.harvard.edu/abs/2018ApJ...867..119F} {867, 119}

\bibitem[\protect\citeauthoryear{{Gair}, {Mandel}, {Miller}  \&
  {Volonteri}}{{Gair} et~al.}{2011}]{gair2011imbhet}
{Gair} J.~R.,  {Mandel} I.,  {Miller} M.~C.,   {Volonteri} M.,  2011, \mn@doi
  [General Relativity and Gravitation] {10.1007/s10714-010-1104-3}, \href
  {https://ui.adsabs.harvard.edu/abs/2011GReGr..43..485G} {43, 485}

\bibitem[\protect\citeauthoryear{Galli \& Palla}{Galli \&
  Palla}{2013}]{galli2013dawn}
Galli D.,  Palla F.,  2013, \araa, 51, 163

\bibitem[\protect\citeauthoryear{Giersz, Leigh, Hypki, L{\"u}tzgendorf  \&
  Askar}{Giersz et~al.}{2015}]{giersz2015mocca}
Giersz M.,  Leigh N.,  Hypki A.,  L{\"u}tzgendorf N.,   Askar A.,  2015,
  \mnras, 454, 3150

\bibitem[\protect\citeauthoryear{Hahn \& Abel}{Hahn \&
  Abel}{2011}]{hahn2011multi}
Hahn O.,  Abel T.,  2011, \mnras, 415, 2101

\bibitem[\protect\citeauthoryear{{Hartwig}, {Volonteri}, {Bromm}, {Klessen},
  {Barausse}, {Magg}  \& {Stacy}}{{Hartwig} et~al.}{2016}]{hartwig2016}
{Hartwig} T.,  {Volonteri} M.,  {Bromm} V.,  {Klessen} R.~S.,  {Barausse} E.,
  {Magg} M.,   {Stacy} A.,  2016, \mn@doi [\mnras] {10.1093/mnrasl/slw074},
  \href {https://ui.adsabs.harvard.edu/abs/2016MNRAS.460L..74H} {460, L74}

\bibitem[\protect\citeauthoryear{Haster, Antonini, Kalogera  \& Mandel}{Haster
  et~al.}{2016}]{haster2016n}
Haster C.-J.,  Antonini F.,  Kalogera V.,   Mandel I.,  2016, \apj, 832, 192

\bibitem[\protect\citeauthoryear{Hild, Chelkowski, Freise, Franc, Morgado,
  Flaminio  \& DeSalvo}{Hild et~al.}{2009}]{hild2009xylophone}
Hild S.,  Chelkowski S.,  Freise A.,  Franc J.,  Morgado N.,  Flaminio R.,
  DeSalvo R.,  2009, Classical and Quantum Gravity, 27, 015003

\bibitem[\protect\citeauthoryear{Hirano \& Bromm}{Hirano \&
  Bromm}{2017}]{hirano2017formation}
Hirano S.,  Bromm V.,  2017, \mnras, 470, 898

\bibitem[\protect\citeauthoryear{Hirano, Hosokawa, Yoshida, Umeda, Omukai,
  Chiaki  \& Yorke}{Hirano et~al.}{2014}]{hirano2014one}
Hirano S.,  Hosokawa T.,  Yoshida N.,  Umeda H.,  Omukai K.,  Chiaki G.,
  Yorke H.~W.,  2014, \apj, 781, 60

\bibitem[\protect\citeauthoryear{Hoang, Naoz, Kocsis, Rasio  \&
  Dosopoulou}{Hoang et~al.}{2018}]{hoang2018black}
Hoang B.-M.,  Naoz S.,  Kocsis B.,  Rasio F.~A.,   Dosopoulou F.,  2018, \apj,
  856, 140

\bibitem[\protect\citeauthoryear{Hopkins}{Hopkins}{2015}]{hopkins2015new}
Hopkins P.~F.,  2015, \mnras, 450, 53

\bibitem[\protect\citeauthoryear{Huerta et~al.,}{Huerta
  et~al.}{2018}]{PhysRevD.97.024031}
Huerta E.~A.,  et~al., 2018, \mn@doi [\prd] {10.1103/PhysRevD.97.024031}, 97,
  024031

\bibitem[\protect\citeauthoryear{Inayoshi, Haiman  \& Ostriker}{Inayoshi
  et~al.}{2016a}]{inayoshi2016hyper}
Inayoshi K.,  Haiman Z.,   Ostriker J.~P.,  2016a, \mnras, 459, 3738

\bibitem[\protect\citeauthoryear{Inayoshi, Kashiyama, Visbal  \&
  Haiman}{Inayoshi et~al.}{2016b}]{inayoshi2016gravitational}
Inayoshi K.,  Kashiyama K.,  Visbal E.,   Haiman Z.,  2016b, \mnras, 461, 2722

\bibitem[\protect\citeauthoryear{{Inayoshi}, {Visbal}  \& {Haiman}}{{Inayoshi}
  et~al.}{2019}]{inayoshi2019}
{Inayoshi} K.,  {Visbal} E.,   {Haiman} Z.,  2019, arXiv e-prints, \href
  {https://ui.adsabs.harvard.edu/abs/2019arXiv191105791I} {p. arXiv:1911.05791}

\bibitem[\protect\citeauthoryear{Jaacks, Thompson, Finkelstein  \&
  Bromm}{Jaacks et~al.}{2018}]{jaacks2018baseline}
Jaacks J.,  Thompson R.,  Finkelstein S.~L.,   Bromm V.,  2018, \mnras, 475,
  4396

\bibitem[\protect\citeauthoryear{{Jaacks}, {Finkelstein}  \& {Bromm}}{{Jaacks}
  et~al.}{2019}]{jaacks2018legacy}
{Jaacks} J.,  {Finkelstein} S.~L.,   {Bromm} V.,  2019, \mn@doi [\mnras]
  {10.1093/mnras/stz1529}, \href
  {https://ui.adsabs.harvard.edu/abs/2019MNRAS.488.2202J} {488, 2202}

\bibitem[\protect\citeauthoryear{Jani, Shoemaker  \& Cutler}{Jani
  et~al.}{2019}]{jani2019detectability}
Jani K.,  Shoemaker D.,   Cutler C.,  2019, arXiv preprint arXiv:1908.04985

\bibitem[\protect\citeauthoryear{Jeon, Pawlik, Bromm  \&
  Milosavljevi{\'c}}{Jeon et~al.}{2014}]{jeon2014radiative}
Jeon M.,  Pawlik A.~H.,  Bromm V.,   Milosavljevi{\'c} M.,  2014, \mnras, 440,
  3778

\bibitem[\protect\citeauthoryear{Johnson}{Johnson}{2013}]{johnson2013formation}
Johnson J.~L.,  2013, in , The First Galaxies.
Springer, pp 177--222

\bibitem[\protect\citeauthoryear{Johnson \& Bromm}{Johnson \&
  Bromm}{2007}]{johnson2007aftermath}
Johnson J.~L.,  Bromm V.,  2007, \mnras, 374, 1557

\bibitem[\protect\citeauthoryear{Johnson \& Haardt}{Johnson \&
  Haardt}{2016}]{johnson2016early}
Johnson J.~L.,  Haardt F.,  2016, \pasa, 33

\bibitem[\protect\citeauthoryear{Johnson, Greif  \& Bromm}{Johnson
  et~al.}{2007}]{johnson2007local}
Johnson J.~L.,  Greif T.~H.,   Bromm V.,  2007, \apj, 665, 85

\bibitem[\protect\citeauthoryear{{Katz}}{{Katz}}{1992}]{1992ApJ...391..502K}
{Katz} N.,  1992, \mn@doi [\apj] {10.1086/171366}, \href
  {https://ui.adsabs.harvard.edu/abs/1992ApJ...391..502K} {391, 502}

\bibitem[\protect\citeauthoryear{Katz, Sijacki  \& Haehnelt}{Katz
  et~al.}{2015}]{katz2015seeding}
Katz H.,  Sijacki D.,   Haehnelt M.~G.,  2015, \mnras, 451, 2352

\bibitem[\protect\citeauthoryear{Kim et~al.,}{Kim
  et~al.}{2018}]{kim2018formation}
Kim J.-h.,  et~al., 2018, \mnras, 474, 4232

\bibitem[\protect\citeauthoryear{Kinugawa, Inayoshi, Hotokezaka, Nakauchi  \&
  Nakamura}{Kinugawa et~al.}{2014}]{kinugawa2014possible}
Kinugawa T.,  Inayoshi K.,  Hotokezaka K.,  Nakauchi D.,   Nakamura T.,  2014,
  \mnras, 442, 2963

\bibitem[\protect\citeauthoryear{Kinugawa, Miyamoto, Kanda  \&
  Nakamura}{Kinugawa et~al.}{2015}]{kinugawa2015detection}
Kinugawa T.,  Miyamoto A.,  Kanda N.,   Nakamura T.,  2015, \mnras, 456, 1093

\bibitem[\protect\citeauthoryear{Kormendy \& Ho}{Kormendy \&
  Ho}{2013}]{kormendy2013coevolution}
Kormendy J.,  Ho L.~C.,  2013, \araa, 51, 511

\bibitem[\protect\citeauthoryear{Kremer et~al.,}{Kremer
  et~al.}{2019}]{kremer2019post}
Kremer K.,  et~al., 2019, \prd, 99, 063003

\bibitem[\protect\citeauthoryear{{Kuns}, {Yu}, {Chen}  \& {Adhikari}}{{Kuns}
  et~al.}{2019}]{tiango}
{Kuns} K.~A.,  {Yu} H.,  {Chen} Y.,   {Adhikari} R.~X.,  2019, arXiv e-prints,
  \href {https://ui.adsabs.harvard.edu/abs/2019arXiv190806004K} {p.
  arXiv:1908.06004}

\bibitem[\protect\citeauthoryear{{Latif}, {Khochfar}  \& {Whalen}}{{Latif}
  et~al.}{2020}]{latif2020}
{Latif} M.~A.,  {Khochfar} S.,   {Whalen} D.,  2020, arXiv e-prints, \href
  {https://ui.adsabs.harvard.edu/abs/2020arXiv200200983L} {p. arXiv:2002.00983}

\bibitem[\protect\citeauthoryear{{Liu} \& {Bromm}}{{Liu} \&
  {Bromm}}{2018}]{lithium}
{Liu} B.,  {Bromm} V.,  2018, \mn@doi [\mnras] {10.1093/mnras/sty350}, \href
  {http://adsabs.harvard.edu/abs/2018MNRAS.476.1826L} {476, 1826}

\bibitem[\protect\citeauthoryear{Liu, Jaacks, Finkelstein  \& Bromm}{Liu
  et~al.}{2019}]{liu2019global}
Liu B.,  Jaacks J.,  Finkelstein S.~L.,   Bromm V.,  2019, \mnras, 486, 3617

\bibitem[\protect\citeauthoryear{Lupi, Colpi, Devecchi, Galanti  \&
  Volonteri}{Lupi et~al.}{2014}]{lupi2014constraining}
Lupi A.,  Colpi M.,  Devecchi B.,  Galanti G.,   Volonteri M.,  2014, \mnras,
  442, 3616

\bibitem[\protect\citeauthoryear{Machida \& Nakamura}{Machida \&
  Nakamura}{2015}]{machida2015accretion}
Machida M.~N.,  Nakamura T.,  2015, \mnras, 448, 1405

\bibitem[\protect\citeauthoryear{{Madau} \& {Dickinson}}{{Madau} \&
  {Dickinson}}{2014}]{madau2014araa}
{Madau} P.,  {Dickinson} M.,  2014, \mn@doi [\araa]
  {10.1146/annurev-astro-081811-125615}, \href
  {https://ui.adsabs.harvard.edu/abs/2014ARA&A..52..415M} {52, 415}

\bibitem[\protect\citeauthoryear{Madau \& Rees}{Madau \&
  Rees}{2001}]{madau2001massive}
Madau P.,  Rees M.~J.,  2001, \apj, 551, L27

\bibitem[\protect\citeauthoryear{Madau, Haardt  \& Dotti}{Madau
  et~al.}{2014}]{madau2014super}
Madau P.,  Haardt F.,   Dotti M.,  2014, \apj, 784, L38

\bibitem[\protect\citeauthoryear{Maio}{Maio}{2019}]{maio2019early}
Maio U.,  2019, arXiv preprint arXiv:1908.04823

\bibitem[\protect\citeauthoryear{Mapelli, Giacobbo, Santoliquido  \&
  Artale}{Mapelli et~al.}{2019}]{mapelli2019properties}
Mapelli M.,  Giacobbo N.,  Santoliquido F.,   Artale M.~C.,  2019, \mnras, 487,
  2

\bibitem[\protect\citeauthoryear{Martynov et~al.,}{Martynov
  et~al.}{2016}]{martynov2016sensitivity}
Martynov D.~V.,  et~al., 2016, \prd, 93, 112004

\bibitem[\protect\citeauthoryear{{Mirocha} \& {Furlanetto}}{{Mirocha} \&
  {Furlanetto}}{2019}]{Mirocha2018}
{Mirocha} J.,  {Furlanetto} S.~R.,  2019, \mn@doi [\mnras]
  {10.1093/mnras/sty3260}, \href
  {https://ui.adsabs.harvard.edu/\#abs/2019MNRAS.483.1980M} {483, 1980}

\bibitem[\protect\citeauthoryear{Negri \& Volonteri}{Negri \&
  Volonteri}{2017}]{negri2017black}
Negri A.,  Volonteri M.,  2017, \mnras, 467, 3475

\bibitem[\protect\citeauthoryear{{O'Leary}, {Kocsis}  \& {Loeb}}{{O'Leary}
  et~al.}{2009}]{oleary2009gwnsc}
{O'Leary} R.~M.,  {Kocsis} B.,   {Loeb} A.,  2009, \mn@doi [\mnras]
  {10.1111/j.1365-2966.2009.14653.x}, \href
  {https://ui.adsabs.harvard.edu/abs/2009MNRAS.395.2127O} {395, 2127}

\bibitem[\protect\citeauthoryear{O'Leary, Meiron  \& Kocsis}{O'Leary
  et~al.}{2016}]{o2016dynamical}
O'Leary R.~M.,  Meiron Y.,   Kocsis B.,  2016, \apj, 824, L12

\bibitem[\protect\citeauthoryear{{Ogiya}, {Hahn}, {Mingarelli}  \&
  {Volonteri}}{{Ogiya} et~al.}{2019}]{ogiya2019}
{Ogiya} G.,  {Hahn} O.,  {Mingarelli} C. M.~F.,   {Volonteri} M.,  2019, arXiv
  e-prints, \href {https://ui.adsabs.harvard.edu/abs/2019arXiv191111526O} {p.
  arXiv:1911.11526}

\bibitem[\protect\citeauthoryear{Omukai, Schneider  \& Haiman}{Omukai
  et~al.}{2008}]{omukai2008can}
Omukai K.,  Schneider R.,   Haiman Z.,  2008, \apj, 686, 801

\bibitem[\protect\citeauthoryear{Perna, Wang, Farr, Leigh  \& Cantiello}{Perna
  et~al.}{2019}]{perna2019constraining}
Perna R.,  Wang Y.-H.,  Farr W.~M.,  Leigh N.,   Cantiello M.,  2019, \apj,
  878, L1

\bibitem[\protect\citeauthoryear{Petrovich \& Antonini}{Petrovich \&
  Antonini}{2017}]{petrovich2017greatly}
Petrovich C.,  Antonini F.,  2017, \apj, 846, 146

\bibitem[\protect\citeauthoryear{Pezzulli, Valiante  \& Schneider}{Pezzulli
  et~al.}{2016}]{pezzulli2016super}
Pezzulli E.,  Valiante R.,   Schneider R.,  2016, \mnras, 458, 3047

\bibitem[\protect\citeauthoryear{Pfister, Volonteri, Dubois, Dotti  \&
  Colpi}{Pfister et~al.}{2019}]{pfister2019erratic}
Pfister H.,  Volonteri M.,  Dubois Y.,  Dotti M.,   Colpi M.,  2019, \mnras,
  486, 101

\bibitem[\protect\citeauthoryear{{Planck Collaboration} et~al.,}{{Planck
  Collaboration} et~al.}{2016}]{planck}
{Planck Collaboration} et~al., 2016, \mn@doi [\aap]
  {10.1051/0004-6361/201525830}, \href
  {http://adsabs.harvard.edu/abs/2016A%26A...594A..13P} {594, A13}

\bibitem[\protect\citeauthoryear{Punturo et~al.,}{Punturo
  et~al.}{2010}]{punturo2010einstein}
Punturo M.,  et~al., 2010, Classical and Quantum Gravity, 27, 194002

\bibitem[\protect\citeauthoryear{Regan, Wise, O'Shea  \& Norman}{Regan
  et~al.}{2019}]{regan2019emergence}
Regan J.~A.,  Wise J.~H.,  O'Shea B.~W.,   Norman M.~L.,  2019, arXiv preprint
  arXiv:1908.02823

\bibitem[\protect\citeauthoryear{Reines \& Volonteri}{Reines \&
  Volonteri}{2015}]{reines2015relations}
Reines A.~E.,  Volonteri M.,  2015, \apj, 813, 82

\bibitem[\protect\citeauthoryear{Repetto, Davies  \& Sigurdsson}{Repetto
  et~al.}{2012}]{repetto2012investigating}
Repetto S.,  Davies M.~B.,   Sigurdsson S.,  2012, \mnras, 425, 2799

\bibitem[\protect\citeauthoryear{Ritter, Safranek-Shrader, Gnat,
  Milosavljevi{\'c}  \& Bromm}{Ritter et~al.}{2012}]{ritter2012confined}
Ritter J.~S.,  Safranek-Shrader C.,  Gnat O.,  Milosavljevi{\'c} M.,   Bromm
  V.,  2012, \apj, 761, 56

\bibitem[\protect\citeauthoryear{Ritter, Sluder, Safranek-Shrader,
  Milosavljevi{\'c}  \& Bromm}{Ritter et~al.}{2015}]{ritter2015metal}
Ritter J.~S.,  Sluder A.,  Safranek-Shrader C.,  Milosavljevi{\'c} M.,   Bromm
  V.,  2015, \mnras, 451, 1190

\bibitem[\protect\citeauthoryear{Robson, Cornish  \& Liug}{Robson
  et~al.}{2019}]{robson2019construction}
Robson T.,  Cornish N.~J.,   Liug C.,  2019, Classical and Quantum Gravity, 36,
  105011

\bibitem[\protect\citeauthoryear{Rodriguez \& Loeb}{Rodriguez \&
  Loeb}{2018}]{rodriguez2018redshift}
Rodriguez C.~L.,  Loeb A.,  2018, \apj, 866, L5

\bibitem[\protect\citeauthoryear{Rodriguez, Chatterjee  \& Rasio}{Rodriguez
  et~al.}{2016a}]{rodriguez2016binary}
Rodriguez C.~L.,  Chatterjee S.,   Rasio F.~A.,  2016a, \prd, 93, 084029

\bibitem[\protect\citeauthoryear{Rodriguez, Haster, Chatterjee, Kalogera  \&
  Rasio}{Rodriguez et~al.}{2016b}]{rodriguez2016dynamical}
Rodriguez C.~L.,  Haster C.-J.,  Chatterjee S.,  Kalogera V.,   Rasio F.~A.,
  2016b, \apj, 824, L8

\bibitem[\protect\citeauthoryear{Rodriguez, Amaro-Seoane, Chatterjee  \&
  Rasio}{Rodriguez et~al.}{2018}]{rodriguez2018post}
Rodriguez C.~L.,  Amaro-Seoane P.,  Chatterjee S.,   Rasio F.~A.,  2018, \prl,
  120, 151101

\bibitem[\protect\citeauthoryear{Ro{\v{s}}kar, Fiacconi, Mayer, Kazantzidis,
  Quinn  \& Wadsley}{Ro{\v{s}}kar et~al.}{2015}]{rovskar2015orbital}
Ro{\v{s}}kar R.,  Fiacconi D.,  Mayer L.,  Kazantzidis S.,  Quinn T.~R.,
  Wadsley J.,  2015, \mnras, 449, 494

\bibitem[\protect\citeauthoryear{Ryu, Tanaka  \& Perna}{Ryu
  et~al.}{2015}]{ryu2015formation}
Ryu T.,  Tanaka T.~L.,   Perna R.,  2015, \mnras, 456, 223

\bibitem[\protect\citeauthoryear{Safarzadeh}{Safarzadeh}{2020}]{safarzadeh2020branching}
Safarzadeh M.,  2020, \apj, 892, L8

\bibitem[\protect\citeauthoryear{Safarzadeh \& Berger}{Safarzadeh \&
  Berger}{2019}]{safarzadeh2019measuring}
Safarzadeh M.,  Berger E.,  2019, \apj, 878, L12

\bibitem[\protect\citeauthoryear{Safranek-Shrader, Bromm  \&
  Milosavljevi{\'c}}{Safranek-Shrader et~al.}{2010}]{safranek2010}
Safranek-Shrader C.,  Bromm V.,   Milosavljevi{\'c} M.,  2010, \apj, 723, 1568

\bibitem[\protect\citeauthoryear{Sakurai, Vorobyov, Hosokawa, Yoshida, Omukai
  \& Yorke}{Sakurai et~al.}{2016}]{sakurai2016supermassive}
Sakurai Y.,  Vorobyov E.~I.,  Hosokawa T.,  Yoshida N.,  Omukai K.,   Yorke
  H.~W.,  2016, \mnras, 459, 1137

\bibitem[\protect\citeauthoryear{{Sakurai}, {Yoshida}, {Fujii}  \&
  {Hirano}}{{Sakurai} et~al.}{2017}]{sakurai2017imbh}
{Sakurai} Y.,  {Yoshida} N.,  {Fujii} M.~S.,   {Hirano} S.,  2017, \mn@doi
  [\mnras] {10.1093/mnras/stx2044}, \href
  {https://ui.adsabs.harvard.edu/abs/2017MNRAS.472.1677S} {472, 1677}

\bibitem[\protect\citeauthoryear{Salvaterra, Haardt, Volonteri  \&
  Moretti}{Salvaterra et~al.}{2012}]{salvaterra2012limits}
Salvaterra R.,  Haardt F.,  Volonteri M.,   Moretti A.,  2012, \aap, 545, L6

\bibitem[\protect\citeauthoryear{Santamaria et~al.,}{Santamaria
  et~al.}{2010}]{santamaria2010matching}
Santamaria L.,  et~al., 2010, \prd, 82, 064016

\bibitem[\protect\citeauthoryear{Schauer, Liu  \& Bromm}{Schauer
  et~al.}{2019}]{schauer2019constraining}
Schauer A.~T.,  Liu B.,   Bromm V.,  2019, \apjl, 877, L5

\bibitem[\protect\citeauthoryear{Schleicher, Palla, Ferrara, Galli  \&
  Latif}{Schleicher et~al.}{2013}]{schleicher2013massive}
Schleicher D.~R.,  Palla F.,  Ferrara A.,  Galli D.,   Latif M.,  2013, \aap,
  558, A59

\bibitem[\protect\citeauthoryear{Schneider, Ferrara, Ciardi, Ferrari  \&
  Matarrese}{Schneider et~al.}{2000}]{schneider2000gravitational}
Schneider R.,  Ferrara A.,  Ciardi B.,  Ferrari V.,   Matarrese S.,  2000,
  \mnras, 317, 385

\bibitem[\protect\citeauthoryear{Schneider, Omukai, Bianchi  \&
  Valiante}{Schneider et~al.}{2011}]{schneider2011}
Schneider R.,  Omukai K.,  Bianchi S.,   Valiante R.,  2011, \mnras, 419, 1566

\bibitem[\protect\citeauthoryear{Sesana \& Khan}{Sesana \&
  Khan}{2015}]{sesana2015scattering}
Sesana A.,  Khan F.~M.,  2015, \mnras, 454, L66

\bibitem[\protect\citeauthoryear{Sesana, Gair, Mandel  \& Vecchio}{Sesana
  et~al.}{2009}]{sesana2009observing}
Sesana A.,  Gair J.,  Mandel I.,   Vecchio A.,  2009, The Astrophysical Journal
  Letters, 698, L129

\bibitem[\protect\citeauthoryear{Sesana, Gair, Berti  \& Volonteri}{Sesana
  et~al.}{2011}]{sesana2011reconstructing}
Sesana A.,  Gair J.,  Berti E.,   Volonteri M.,  2011, \prd, 83, 044036

\bibitem[\protect\citeauthoryear{Smith \& Bromm}{Smith \&
  Bromm}{2019}]{smith2019supermassive}
Smith A.,  Bromm V.,  2019, Contemporary Physics, pp 1--16

\bibitem[\protect\citeauthoryear{{Smith}, {Regan}, {Downes}, {Norman}, {O'Shea}
   \& {Wise}}{{Smith} et~al.}{2018}]{smith2018growth}
{Smith} B.~D.,  {Regan} J.~A.,  {Downes} T.~P.,  {Norman} M.~L.,  {O'Shea}
  B.~W.,   {Wise} J.~H.,  2018, \mn@doi [\mnras] {10.1093/mnras/sty2103}, \href
  {https://ui.adsabs.harvard.edu/abs/2018MNRAS.480.3762S} {480, 3762}

\bibitem[\protect\citeauthoryear{Springel}{Springel}{2005}]{springel2005cosmological}
Springel V.,  2005, \mnras, 364, 1105

\bibitem[\protect\citeauthoryear{{Springel} \& {Hernquist}}{{Springel} \&
  {Hernquist}}{2003}]{springel2003wind}
{Springel} V.,  {Hernquist} L.,  2003, \mn@doi [\mnras]
  {10.1046/j.1365-8711.2003.06206.x}, \href
  {https://ui.adsabs.harvard.edu/abs/2003MNRAS.339..289S} {339, 289}

\bibitem[\protect\citeauthoryear{Springel, Di~Matteo  \& Hernquist}{Springel
  et~al.}{2005}]{springel2005modelling}
Springel V.,  Di~Matteo T.,   Hernquist L.,  2005, \mnras, 361, 776

\bibitem[\protect\citeauthoryear{Stacy \& Bromm}{Stacy \&
  Bromm}{2013}]{stacy2013constraining}
Stacy A.,  Bromm V.,  2013, \mnras, 433, 1094

\bibitem[\protect\citeauthoryear{Stacy, Bromm  \& Lee}{Stacy
  et~al.}{2016}]{stacy2016building}
Stacy A.,  Bromm V.,   Lee A.~T.,  2016, \mnras, 462, 1307

\bibitem[\protect\citeauthoryear{Stinson, Seth, Katz, Wadsley, Governato  \&
  Quinn}{Stinson et~al.}{2006}]{stinson2006star}
Stinson G.,  Seth A.,  Katz N.,  Wadsley J.,  Governato F.,   Quinn T.,  2006,
  \mnras, 373, 1074

\bibitem[\protect\citeauthoryear{Sugimura, Matsumoto, Hosokawa, Hirano  \&
  Omukai}{Sugimura et~al.}{2020a}]{sugimura2020birth}
Sugimura K.,  Matsumoto T.,  Hosokawa T.,  Hirano S.,   Omukai K.,  2020a,
  arXiv preprint arXiv:2002.00012

\bibitem[\protect\citeauthoryear{{Sugimura}, {Matsumoto}, {Hosokawa}, {Hirano}
  \& {Omukai}}{{Sugimura} et~al.}{2020b}]{hosokawa2020}
{Sugimura} K.,  {Matsumoto} T.,  {Hosokawa} T.,  {Hirano} S.,   {Omukai} K.,
  2020b, \mn@doi [\apjl] {10.3847/2041-8213/ab7d37}, \href
  {https://ui.adsabs.harvard.edu/abs/2020ApJ...892L..14S} {892, L14}

\bibitem[\protect\citeauthoryear{Susa, Hasegawa  \& Tominaga}{Susa
  et~al.}{2014}]{susa2014mass}
Susa H.,  Hasegawa K.,   Tominaga N.,  2014, \apj, 792, 32

\bibitem[\protect\citeauthoryear{Tamfal, Capelo, Kazantzidis, Mayer, Potter,
  Stadel  \& Widrow}{Tamfal et~al.}{2018}]{tamfal2018formation}
Tamfal T.,  Capelo P.~R.,  Kazantzidis S.,  Mayer L.,  Potter D.,  Stadel J.,
  Widrow L.~M.,  2018, \apj, 864, L19

\bibitem[\protect\citeauthoryear{Tang, Eldridge, Stanway  \& Bray}{Tang
  et~al.}{2020}]{tang2020dependence}
Tang P.~N.,  Eldridge J.,  Stanway E.~R.,   Bray J.~C.,  2020, \mnras, 493, L6

\bibitem[\protect\citeauthoryear{Thompson}{Thompson}{2014}]{thompson2014pygadgetreader}
Thompson R.,  2014, Astrophysics Source Code Library

\bibitem[\protect\citeauthoryear{Toyouchi, Hosokawa, Sugimura, Nakatani  \&
  Kuiper}{Toyouchi et~al.}{2019}]{toyouchi2019super}
Toyouchi D.,  Hosokawa T.,  Sugimura K.,  Nakatani R.,   Kuiper R.,  2019,
  \mnras, 483, 2031

\bibitem[\protect\citeauthoryear{Tremmel, Governato, Volonteri  \&
  Quinn}{Tremmel et~al.}{2015}]{tremmel2015off}
Tremmel M.,  Governato F.,  Volonteri M.,   Quinn T.~R.,  2015, \mnras, 451,
  1868

\bibitem[\protect\citeauthoryear{Tremmel, Karcher, Governato, Volonteri, Quinn,
  Pontzen, Anderson  \& Bellovary}{Tremmel et~al.}{2017}]{tremmel2017romulus}
Tremmel M.,  Karcher M.,  Governato F.,  Volonteri M.,  Quinn T.,  Pontzen A.,
  Anderson L.,   Bellovary J.,  2017, \mnras, 470, 1121

\bibitem[\protect\citeauthoryear{Trenti \& Stiavelli}{Trenti \&
  Stiavelli}{2009}]{trenti2009formation}
Trenti M.,  Stiavelli M.,  2009, \apj, 694, 879

\bibitem[\protect\citeauthoryear{Turk, Smith, Oishi, Skory, Skillman, Abel  \&
  Norman}{Turk et~al.}{2010}]{turk2010yt}
Turk M.~J.,  Smith B.~D.,  Oishi J.~S.,  Skory S.,  Skillman S.~W.,  Abel T.,
  Norman M.~L.,  2010, \apjs, 192, 9

\bibitem[\protect\citeauthoryear{{Visbal}, {Haiman}  \& {Bryan}}{{Visbal}
  et~al.}{2015}]{visbal2015}
{Visbal} E.,  {Haiman} Z.,   {Bryan} G.~L.,  2015, \mn@doi [\mnras]
  {10.1093/mnras/stv1941}, \href
  {https://ui.adsabs.harvard.edu/abs/2015MNRAS.453.4456V} {453, 4456}

\bibitem[\protect\citeauthoryear{Vitale, Farr, Ng  \& Rodriguez}{Vitale
  et~al.}{2019}]{vitale2019measuring}
Vitale S.,  Farr W.~M.,  Ng K.~K.,   Rodriguez C.~L.,  2019, \apj, 886, L1

\bibitem[\protect\citeauthoryear{Volonteri}{Volonteri}{2010}]{volonteri2010formation}
Volonteri M.,  2010, \araa, 18, 279

\bibitem[\protect\citeauthoryear{Volonteri, Silk  \& Dubus}{Volonteri
  et~al.}{2015}]{volonteri2015case}
Volonteri M.,  Silk J.,   Dubus G.,  2015, \apj, 804, 148

\bibitem[\protect\citeauthoryear{Wei, Wu, Melia, Wei  \& Feng}{Wei
  et~al.}{2014}]{wei2014cosmological}
Wei J.-J.,  Wu X.-F.,  Melia F.,  Wei D.-M.,   Feng L.-L.,  2014, \mnras, 439,
  3329

\bibitem[\protect\citeauthoryear{Whalen \& Fryer}{Whalen \&
  Fryer}{2012}]{whalen2012formation}
Whalen D.~J.,  Fryer C.~L.,  2012, \apjl, 756, L19

\bibitem[\protect\citeauthoryear{{Wise}, {Regan}, {O'Shea}, {Norman}, {Downes}
  \& {Xu}}{{Wise} et~al.}{2019}]{wise2019formation}
{Wise} J.~H.,  {Regan} J.~A.,  {O'Shea} B.~W.,  {Norman} M.~L.,  {Downes}
  T.~P.,   {Xu} H.,  2019, \mn@doi [\nat] {10.1038/s41586-019-0873-4}, \href
  {https://ui.adsabs.harvard.edu/abs/2019Natur.566...85W} {566, 85}

\bibitem[\protect\citeauthoryear{Wolcott-Green \& Haiman}{Wolcott-Green \&
  Haiman}{2011}]{wolcott2011suppression}
Wolcott-Green J.,  Haiman Z.,  2011, \mnras, 412, 2603

\bibitem[\protect\citeauthoryear{Wolcott-Green, Haiman  \& Bryan}{Wolcott-Green
  et~al.}{2011}]{wolcott2011photodissociation}
Wolcott-Green J.,  Haiman Z.,   Bryan G.~L.,  2011, \mnras, 418, 838

\bibitem[\protect\citeauthoryear{Yoshida, Abel, Hernquist  \& Sugiyama}{Yoshida
  et~al.}{2003}]{yoshida2003simulations}
Yoshida N.,  Abel T.,  Hernquist L.,   Sugiyama N.,  2003, \apj, 592, 645

\bibitem[\protect\citeauthoryear{{de Souza}, {Yoshida}  \& {Ioka}}{{de Souza}
  et~al.}{2011}]{deS2011}
{de Souza} R.~S.,  {Yoshida} N.,   {Ioka} K.,  2011, \mn@doi [\aap]
  {10.1051/0004-6361/201117242}, \href
  {https://ui.adsabs.harvard.edu/abs/2011A&A...533A..32D} {533, A32}

\makeatother
\end{thebibliography}

\appendix
\section{Effect of stellar feedback}
\label{a1}

\begin{figure}
\includegraphics[width=1\columnwidth]{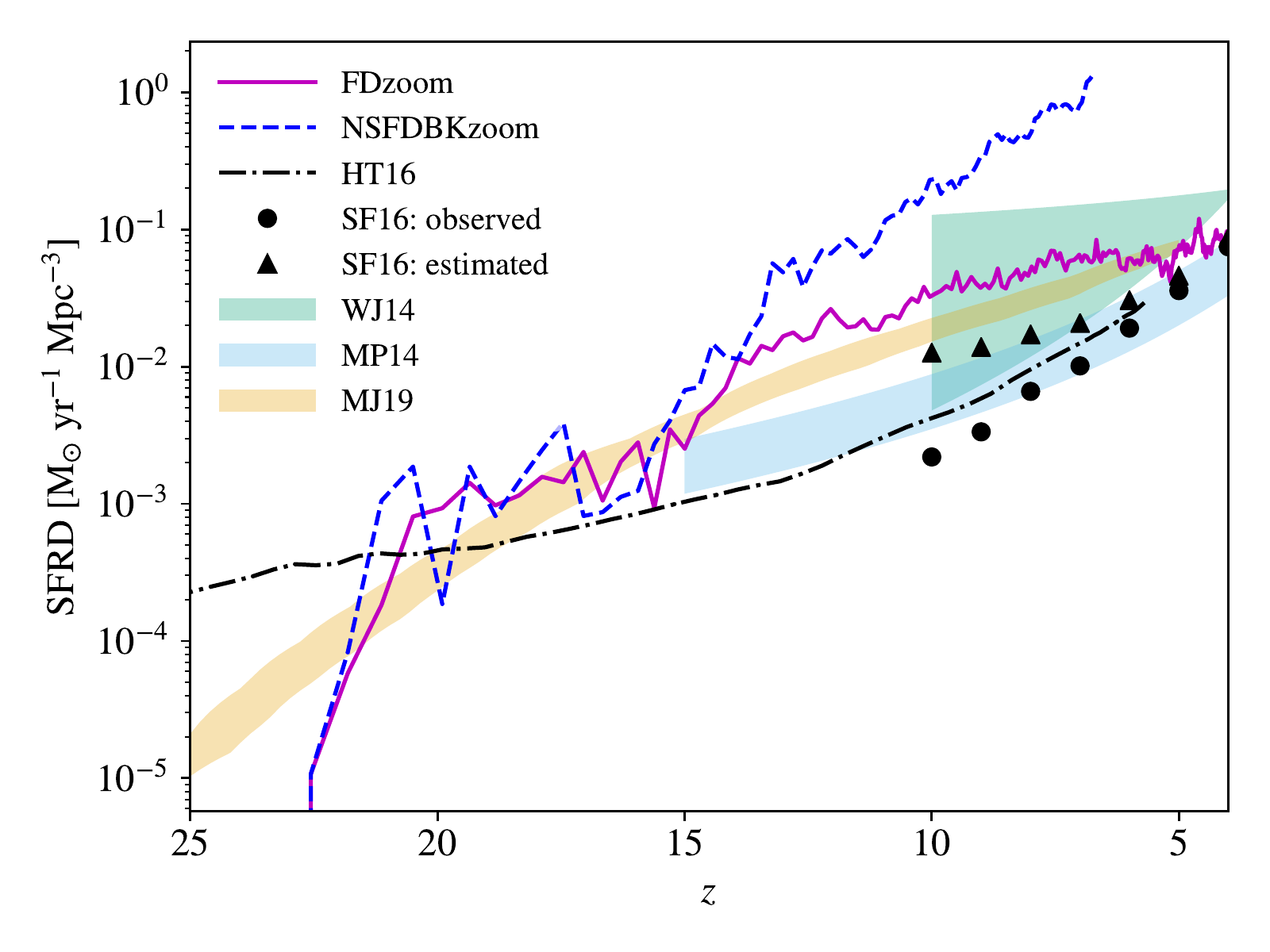}
\caption{Total (co-moving) SFRD from \texttt{FDzoom\_Hseed} (solid) and \texttt{NSFDBKzoom\_Hseed} (dashed). We also plot the corresponding results in the literature for comparison, which are described in detail in the caption of Fig.~\ref{fsfh}. Without PI heating and SN-driven winds from Pop~II stars, the total SFRD is enhanced by a factor of $\sim$10 at $z\lesssim 14$, becoming much higher than observational constraints at $z\lesssim 10$.}%
\label{fa1}
\end{figure}

To demonstrate the effect of stellar feedback, we focus on the star formation and BH accretion histories, as well as GW signals, in the \texttt{zoom} setup with the \texttt{Hseed} seeding scenario. 
Without PI heating and SN-driven winds from Pop~II stars, the total SFRD is enhanced by about one order of magnitude at $z\lesssim 14$, becoming much higher than the observational constraints at $z\lesssim 10$, as shown in Fig.~\ref{fa1}. This leads to a factor of 10 increase in the Pop~II stellar density by $z=7$. However, the Pop~III SFRD remains almost unchanged (thus not shown). This is reasonable as we do not turn off the Pop~III feedback and LW field for \texttt{NSFDBK} which play the most important roles in regulating Pop~III SF. Another factor is that the LW radiation is enhanced by the increased Pop~II stars for \texttt{NSFDBK}, which compensates the lack of Pop~II winds and PI heating. 

\begin{figure}
\includegraphics[width=1\columnwidth]{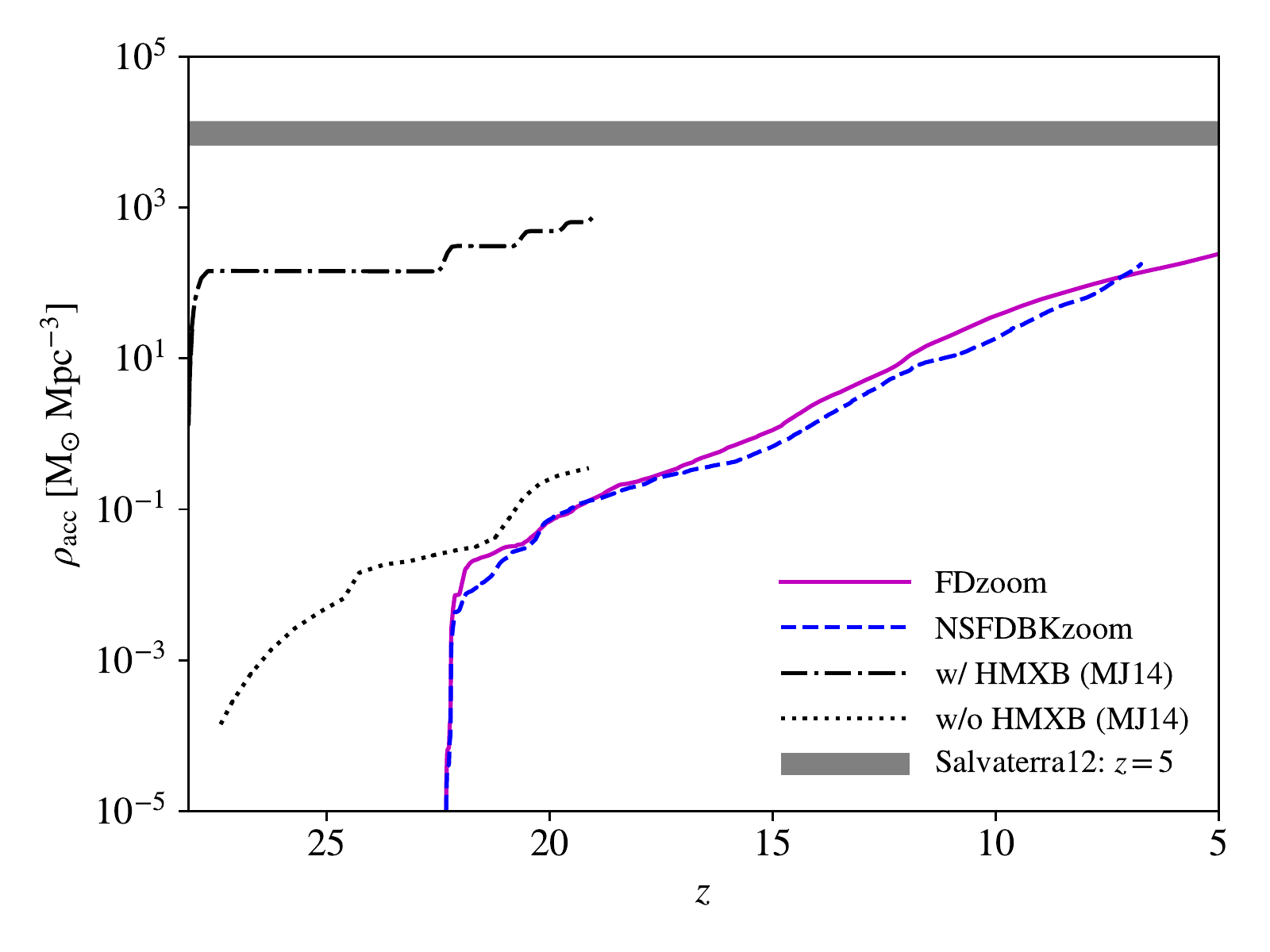}
\caption{Total (co-moving) accreted mass densities of BHs as functions of redshift in \texttt{FDzoom\_Hseed} (solid) and \texttt{NSFDBKzoom\_Hseed} (dashed). The shaded region shows the upper limit on $\rho_{\mathrm{acc}}$ at $z=5$, placed by the unresolved cosmic x-ray background, from \citealt{salvaterra2012limits} (Salvaterra12). For comparison, we also show the results of the zoom-in simulation in \citealt{jeon2014radiative} (MJ14) with (dashed-dotted) and without (dotted) HMXBs.}
\label{fa2}
\end{figure}

Fig.~\ref{fa2} shows the global accretion histories of BHs in \texttt{FDzoom\_Hseed} (solid) and \texttt{NSFDBKzoom\_Hseed} (dashed). Interestingly, BH accretion also remains almost unchanged when Pop~II stellar feedback is turned off. As mentioned before, since Pop~III SFRD is identical in \texttt{NSFDBKzoom\_Hseed}, the gaseous environments of BHs must also be similar to those in \texttt{FDzoom\_Hseed} such that cold gas is unavailable for BHs, and accretion is slow. Our explanation is that without Pop~II winds and PI heating, even though the cold gas is not heated and blown away, it is efficiently turned into stars, still unable to enhance BH accretion. 

\begin{figure}
\includegraphics[width=1.0\columnwidth]{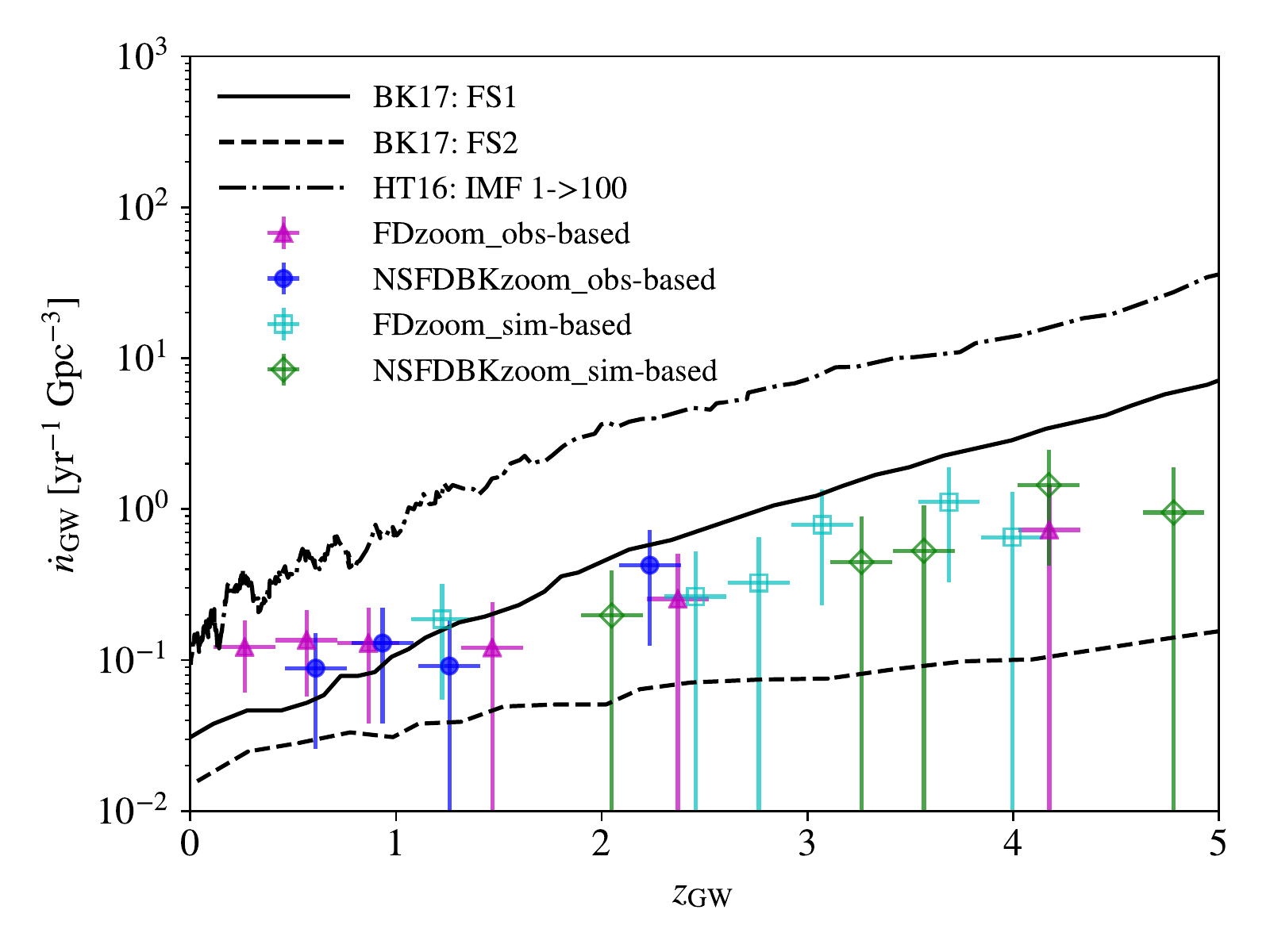}
\caption{Co-moving rest-frame rate densities of GW events from ex-situ BBHs formed at $z_{\mathrm{BBH}}> 7$ in \texttt{FDzoom\_Hseed} and \texttt{NSFDBKzoom\_Hseed}, measured with a bin size $\Delta z_{\mathrm{GW}}=0.3$. The results under the \texttt{obs-based} scaling relations with $\gamma=1.5$ are shown with solid data points (triangles and circles for \texttt{FDzoom\_Hseed} and \texttt{NSFDBKzoom\_Hseed}). While the results under the \texttt{sim-based} scaling relations with $\gamma=1.5$ and $f_{\mathrm{bulge}}=1.0$ are shown with empty data points (squares and diamonds for \texttt{FDzoom\_Hseed} and \texttt{NSFDBKzoom\_Hseed}). Following Fig.~\ref{fgw1}, we also show the original rate densities for in-situ BBHs in the literature.}
\label{fa3}
\end{figure}

Is has been shown above that the (Pop~II) star formation history for \texttt{NSFDBKzoom\_Hseed} is in conflict with observational constraints, which renders the corresponding GW signals as unphysical. However, it is still interesting to evaluate the impact on GW signals from Pop~II feedback. Here we look into the optimistic cases of \texttt{obs-based} scaling relations with $\gamma=1.5$ and \texttt{sim-based} scaling relatoins with $\gamma=1.5$ and $f_{\mathrm{bulge}}=1.0$. Fig.~\ref{fa3} shows the intrinsic rate densities of GW events from ex-situ BBHs formed at $z>7$. The corresponding detection rates for AdLIGO (by design), ETxylophone and LISA are shown in Table~\ref{t3}. It turns out that the GW rates are not sensitive to Pop~II feedback. The differences in detection rates are generally within $\sim 50\%$ between \texttt{NSFDBKzoom\_Hseed} and \texttt{FDzoom\_Hseed}, and almost negligible for the case of \texttt{obs-based} scaling relations. In principle, there is additional dynamical friction from Pop~II stellar particles (whose numbers are increased by 10 times at $z=7$) in \texttt{NSFDBKzoom\_Hseed}, that may facilitate dynamical capture. However, this effect appears to be weak. The reason is that more than half of the ex-situ BBHs in our simulations are formed inside the same minihaloes, reflecting dynamical captures of the stellar remnants formed in the primary and secondary collapsed gas clumps, such that dynamical fraction from Pop~II stars does not play an important role in BBH formation.

\begin{table}
\caption{Detection rates per year and percentages (in brackets) of simulated GW events with $\mathrm{SNR}>10$ from ex-situ BBHs formed at $z_{\mathrm{BBH}}> 7$ in \texttt{FDzoom\_Hseed} and \texttt{NSFDBKzoom\_Hseed}, under the \texttt{obs-based} scaling relations with $\gamma=1.5$ and \texttt{sim-based} scaling relations with $\gamma=1.5$ and $f_{\mathrm{bulge}}=1.0$. The detection percentage reflects the ratio of the number of sources with $\mathrm{SNR}>10$ and the total number of sources (at $z_{\mathrm{GW}}\ge 0$). The first column is the flag \texttt{FDBKPopII}, indicating whether Pop~II feedback is included. See Sec.~\ref{s4.4} for descriptions of the detectors considered here.
}
\begin{tabular}{ccccc}
\hline
& \texttt{obs-based} & $\gamma=1.5$\\
\texttt{FDBKPopII} & AdLIGO & ETxylophone & LISA\\
\cmark & 5.9 (54\%) & 30 (85\%) & 30 (85\%) \\
\xmark & 7 (43\%) & 39 (100\%) & 39 (100\%)\\
\hline
& \texttt{sim-based} & $\gamma=1.5$ & $f_{\mathrm{bulge}}=1.0$\\
\texttt{FDBKPopII} & AdLIGO & ETxylophone & LISA\\
\cmark & 0 & 143 (77\%) & 11 (15\%)\\
\xmark & 0 & 104 (86\%) & 10.1 (14\%) \\
\hline
\end{tabular}
\label{t3}
\end{table}


\section{Numerical convergence}
\label{a2}
\begin{figure}
\includegraphics[width=1\columnwidth]{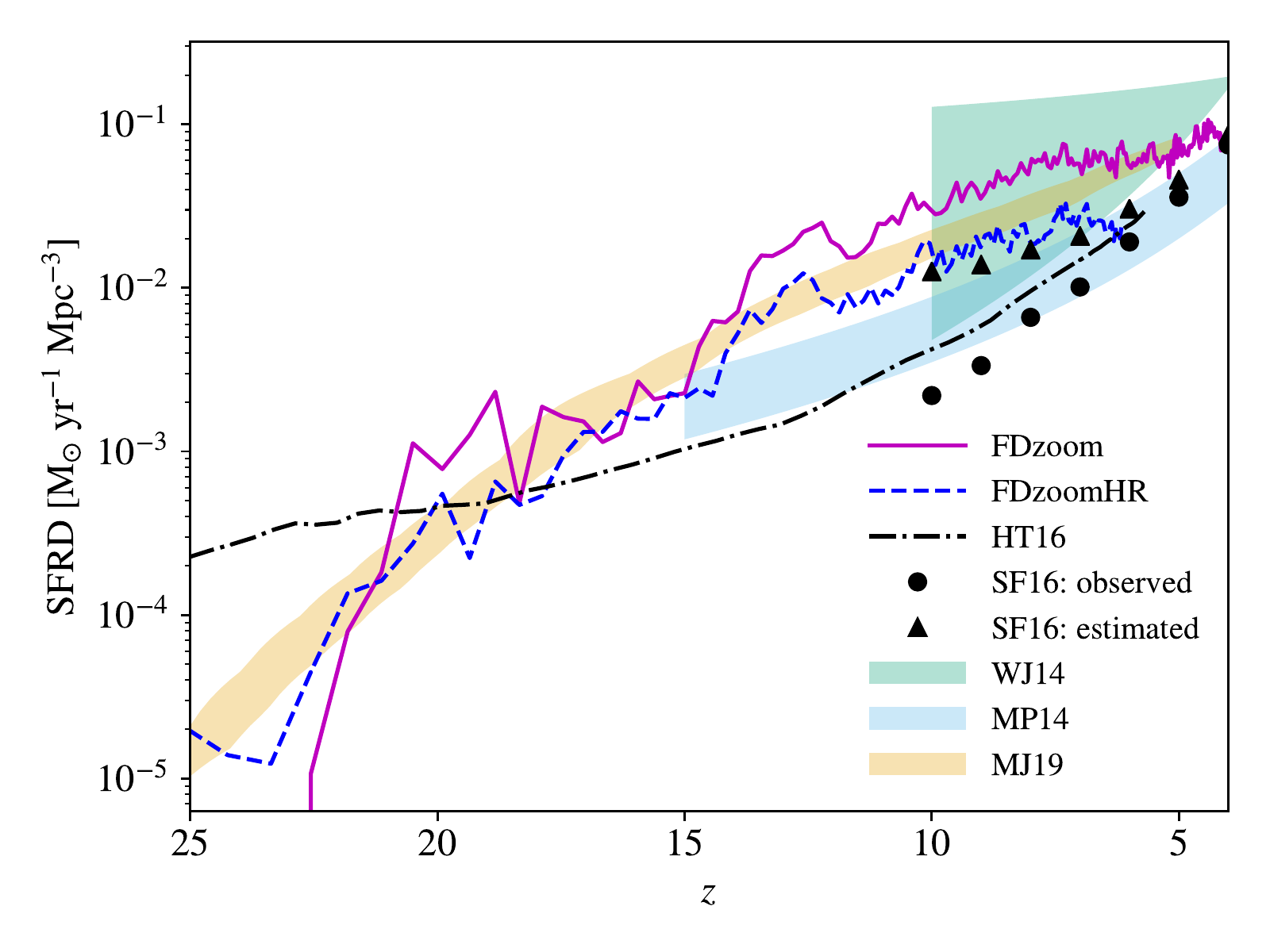}
\caption{Total (co-moving) SFRD from \texttt{FDzoom\_Lseed} (solid) and \texttt{FDzoomHR\_Lseed} (dashed). We also plot the corresponding results in the literature for comparison, which are described in detail in the caption of Fig.~\ref{fsfh}. With higher resolution for gas and DM, the total SFRD is reduced by a factor of 2 at $z\lesssim 14$.}%
\label{fa41}
\end{figure}

\begin{figure}
\includegraphics[width=1\columnwidth]{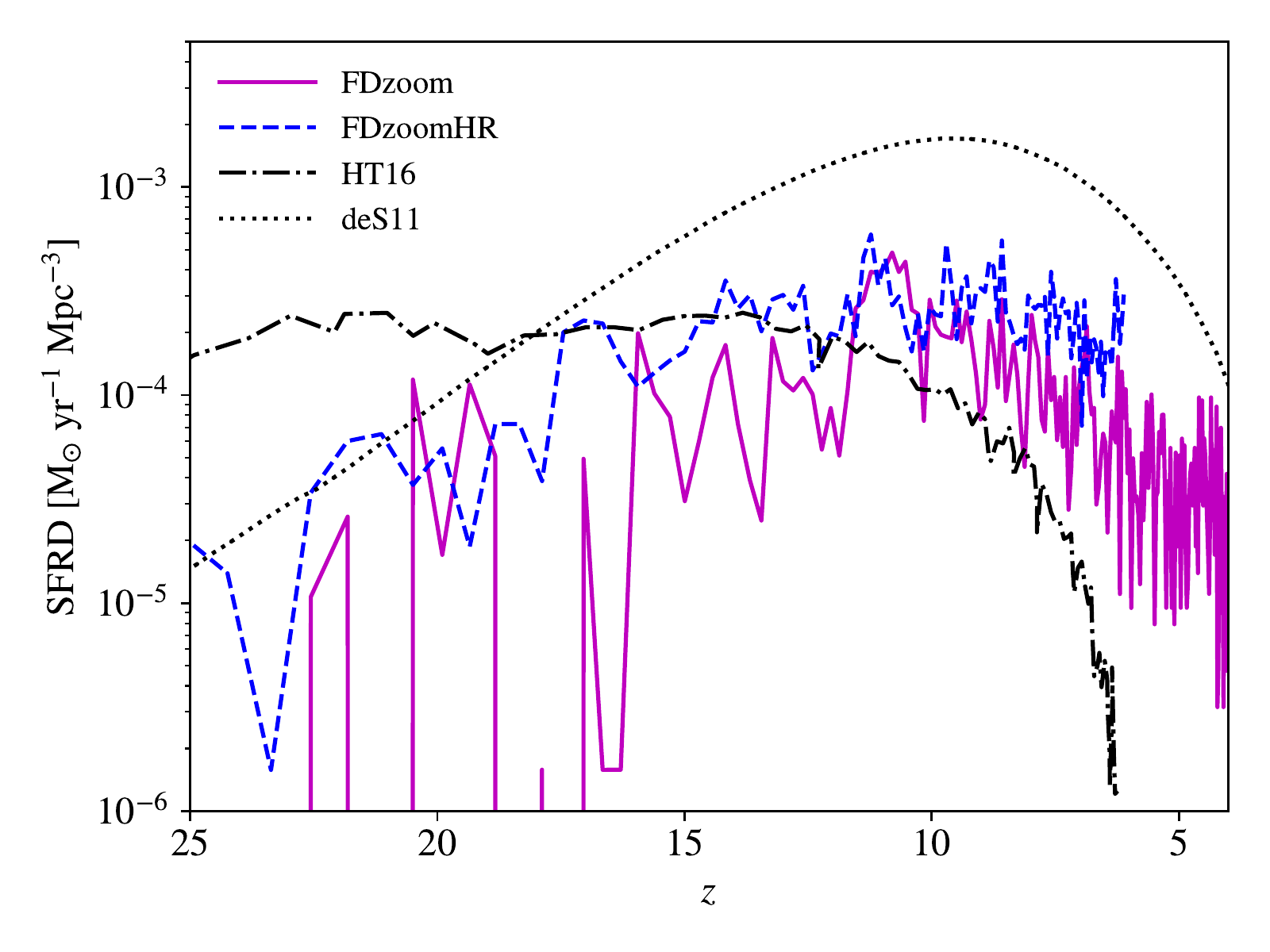}
\caption{Pop~III (co-moving) SFRD from \texttt{FDzoom\_Lseed} (solid) and \texttt{FDzoomHR\_Lseed} (dashed). We also plot the corresponding results in the literature for comparison, which are described in detail in the caption of Fig.~\ref{fsfh}. With higher resolution for gas and DM, the Pop~III SFRD is moderately increased, especially at $z\gtrsim 12$, such that the total density of Pop~III stars formed by $z=7$ is enhanced by a factor of 2.}%
\label{fa42}
\end{figure}

\begin{figure}
\includegraphics[width=1\columnwidth]{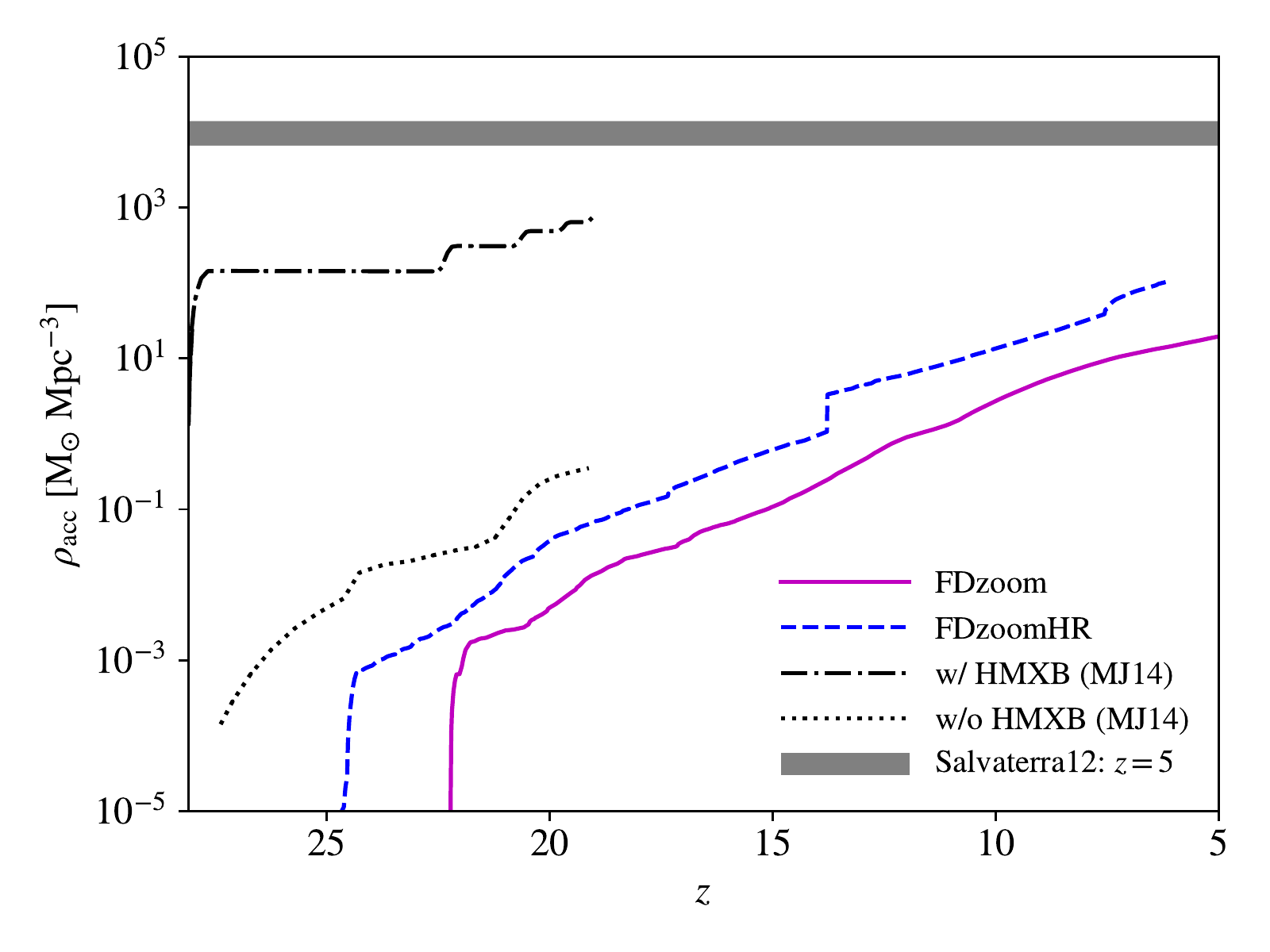}
\caption{Total (co-moving) accreted mass densities of BHs as functions of redshift in \texttt{FDzoom\_Lseed} (solid) and \texttt{FDzoomHR\_Lseed} (dashed). The shaded region shows the upper limit on $\rho_{\mathrm{acc}}$ at $z=5$, placed by the unresolved cosmic x-ray background, from \citealt{salvaterra2012limits} (Salvaterra12). For comparison, we also show the results of the zoom-in simulation in \citealt{jeon2014radiative} (MJ14) with (dashed-dotted) and without (dotted) HMXBs.}
\label{fa5}
\end{figure}

Similar comparisons as those in appendix~\ref{a1} are made for \texttt{FDzoom\_Lseed} and \texttt{FDzoomHR\_Lseed} to test numerical convergence, as shown in Fig.~\ref{fa41}-\ref{fa5} and Table~\ref{t4}. Generally speaking, the code shows good enough convergence in star formation and BH accretion histories, as well as GW signals. 
With higher resolution for gas and DM, the total SFRD is reduced by a factor of 2 at $z\lesssim 14$ (see Fig.~\ref{fa41}), showing better agreement with observations, while the Pop~III SFRD is moderately increased, especially at $z\gtrsim 12$, such that the total density of Pop~III stars formed by $z=7$ is enhanced by a factor of 2 (see Fig.~\ref{fa42}). These rather small differences can be explained by more efficient winds, especially at small scales\footnote{As indicated in equation~(\ref{e9}), the wind launching probability is approximately proportional to $m_{\star}/m_{\mathrm{SF}}$. In the \texttt{HR} run, $m_{\star}/m_{\mathrm{SF}}$ is increased by up to a factor of 8, so that wind launching becomes more immediate and efficient.}, more resolved minihaloes hosting Pop~III stars at the low-mass end\footnote{Under the fiducial resolution, the smallest resolved structures with $\sim$32 DM particles have masses $\sim 2\times 10^{6}\ \mathrm{M_{\odot}}$, comparable to the $\mathrm{H}_{2}$ cooling threshold for Pop~III SF, which means that some low-mass Pop~III-hosting minihaloes are likely unresolved. In the \texttt{HR} run, these haloes will be resolved, leading to additional Pop~III star formation.} and weaker LW feedback due to the reduced Pop~II SF. 
The \texttt{HR} accreted mass density is higher than its low-resolution counterpart by a factor of $\sim5$ for $z\lesssim 22$ (see Fig.~\ref{fa5}), caused by the enhancement in Pop~III SF and the nature of sub-grid Bondi accretion.

In the optimistic cases of \texttt{obs-based} scaling relations with $\gamma=1.5$ and \texttt{sim-based} scaling relatoins with $\gamma=1.5$ and $f_{\mathrm{bulge}}=1.0$, 
the total number of GW sources in \texttt{FDzoomHR\_Lseed} is $N_{\mathrm{merger,HR}}=12$, 3 times the number $N_{\mathrm{merger}}=4$ in \texttt{FDzoom\_Lseed}. The detection rates for ETxylophone and DOoptimal are also higher by up to a factor of $\sim 3$, which is reasonable since these instruments can reach almost all sources. While for AdLIGO and LISA, the differences can be larger but suffer from large statistical uncertainties. The samples of GW sources are too small to tell whether such differences in detection rates imply poor numerical convergence. The increase of GW events in the high-resolution run can be explained by the enhanced Pop~III SF and the better resolved dynamical friction from gas.

\begin{table}
\caption{Detection rates per year and percentages (in brackets) of simulated GW events with $\mathrm{SNR}>10$ from ex-situ BBHs formed at $z_{\mathrm{BBH}}> 7$ in \texttt{FDzoom\_Lseed} and \texttt{FDzoomHR\_Lseed}, under the \texttt{obs-based} scaling relations with $\gamma=1.5$ and \texttt{sim-based} scaling relations with $\gamma=1.5$ and $f_{\mathrm{bulge}}=1.0$. The detection percentage reflects the ratio of the number of sources with $\mathrm{SNR}>10$ and the total number of sources (at $z_{\mathrm{GW}}\ge 0$). The first column is the flag \texttt{HR}, indicating whether the resolution for gas and dark matter is increased. See Sec.~\ref{s4.4} for descriptions of the detectors considered here.
}
\begin{tabular}{ccccc}
\hline
& \texttt{obs-based} & $\gamma=1.5$\\
\texttt{HR} & AdLIGO & ETxylophone & DOoptimal & LISA\\
\xmark & 16 (75\%) & 27 (100\%) & 27 (100\%) & 1.5 (25\%) \\
\cmark & 34 (83\%) & 66 (100\%) & 66 (100\%) & 8.4 (50\%)\\
\hline
& \texttt{sim-based} & $\gamma=1.5$ & $f_{\mathrm{bulge}}=1.0$\\
\texttt{HR} & AdLIGO & ETxylophone & DOoptimal & LISA\\
\xmark & 0 & 93 (100\%) & 93 (100\%) & 0\\
\cmark & 0 & 337 (100\%) & 337 (100\%) & 0 \\
\hline
\end{tabular}
\label{t4}
\end{table}

\bsp

\label{lastpage}
\end{document}